\documentclass[10pt]{article}
\usepackage{graphicx}

\newcommand\pubnumber{SLAC-PUB-13079}
\newcommand\pubdate{January, 2008}

\def\SLAC{Stanford Linear Accelerator Center, Stanford University\\
    2575 Sand Hill Road, Menlo Park, California 94025  USA}
\def\doeack{\footnote{Work supported by the US Department of Energy,
                     contract DE--AC02--76SF00515.}}

\def\Title#1{\begin{center} {\Large #1 } \end{center}}
\def\Author#1{\begin{center}{ \sc #1} \end{center}}
\def\Address#1{\begin{center}{ \it #1} \end{center}}

\newcommand\pubblock{\rightline{\begin{tabular}{l} \pubnumber\\
         \pubdate \end{tabular}}}
\newenvironment{Abstract}{\begin{quotation} \begin{center}
                       ABSTRACT
     \end{center}\bigskip  }{\end{quotation}}
\newenvironment{Presented}{\begin{quotation} 
      \begin{center}}{\end{center} \end{quotation}}

\def\Acknowledgements{\bigskip  \bigskip \begin{center} \begin{large}
             \bf ACKNOWLEDGEMENTS \end{large}\end{center}}

\def\beq{\begin{equation}}
\def\eeq#1{\label{#1}\end{equation}}
\def\eeqn{\end{equation}}

\newenvironment{Eqnarray}%
   {\arraycolsep 0.14em\begin{eqnarray}}{\end{eqnarray}}
\def\beqa{\begin{Eqnarray}}
\def\eeqa#1{\label{#1}\end{Eqnarray}}
\def\eeqan{\end{Eqnarray}}
\def\CR{\nonumber \\ }

\def\leqn#1{(\ref{#1})}

\let\bar=\overbar

\def\etal{{\it et al.}}

\def\eg{{\it e.g.}}

\def\VEV#1{\left\langle{ #1} \right\rangle}
\def\bra#1{\left\langle{ #1} \right|}
\def\ket#1{\left| {#1} \right\rangle}

\def\lsim{\mathrel{\raise.3ex\hbox{$<$\kern-.75em\lower1ex\hbox{$\sim$}}}}
\def\gsim{\mathrel{\raise.3ex\hbox{$>$\kern-.75em\lower1ex\hbox{$\sim$}}}}

\def\D{{\cal D}}
\def\L{{\cal L}}
\def\M{{\cal M}}

\def\tr{{\mbox{\rm tr}}}
\def\half{\frac{1}{2}}

\def\third{\frac{1}{3}}

\def\del{\partial}
\def\Dslash{\not{\hbox{\kern-4pt $D$}}}
\def\Eslash{\not{\hbox{\kern-4pt $E$}}}
\def\dslash{\not{\hbox{\kern-2pt $\del$}}}

\def\Re{{ \mbox{\rm Re} }}

\def\Pl{{\mbox{\scriptsize Pl}}}

\def\CM{{\mbox{\scriptsize CM}}}

\def\ee{e^+e^-}

\def\mz{m_Z}

\def\mw{m_W}
\def\mt{m_t}

\def\msb{{\bar{\ssstyle M \kern -1pt S}}}

\def\s#1{\widetilde{#1}}

\makeatletter
\def\section{\@startsection{section}{0}{\z@}{5.5ex plus .5ex minus
 1.5ex}{2.3ex plus .2ex}{\large\bf}}
\def\subsection{\@startsection{subsection}{1}{\z@}{3.5ex plus .5ex minus
 1.5ex}{1.3ex plus .2ex}{\normalsize\bf}}
\def\subsubsection{\@startsection{subsubsection}{2}{\z@}{-3.5ex plus
-1ex minus  -.2ex}{2.3ex plus .2ex}{\normalsize\sl}}

\renewcommand{\@makecaption}[2]{%
   \vskip 10pt
   \setbox\@tempboxa\hbox{\small #1: #2}
   \ifdim \wd\@tempboxa >\hsize     
       \small #1: #2\par          
     \else                        
       \hbox to\hsize{\hfil\box\@tempboxa\hfil}
   \fi}

 \def\citenum#1{{\def\@cite##1##2{##1}\cite{#1}}}
 
\newcount\@tempcntc
\def\@citex[#1]#2{\if@filesw\immediate\write\@auxout{\string\citation{#2}}\fi
  \@tempcnta\z@\@tempcntb\m@ne\def\@citea{}\@cite{\@for\@citeb:=#2\do
    {\@ifundefined
       {b@\@citeb}{\@citeo\@tempcntb\m@ne\@citea\def\@citea{,}{\bf ?}\@warning
       {Citation `\@citeb' on page \thepage \space undefined}}%
    {\setbox\z@\hbox{\global\@tempcntc0\csname b@\@citeb\endcsname\relax}%
     \ifnum\@tempcntc=\z@ \@citeo\@tempcntb\m@ne
       \@citea\def\@citea{,}\hbox{\csname b@\@citeb\endcsname}%
     \else
      \advance\@tempcntb\@ne
      \ifnum\@tempcntb=\@tempcntc
      \else\advance\@tempcntb\m@ne\@citeo
      \@tempcnta\@tempcntc\@tempcntb\@tempcntc\fi\fi}}\@citeo}{#1}}
\def\@citeo{\ifnum\@tempcnta>\@tempcntb\else\@citea\def\@citea{,}%
  \ifnum\@tempcnta=\@tempcntb\the\@tempcnta\else
  {\advance\@tempcnta\@ne\ifnum\@tempcnta=\@tempcntb \else\def\@citea{--}\fi
    \advance\@tempcnta\m@ne\the\@tempcnta\@citea\the\@tempcntb}\fi\fi}
\makeatother

\begin{document}
\begin{titlepage} 
\pubblock

\vfill
\Title{Supersymmetry in Elementary Particle Physics}
\vfill
\Author{Michael E. Peskin\doeack}
\Address{\SLAC}
\vfill
\begin{Abstract}
These lectures give a general introduction to supersymmetry, emphasizing 
its application to models of elementary particle physics at the 100~GeV
energy scale.  I discuss the following topics: 
 the construction of supersymmetric Lagrangians 
with scalars, fermions, and gauge bosons, the structure and 
mass spectrum of the 
Minimal Supersymmetric Standard Model (MSSM), 
the measurement of the parameters
of the MSSM at high-energy colliders, and the solutions that the 
MSSM gives to the problems of electroweak symmetry breaking and dark 
matter.
\end{Abstract}
\vfill
\begin{Presented}
lectures presented at the 2006 TASI Summer School \\
 Boulder, Colorado, June 4--30, 2006
\end{Presented}
\vfill
\end{titlepage}
\def\thefootnote{\fnsymbol{footnote}}
\setcounter{footnote}{0}
\mbox{\quad }
\newpage
\tableofcontents
\newpage
\setcounter{page}{1}

\section{Introduction}

\subsection{Overview}

It is an exciting time now in high-energy physics.  For many years, ever
since the Standard Model was established in the late 1970's, the next 
logical question in the search for the basic laws of physics has been that 
of the mechanism by which the weak interaction gauge symmetry is 
spontaneously broken.  This seemed at the time the one important gap 
that kept the Standard Model from being a complete theory of the  strong,
weak,and electromagnetic interactions~\cite{Linde,Susskind,Weinberg}.  
Thirty years later, after many
precision experiments at high-energy $\ee$ and hadron colliders, this is  
still our situation. In the meantime, another important puzzle has been 
recognized, the fact that 80\% of the mass in the universe is composed of
`dark matter',
a particle species not included in the Standard Model.  Both problems
are likely to be solved by new fundamental interactions operating in the
energy range of a few hundred GeV. 
Up to now, there is no evidence from particle physics
for such new interactions.
But, in 
the next few years, this situation should change 
dramatically.  Beginning in 2008,
 the CERN Large Hadron Collider (LHC) should give us access
to physics at energies well above 1~TeV and thus should probe the energy
region responsible for electroweak symmetry breaking.  Over a longer term,
we can look forward to precision experiments in $\ee$ annihilation in this 
same energy region at the proposed International Linear Collider (ILC).

Given this expectation, it is important for all students of elementary particle
physics to form concrete ideas of what new phenomena we might find as we 
explore this new energy region.  Of course, we have no way of knowing 
exactly what we will find there.  But this makes it all the more important 
to study the alternative theories that have been put forward and to understand
their problems and virtues.

Many different models of new physics relevant to electroweak symmetry 
breaking are being discussed at this TASI school.  Among these, 
supersymmetry has pride of place.   Supersymmetry (or SUSY) 
provides an explicit
realization of all of the aspects of new physics expected in the 
hundred GeV energy region. Because SUSY requires only weak 
interactions to build a realistic theory, it is  possible in a model
with  SUSY to carry out explicit calculations and find the 
answers that the model gives 
to all relevant phenomenological questions.

In these lectures, I will give an introduction to supersymmetry as a 
context for building models
of new physics associated with electroweak symmetry breaking.
Here is an outline of the material:  In Section 2, I will develop
appropriate notation and then  construct supersymmetric Lagrangians for 
scalar, spinor, and vector fields.  In Section 3, I  will define the canonical 
phenomenological model of supersymmetry, the Minimal Supersymmetric
Standard Model (MSSM).   I will discuss the quantum numbers of 
new particles in the 
MSSM and the connection of the MSSM to the idea of 
grand unification.

The remaining sections of these lectures will map out the phenomenology 
of the new particles and interactions expected in models of supersymmetry.
I caution you that I will draw only those parts of the map that cover 
the simplest and most well-studied class of models.  Supersymmetry has an
enormous parameter space which contains many different scenarios for particle
physics, more than I have room to cover here.  I will at least try to 
indicate the possible branches in the path and give references that will 
help you follow some of the alternative routes.

With this restriction, the remaining sections will proceed as follows:
In Section 4, I will 
compute the mass spectrum of the MSSM from its parameters.  
I will also discuss the parameters of the MSSM that characterize
supersymmetry breaking.  In Section 5, I will describe how the MSSM 
parameters will be measured at the LHC and the ILC.
Finally, 
Section 6 will discuss the answers that supersymmetry
gives to the two major questions posed at the beginning of this discussion,
the origin of electroweak symmetry breaking, and the origin of cosmic 
dark matter.

Although I hope that these lectures will be useful to students in 
studying supersymmetry, there are many other excellent treatments of the
subject available.  A highly recommended introduction to SUSY
is the `Supersymmetry Primer' by Steve Martin~\cite{Martin}.  An excellent
presentation of the formalism of supersymmetry is given in the texbook
of Wess and Bagger~\cite{WessBagger}.  Supersymmetry has been reviewed
at previous TASI schools by Bagger~\cite{BaggerTASI},
Lykken~\cite{LykkenTASI}, and Kane~\cite{KaneTASI},
among others. Very
recently, three textbooks of phenomenological supersymmetry have appeared,
by Drees, Godbole, and Roy~\cite{DGR}, Binetruy~\cite{Binetruy}, and
Baer and 
Tata~\cite{BaerTata}.  A fourth textbook, by Dreiner, Haber, and 
Martin~\cite{DHM}, is expected soon.

It would be wonderful if all of these articles and 
books used the same conventions, but
that is too much to expect.  In these lectures, I will use my own, somewhat
ideosyncratic conventions.  These are explained in Section~2.1.  Where 
possible, within the philosophy of that section, I have chosen conventions
that agree with those of Martin's primer~\cite{Martin}.

\subsection{Motivation and Structure of Supersymmetry}

If we propose supersymmetry as a model of electroweak symmetry breaking,
we might begin by asking: What is the problem of electroweak symmetry
breaking, and what are the alternatives for solving it?

Electroweak symmetry is spontaneously broken 
in the minimal form of the Standard Model, which I will refer to as the
MSM.  However, the explanation that the MSM
gives for this
phenomenon is 
not satisfactory.
The sole source of symmetry breaking is a single elementary Higgs boson
field.  All mass of quarks, leptons, and gauge bosons arise from the 
couplings of those particles to the Higgs field.

\begin{figure}
\begin{center}
\includegraphics[height=1.5in]{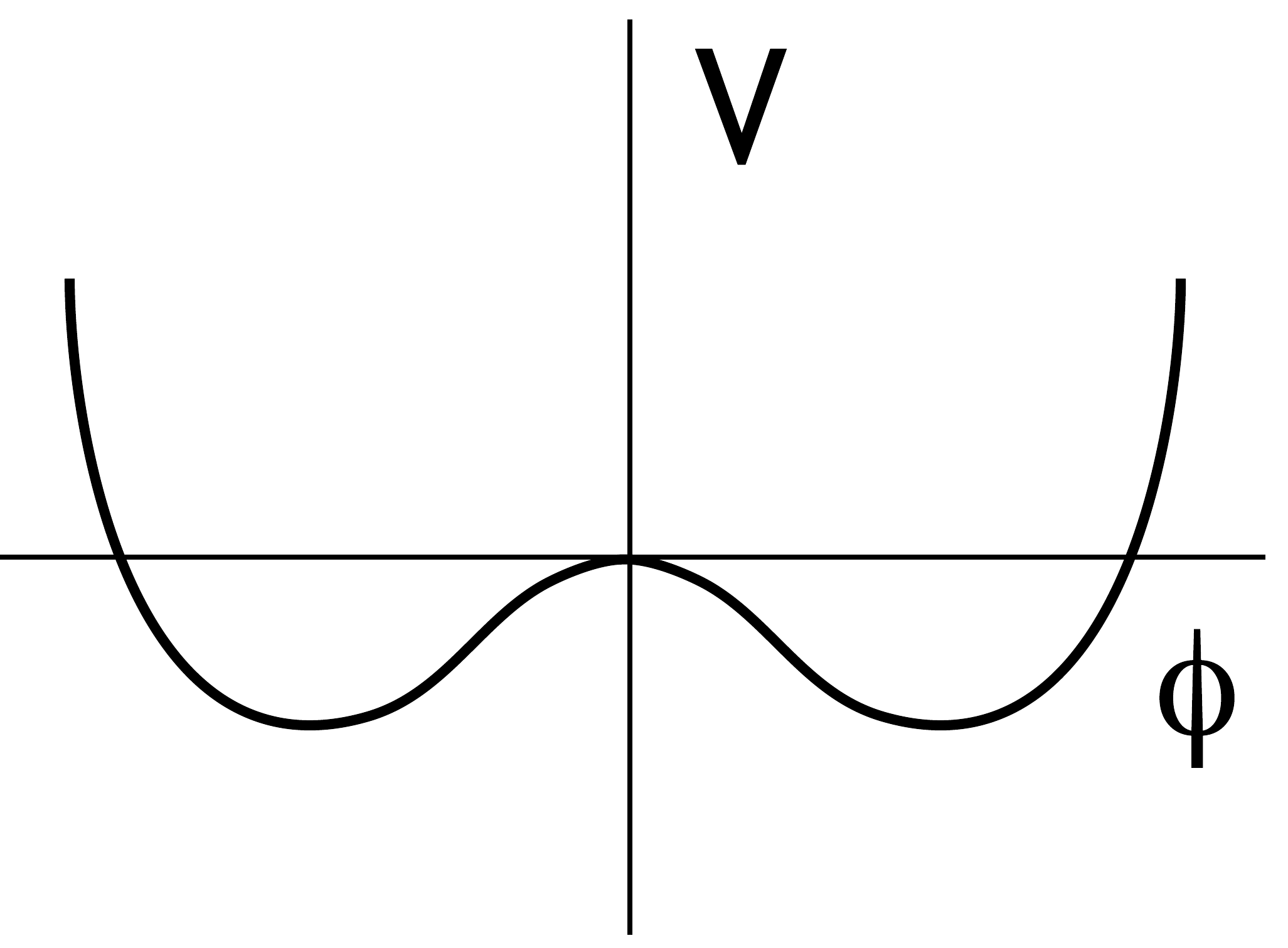}
\caption{The Standard Model Higgs potential \leqn{Higgspot}.}
\label{fig:Higgspot}
\end{center}
\end{figure}

To generate symmetry breaking, we postulate a potential for the Higgs
field 
\beq
    V = \mu^2 |\varphi|^2 + \lambda |\varphi|^4\ ,
\eeq{Higgspot} 
shown in  Fig.~\ref{fig:Higgspot}.
The assumption that 
$\mu^2 < 0$ is the complete explanation for electroweak symmetry 
breaking in the MSM.
Since $\mu$ is a renormalizable coupling of this theory, the value
of $\mu$ cannot
be computed from first principles, and even its sign cannot be 
predicted.

\begin{figure}
\begin{center}
\includegraphics[height=0.8in]{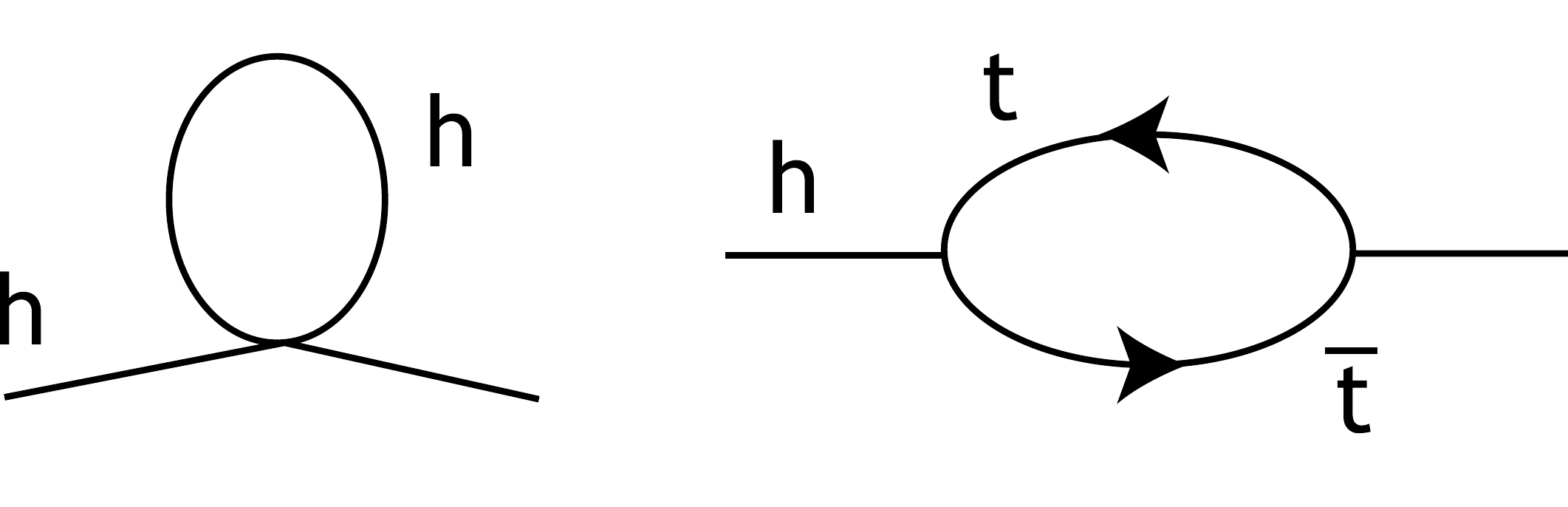}
\caption{Two Standard Model diagrams that give divergent corrections 
to the Higgs mass parameter $\mu^2$.}
\label{fig:Higgsdiags}
\end{center}
\end{figure}

In fact, this explanation has an even worse feature.  
The parameter $\mu^2$ receives
large additive radiative corrections from loop diagrams.  For example, the two 
diagrams shown in Fig.~\ref{fig:Higgsdiags} are ultraviolet divergent.
Supplying a momentum cutoff $\Lambda$, the two diagrams contribute
\beq
    \mu^2 =  \mu^2_{\mbox{bare}} + {\lambda\over 8\pi^2 } \Lambda^2
               -   {3y_t^2\over 8\pi^2} \Lambda^2 + \cdots
\eeq{Higgsadd}
If we view the MSM as an effective theory, $\Lambda$ should be taken to be the 
largest momentum scale at which this theory is still valid.  The presence of 
large additive corrections implies that the criterion $\mu^2 < 0$ is
not a simple condition on the underlying parameters of the effective theory.
The radiative corrections can easily change the sign of $\mu^2$.
Further, if we insist that the MSM has a large range of validity, the 
corrections become much larger than the desired result.  To obtain the Higgs
field vacuum expectation value required for the weak interactions, $|\mu|$
should be about 100 GeV.  If we insist at the same time that the MSM is 
valid up to the Planck scale, $\Lambda \sim 10^{19}$~GeV, the formula
\leqn{Higgsadd} requires a cancellation between the bare value of $\mu$
and the radiative corrections in the first 36 decimal places.  This problem
has its own name, the `gauge hierarchy problem'.  But, to my mind, the 
absence of a logical explantion for electroweak symmetry breaking in the
MSM is already
problem enough.

How could we solve this problem?  There are two different strategies.
One is to look for new strong-couplings dynamics at an energy scale of 
1~TeV or below.  Then the Higgs field could be composite and its 
potential could be the result, for example, of pair condensation of
fermion constituents.  Higgs actually proposed that his field was a 
phenomenological description of a fermion pair condensation mechanism
similar to that in superconductivity~\cite{origHiggs}.  Sometime 
later, Susskind~\cite{Susskind}
and Weinberg~\cite{Weinberg} proposed an explicit model of electroweak 
symmetry breaking by new strong interactions, called `technicolor'.

Today, this approach is disfavored.  Technicolor typically leads to 
flavor-changing neutral currents at an observable level, and also typically
conflicts with the accurate agreement of precision electroweak theory with
experiment.  Specific models do evade these difficulties, but they are 
highly constrained~\cite{TCreview}.

The alternative is to postulate that the electroweak symmetry is broken by a 
weakly-coupled Higgs field, but that this field is part of a model in 
which the Higgs potential is computable.  In particular, 
the Higgs mass term $\mu^2|\varphi|^2$ 
should be generated by well-defined physics within
the model.  A prerequisite for this is that the $\mu^2$ term not receive
additive radiative corrections.  This requires that, at high energy,
the appearance of a nonzero $\mu^2$ in the Lagrangian 
should be forbidden by a symmetry 
of the theory.

There are three ways to arrange a symmetry that forbids the term $\mu^2 
|\varphi|^2$.  We can postulate a symmetry that shifts $\varphi$
\beq 
       \delta \varphi = \epsilon v \ .
\eeq{optionone}
We can postulate a symmetry that connects $\varphi$ to a gauge field, whose
mass can then forbidden by gauge symmetry
\beq 
       \delta \varphi = \epsilon \cdot A \ . 
\eeq{optiontwo}
We can postulate a symmetry that connects $\varphi$ to a fermion field, 
whose mass can then be forbidden by a chiral symmetry.
\beq 
       \delta \varphi = \epsilon \cdot \psi \ . 
\eeq{optionthree}
The options \leqn{optionone} and \leqn{optiontwo} lead, respectively, 
to `little Higgs' models~\cite{ArkaniHamed,LHreview,PLHgg} 
and to models with extra 
space dimensions~\cite{UED,UEDreview}.   
The third option leads to supersymmetry.
This is the route we will now follow.

The symmetry \leqn{optionthree} looks quite innocent, but it is not.
In quantum theory, a symmetry that links a boson with a fermion is 
generated by a conserved charge $Q_\alpha$ that carries spin-1/2
\beq
        [ Q_\alpha, \varphi ] = \psi_\alpha\ ,  \qquad   [Q_\alpha, H] = 0\ .
\eeq{Qcomms}
Such a $Q_\alpha$ implies the existence of a conserved 4-vector charge $R_m$
defined by 
\beq
       \{ Q_\alpha, Q_\beta^\dagger \}  = 2 \gamma_{\alpha \beta}^m  R_m
\eeq{Rdefin}
(It may not be obvious to you that there is no Lorentz scalar component in
this anticommutator, but I will show this in Section~2.1.)  The charge $R_m$
is conserved, because both $Q$ and $Q^\dagger$ commute with $H$.  It is 
nonzero, as we can see by taking the expectation value of \leqn{Rdefin}
in any state
and setting $\alpha = \beta$
\beqa
  \bra{A}   \{ Q_\alpha, Q_\alpha^\dagger \}\ket{A} & = & 
     \bra{A} Q_\alpha Q_\alpha^\dagger \ket{A} + 
  \bra{A} Q_\alpha Q_\alpha^\dagger \ket{A} \CR
     &=& \| Q_\alpha\ket{A} \|^2 + \| Q^\dagger_\alpha \ket{A} \|^2 \ .
\eeqa{positivity}
This expression is non-negative; it
can be zero only if $Q_\alpha$ and $Q_\alpha^\dagger$ 
annihilate every state in the theory.

However, in a relativistic quantum field theory, we do not have the 
freedom to introduce arbitrary charges that have nontrivial Lorentz
transformation properties.  Conservation of energy-momentum and 
angular momentum are already very constraining.  For example, in 
two-body scattering, the scattering amplitude for fixed center of mass
energy can only be a function of one variable, the center of mass scattering
angle $\theta$.  If one adds a second conserved 4-vector charge, almost
all values of $\theta$ will also be forbidden.  Coleman and Mandula 
proved a stronger version of this statement:   In a theory with an addtional
conserved 4-vector charge, there can be no scattering at all, and so the
theory is trivial~\cite{ColemanM}.

If we would like to have \leqn{optionthree} as an exact symmetry, then, 
the only possibility is to set $R_m = P_m$.  That is, the square of the 
fermionic charge $Q_\alpha$ must be {\it the total energy-momentum of 
everything}.  We started out trying to build a theory in which the 
fermionic charge acted only on the Higgs field.  But now, it seems, the 
fermionic charge must act on every field in the theory.  Everything---quarks,
leptons, gauge bosons, even gravitons---must have partners under the symmetry
generated by $Q_\alpha$.  $Q_\alpha$ is fermionic and carries spin 
$\half$. Then every particle
in the theory must have a partner with the opposite statistics and spin 
differing by $\half$ unit.

The idea that the transformation \leqn{optionthree} leads to a profound
generalization of space-time symmetry was discovered
independently several times in the early 1970's~\cite{Golfand,Volkov}.
The 1974 paper by Wess and Zumino~\cite{WessZumino} which gave 
simple linear realizations
of this algebra on multiplets of fields launched the detailed exploration
of this symmetry and its application to particle physics.

The  pursuit of \leqn{optionthree} then necessarily leads
 us to introduce a very 
large number of new particles.  This seems quite daunting.  It might be
a reason to choose one of the other paths, except that these also
lead to new physics models of similarly high complexity.
  I encourage you to carry on with this line of 
analysis a bit longer.  It will lead to a beautiful structure with many
interesting implications for the theory of Nature.

\section{Formalism of Supersymmetry}

\subsection{Fermions in 4 Dimensions}

To work out the full consequences of \leqn{optionthree}, we will need to 
write this equation more precisely. 
To do this, we need to set up a formalism that describes relativistic
fermions in four dimensions in the most general way.
There is no general agreement on the  best 
conventions to use, but every discussion of supersymmetry leans heavily on
the particular choices made.  I will give my choice of conventions in 
this section.

There are two basic spin-$\half$ representations of the Lorentz group.
Each is two-dimensional.  The transformation laws are those of left- and 
right-handed Weyl (2-component) fermions, 
\beqa
  \psi_L &\to& (1- i\vec \alpha \cdot \vec \sigma/2 - \vec \beta\cdot \vec
 \sigma/2)\, \psi_L \CR
  \psi_R &\to& (1- i\vec \alpha \cdot \vec \sigma/2 + \vec \beta\cdot \vec
 \sigma/2)\, \psi_R \ ,
\eeqa{psilaws}
where $\vec \alpha$ is an infinitesimal rotation angle and $\vec\beta$ is 
an infinitesimal boost.  The four-component spinor built from these
ingredients, $\Psi = (\psi_L, \psi_R)$, is a Dirac fermion.

Define the matrix
\beq
            c =  - i \sigma^2  =   
   \pmatrix{ 0 & -1 \cr 1 & 0\cr } \ .
\eeq{cdef}
This useful matrix satisfies $c^2 = -1$, $c^T = - c$.  The combination
\beq
     \psi^T_{1L} c \psi_{2L}   = - \epsilon_{\alpha \beta} 
           \psi_{1L\alpha} \psi_{2L\beta}
\eeq{cinvariant}
is the basic Lorentz invariant product of spinors. 
Many treatments of supersymmetry, for example, that 
in Wess and Bagger's book~\cite{WessBagger}, represent $c$ implicitly 
by raising and 
lowering of spinor indices.  I will stick to this more prosaic approach.

Using the identity $\vec \sigma c = - c (\vec \sigma)^T$,
it is easy to show that  the quantity  $( -c \psi^*_L )$
transforms like $\psi_R$.  So if we wish, we can replace every $\psi_R$
by a $\psi_L$ and write all fermions in the theory as left-handed Weyl
fermions.  
With this notation, for example, we would call $e^-_L$ and $e^+_L$ 
fermions and $e^-_R$ and $e^+_R$ antifermions.  This convention does not 
respect parity, but parity is not a symmetry of the Standard 
Model.  The convention of representing all fermions in terms of 
left-handed Weyl fermions
turns out to be very useful for not only for supersymmetry but also
for other theories of physics beyond the Standard Model. 

Applying this convention, a Dirac fermion takes the form
\beq
        \Psi = \pmatrix{\psi_{1L} \cr  - c \psi_{2L}^* \cr 
                          }
\eeq{Diracdecomp}
Write the Dirac matrices in terms of $2\times 2$ matrices as
\beq
   \gamma^m = \pmatrix{ 0 &  \sigma^m \cr \bar\sigma^m & 0 \cr 
                           }
\eeq{gammadecomp}
with
\beq
  \sigma^m = (1, \vec \sigma)^m \qquad 
  \bar \sigma^m = (1, -\vec \sigma)^m \qquad   c \sigma^m = (\bar\sigma^m)^T c
\eeq{sigmaids}
Then the Dirac Lagrangian can be rewritten in the form 
\beqa 
      \L &=& \bar \Psi i \gamma\cdot \del \Psi - M \bar \Psi \Psi \CR
      &=&  \psi^\dagger_{1L} i \bar\sigma\cdot \del \psi_{1L} + 
        \psi^\dagger_{2L} i \bar\sigma\cdot \del \psi_{2L} \CR
     & & \hskip 0.2in - (m \psi^T_{1L} c \psi_{2L} 
           - m^*  \psi^\dagger_{1L} c \psi^*_{2L} )\ .
\eeqa{Diraceqdecomp}
For the bilinears in the last line, we can use fermion anticommutation 
and the antisymmetry of $c$ to show 
\beq
    \psi^T_{1L} c \psi_{2L} = +   \psi^T_{2L} c \psi_{1L}\ .
\eeq{psibilinear}
and, similarly,
\beq
  (  \psi^T_{1L} c \psi_{2L})^\dagger = \psi^\dagger_{2L}(- c) 
   \psi^*_{1L} =  -  \psi^\dagger_{1L}c \psi^*_{2L}\ .
\eeq{psibilineartwo}
The mass term looks odd, because it is fermion number violating.  However,
the definition of fermion number is that given in the previous paragraph.
The fields $\psi_{1L}$ and $\psi_{2L}$ annihilate, respectively, $e^-_L$ and
$e^+_L$.  So this mass term generates the conversion of $e^-_L$ to
 $e^-_R$, which
is precisely what we would expect a mass term to do.

If we write all fermions as left-handed Weyl fermions, the possibilities
for fermion actions are highly restricted.  The most general Lorentz-invariant
free field Lagrangian takes the form
\beq 
      \L = \psi^\dagger_{k} i \bar\sigma\cdot \del \psi_{k}
             - \half (m_{jk} \psi^T_{j} c \psi_{k} 
           - m^*_{jk}  \psi^\dagger_{j} c \psi^*_{k} ) \ .
\eeq{genfermionL}
where $j,k$ index the fermion fields.  Here and in the rest of these
lectures, I drop the 
subscript $L$.  The matrix $m_{jk}$ is a complex symmetric matrix.  
For a Dirac fermion,
\beq
        m_{jk} = \pmatrix {0 & m \cr m & 0\cr}_{jk}
\eeq{Diracm}
as we have seen in \leqn{Diraceqdecomp}. This matrix respects the
charge
\beq
        Q \psi_1 = + \psi_1  \ ,    \qquad   Q \psi_2 = - \psi_2   \ ,
\eeq{Qfermion}
which is equivalent to the original Dirac fermion number. 
 A Majorana fermion is described
in the same formalism by the mass matrix
\beq
          m_{jk} =  m \delta_{jk}\ .
\eeq{Majoranam}
The most general fermion mass is a mixture of Dirac and Majorana terms.
We will meet such fermion masses in our study of supersymmetry.  These 
more general mass matrices also occur in other new physics models and in models
of the masses of neutrinos.

The SUSY charges are four-dimensional fermions.  The minimum set of 
SUSY charges thus includes one Weyl fermion $Q_\alpha$ and its Hermitian 
conjugate $Q^\dagger_\alpha$.  We can now analyze the anticommutator
$  \{ Q_\alpha , Q^\dagger_\beta \}$.  Since the indices belong to 
different Lorentz representations, this object does not contain a scalar.
The indices transform as do the spinor indices of $\sigma^m$, and so 
we can rewrite \leqn{Rdefin} with $R^m = P^m$ as
\beq
    \{ Q_\alpha , Q^\dagger_\beta \} = 2 \sigma^m_{\alpha \beta} P_m \ .
\eeq{basicACR}

It is possible to construct quantum field theories with larger supersymmetry
algebras.  These must include \leqn{basicACR}, and so the general form 
is~\cite{LHS}
\beq
    \{ Q_\alpha^i , Q^{\dagger j}_\beta \} = 
           2 \sigma^m_{\alpha \beta} P_m \delta^{ij}\ , 
\eeq{basicACRwN}
for $i, j = 1 \ldots N$. 
This relation can be supplemented by a nontrivial anticommutator 
\beq
    \{ Q_\alpha^i , Q_\beta^j \} = 
           2 \epsilon_{\alpha\beta} {\cal Q}^{ij}
\eeq{basicACRwcentral}
where the {\it central charge} $ {\cal Q}^{ij}$ is antisymmetric in 
$[ij]$.   Theories with $N > 4$ necessarily contain particles of spin 
greater than 1.  Yang-Mills theory with $N = 4$ supersymmetry is 
an especially beautiful model with exact scale invariance and many other 
attractive formal properties~\cite{Neqfour}.  In these lectures, however, 
I will restrict myself to the minimal case of $N=1$ supersymmetry.

I will discuss supersymmetry transformations using the 
operation on fields
\beq 
   \delta_\xi  \Phi = [ \xi^T c Q + Q^\dagger c \xi^*, \Phi] \ . 
\eeq{deltaxidef}
Note that the operator $\delta_\xi$
 contains pairs of anticommuting objects and so 
obeys commutation rather than anticommutation relations.
The operator $P_m$ acts on fields as the generator of translations,
$P_m = i \del_m$.
Using this, we can rewrite \leqn{basicACR} as
\beq
   [ \delta_\xi,  \delta_\eta ]  = 2i \left( \xi^\dagger \bar \sigma^m \eta
          - \eta^\dagger \bar \sigma^m \xi \right) \ \del_m
\eeq{deltaACR}
I will take this equation as the basic (anti)-commutation
relation of supersymmetry.  In the next 
two sections, I will construct some representations
of this commutation relation on multiplets of fields.

\subsection{Supersymmetric Lagrangians with Scalars and Fermions}

The simplest representation of the supersymmetry algebra \leqn{deltaACR}
directly generalizes the transformation \leqn{optionthree} from which we 
derived the idea of supersymmetry.  The full set of fields required includes
a complex-valued boson field $\phi$ and a Weyl fermion field $\psi$. 
These fields create and destroy a scalar particle and its antiparticle, 
a left-handed massless fermion, and its right-handed antiparticle.  Note that
the particle content has an equal number of fermions and bosons.  This 
particle content is called a {\it chiral supermultiplet}.

I will now write out the transformation laws for the fields corresponding
to a chiral supermultiplet.  It is convenient to add a second complex-valued
boson field $F$ that will have no associated particles.  Such a 
field is called an {\it auxiliary field}.  We can then write the 
transformations that generalize \leqn{optionthree} as 
\beqa
    \delta_\xi \phi &=& \sqrt{2} \xi^T c \psi  \CR
    \delta_\xi \psi & =& \sqrt{2} i \sigma^n c \xi^* \del_n \phi + 
        \sqrt{2} F\, \xi \CR
    \delta_\xi F &=&  - \sqrt{2} i \xi^\dagger \bar \sigma^m \del_m \psi\ .
\eeqa{chiralrep}
The conjugates of these transformations are
\beqa
    \delta_\xi \phi^* &=& -\sqrt{2} \psi^\dagger  c  \xi^*  \CR
    \delta_\xi \psi^\dagger & =& \sqrt{2} i \xi^T c \sigma^n  \del_n \phi^* + 
        \sqrt{2}\xi^\dagger  F^* \CR
    \delta_\xi F^* &=&  \sqrt{2} i \del_m \psi^\dagger \bar \sigma^m \xi\ .
\eeqa{antichiralrep}
These latter transformations define the {\it antichiral supermultiplet}.
I claim that the transformations \leqn{chiralrep} and \leqn{antichiralrep}, 
first, satisfy the fundamental commutation 
relation \leqn{deltaACR} and, second, leave a suitable Lagrangian invariant.
Both properties are necessary, and both must be checked, in order for a 
set of transformations to generate a symmetry group of a field theory.

The transformation laws \leqn{chiralrep} seem complicated. You might 
wonder if there is a formalism that generates these relations automatically
and manipulates them more easily than working with the three distinct
component 
fields $(\phi, \psi, F)$.  In the next section, I will introduce a 
formalism called {\it superspace} that makes it almost automatic  to 
work with the chiral supermultiplet.  However, the superspace description 
of the multiplet containing gauge fields is more complicated, and the 
difficulty of working with superspace becomes exponentially greater
in theories that include gravity, higher dimensions, or $N > 1$ supersymmetry.
At some stage, one must go back to components.
I strongly recommend that you gain experience by working
 through the component field calculations
described in these notes in full detail, however many large pieces of paper
that might require.

To verify each of the two claims I have made for \leqn{chiralrep} 
requires a little calculation.  Here is the 
check of the commutation relation applied to the field $\phi$:
\beqa
  [\delta_\xi, \delta_\eta] \phi & = & \delta_\xi(\sqrt{2} \eta^T c \psi) - 
         (\xi \leftrightarrow  \eta) \CR
    & = & \sqrt{2} \eta^T c (\sqrt{2}i \sigma^n c \xi^* \del_n \phi) - 
  (\xi \leftrightarrow  \eta) \CR
    & = & -2i \eta^T (\bar \sigma^n)^T  \xi^* \del_n \phi - 
  (\xi \leftrightarrow  \eta) \CR
    & = & 2i [ \xi^\dagger \bar \sigma^n \eta - \eta^\dagger \bar\sigma^n \xi]
             \del_n \phi \CR
\eeqa{phiACR}
The check of the commutation relation applied to $F$ is 
equally straightforward.  The check on $\psi$ is a bit lengthier.  It requires
a {\it Fierz identity}, that is, a spinor index rearrangment identity.  
Specifically, we need
\beq
     \eta_\alpha \xi_\beta^\dagger = -  \half (\xi^\dagger \bar\sigma_m \eta) 
           \sigma^m_{\alpha\beta}\ ,
\eeq{firstFierz}
which you can derive by writing out the four components explicitly.
After some algebra that involves the use of this identity, you can see
that the SUSY
commutation relation applied to $\psi$ also takes the correct form.

Next, I claim that the Lagrangian
\beq
  \L = \del^m \phi^* \del_m\phi  + \psi^\dagger i \bar\sigma\cdot \del \psi
       + F^* F \
\eeq{Lchiral}
is invariant to the transformation  \leqn{chiralrep}.  I will assume that 
the Lagrangian \leqn{Lchiral} is integrated $\int d^4x$ and use integration
by parts freely. Then
\beqa
   \delta_\xi \L & = & \del^m \phi^* \del_m(\sqrt{2} \xi^T c \psi) + 
           (-\sqrt{2} \del^m \psi^\dagger c \xi^*) \del\phi \CR
      & & \quad   + \psi^\dagger i \bar\sigma\cdot \del [ \sqrt{2} i \sigma^n
       c \xi^* \del_m \phi + \sqrt{2}\xi F]  \CR
  & & \quad + [ \sqrt{2} i \del_n \phi^* \xi^T
           c \sigma^n + \sqrt{2} \xi^\dagger F^*] i \bar\sigma\cdot \del \psi
       \CR & & \quad + F^* [ - \sqrt{2}i \xi^\dagger \bar \sigma^m \del_m
      \psi] + [\sqrt{2} i \del_m \psi^\dagger \bar \sigma^m \xi] F \CR
     & = & - \phi^* \sqrt{2} \xi^T c \del^2 \psi + \sqrt{2} \del_n \phi^*
      \xi^T c \sigma^n \bar\sigma^m \del_n \del_m \psi \CR
    & & \quad + 
       \sqrt{2} \psi^\dagger c \xi^* \del^2 \phi - \sqrt{2} \psi^\dagger
 \bar \sigma^m \sigma^n c \xi^* \del_m \del_n \phi \CR
   & & \quad + \sqrt{2} i \psi^\dagger \bar \sigma^m \del_m F \xi + 
        \sqrt{2} i \del_m \psi^\dagger \bar \sigma^m \xi F \CR
  & & \quad  -\sqrt{2} i\xi^\dagger F^* \bar \sigma^m \del_m \psi 
+ \sqrt{2}i F^*
         \xi^\dagger \bar \sigma^m \del_m \psi \CR
   & = &  0 \ .
\eeqa{Lchiralcheck}
    In the final expression, the four lines cancel line by line.  In the 
first two lines, the cancellation is made by using the 
identity $(\bar \sigma\cdot \del)( \sigma \cdot \del) = \del^2$.

So far, our supersymmetry Lagrangian is  just a massless
free field theory.  However, it is possible to add
rather general interactions that respect the symmetry.  Let $W(\phi)$ be an
analytic function of $\phi$, that is, a function that depends on $\phi$
but not on $\phi^*$.   Let
\beq
  \L_W = F {\del W\over \del \phi} - \half \psi^T c \psi {\del^2 W\over \del
      \phi^2}
\eeq{WLagrangian}
I claim that
 $\L_W$ is invariant to \leqn{chiralrep}. Then
 we can add $(\L_W + \L_W^\dagger)$
 to the free field Lagrangian to introduce interactions into the theory.
 The function $W$ is called the {\it superpotential}.

We can readily check that $\L_W$ is indeed invariant:
\beqa
   \delta_\xi \L_W &= & F  {\del^2 W\over \del \phi^2} (\sqrt{2} \xi^T c \psi)
             - \sqrt{2} F \xi^T c \psi  {\del^2 W\over \del \phi^2}\CR
        & & \qquad -\sqrt{2} i \xi^\dagger \bar \sigma^m \del_m  \psi 
 {\del W\over \del \phi}  - \psi^T c \sqrt{2} i \sigma^n c \xi^* \del_n \phi
             {\del^2 W\over \del \phi^2} \CR
     & & \qquad 
     - \psi^T c \psi {\del^3 W\over \del \phi^3}\sqrt{2} \xi^T c \psi\ .
\eeqa{LWcheck}
The second line rearranges to 
\beq
     -\sqrt{2} i \xi^\dagger \bar \sigma \left( \del_n\psi 
 {\del W\over \del \phi} + \psi \del_n\phi   {\del^2 W\over \del \phi^2}
\right) \ , 
\eeq{rearrLWcheck}
which is a total derivative.  The third line is proportional to $\psi_\alpha
\psi_\beta\psi_\gamma$, which vanishes by fermion antisymmetry since the 
spinor indices take only two values.  Thus it is true that 
\beq
  \delta_\xi \L_W = 0 \ .
\eeq{LWcheckfinal}

The proofs of invariance that I have just given generalize straightforwardly
to systems of several chiral supermultiplets.  The requirement on the 
superpotential is that it should be an analytic function of the complex 
scalar fields $\phi_k$.  Then the following Lagrangian is supersymmetric:
\beq
\L = \del^m \phi^*_k \del_m\phi_k  
+ \psi^\dagger_k i \bar\sigma\cdot \del \psi_k
       + F^*_k F_k  + \L_W + \L_W^\dagger \ , 
\eeq{Lchiralgen}
where
\beq
  \L_W = F_k {\del W\over \del \phi_k} 
      - \half \psi^T_j c \psi_k {\del^2 W\over \del \phi_j \del\phi_k} \ .
\eeq{WkLagrangian}
In this Lagrangian, the fields $F_k$ are Lagrange multipliers.  They obey 
the constraint equations
\beq
           F^*_k = - {\del W\over \del\phi_k} \ .
\eeq{Feqs}
Using these equations to eliminate the $F_k$, we find an interacting theory
with the fields $\phi_k$ and $\psi_k$, a Yukawa coupling term proportional
to the second derivative of $W$, as given in \leqn{WkLagrangian}, and the 
potential energy 
\beq
      V_F =  \sum_k \left| {\del W \over \del \phi_k } \right|^2 \ .
\eeq{VF}
I will refer to $V_F$ as the {\it F-term potential}.  Later we will 
meet a second contribution $V_D$, the {\it D-term potential}.  These two 
terms, both obtained by integrating out auxiliary fields, make up the 
classical potential energy of a general supersymmetric field theory of 
scalar, fermion, and gauge fields.

The simplest example of the F-term potential appears in the theory with one 
chiral supermultiplet and the superpotential $W = \half m \phi^2$.
The constraint equation for $F$ is~\cite{NWeqn}
\beq
       F^* = -  m \phi \ .
\eeq{NWeq}
After eliminating $F$, we find the Lagrangian
\beq
   \L = \del^n\phi^* \del_n \phi - |m|^2 \phi^*\phi + \psi^\dagger i 
 \bar\sigma \cdot \del \psi - \half (m \psi^T c \psi - m^* \psi^\dagger c 
     \psi^*)
\eeq{freem}
This is a theory of two free scalar bosons of mass $|m|$ and a free Majorana
fermion with the same mass $|m|$.  The Majorana fermion has two spin states,
so the number of boson and fermion physical states is equal, as required. 

The form of the expression \leqn{VF} implies that $V_F \geq 0$, and that
$V_F = 0$ only if all $F_k = 0$.  This constraint on the potential energy 
follows from a deeper consideration about supersymmetry.  Go back to the 
anticommutation relation \leqn{basicACR}, evaluate it for $\alpha = \beta$,
and take the vacuum expectation value.  This gives
\beq
  \bra{0}  \{ Q_\alpha , Q^\dagger_\alpha \} \ket{0} = 
 \bra{0} (H-P^3) \ket{0} =  \bra{0} H \ket{0} \ ,
\eeq{vacuumeq}
since the vacuum expectation value of $P^3$ vanishes by rotational 
invariance.
Below \leqn{Rdefin}, I argued that the left-hand side of this equation is 
greater than or equal to zero. It is equal to zero if and only if
\beq
              Q_\alpha \ket{0} = Q^\dagger_\alpha \ket{0} = 0 
\eeq{Qannih}
The formulae \leqn{Qannih} give
 the criterion than the vacuum is invariant under
supersymmetry.  If this relation is not obeyed, supersymmetry is 
spontaneously broken.  Taking the vacuum expectation value 
of the transformation
law for the chiral representation, we find
\beqa
  \bra{0} [\xi^T c Q + Q^\dagger c \xi^*, \psi_k ] \ket{0} & = & 
         \bra{0} \sqrt{2} i \sigma^n \xi^* \del_n \phi_k + \xi F_k \ket{0} \CR
      & = &  \xi \bra{0} F_k \ket{0} \ .
\eeqa{psitoF}
In the last line I have used the fact that the vacuum expectation value
of $\phi(x)$ is translation invariant, so its derivative vanishes.
The left-hand side of \leqn{psitoF} vanishes if the vacuum state is 
invariant under supersymmetry.

The results of the previous paragraph can be summarized in the following 
way:   If supersymmetry is a manifest symmetry of a quantum field theory, 
\beq
         \bra{0} H \ket{0} = 0\ , \ \mbox{and} \ \bra{0} F_k \ket{0} = 0
\eeq{susyzeros}
for every $F$ field of a chiral multiplet.  In complete generality,
\beq
         \bra{0} H \ket{0} \geq 0 \ .
\eeq{Hpositive}
The case where $\VEV{H}$ is positive and nonzero corresponds to spontaneously
broken supersymmetry.  If the theory has a state satisfying \leqn{Qannih},
this is
necesssarily the state in the theory with lowest energy. Thus, supersymmetry
can be spontaneously broken only if a supersymmetric vacuum state does not
exist\footnote{It is possible that a supersymmetric vacuum state might 
exist but that a higher-energy vacuum state might be metastable.  A model
built on this metastable state would show spontaneous breaking of 
supersymmetry~\cite{ISS}.}

For the moment, we will work with theories that preserve supersymmetry.  I 
will give examples of theories with spontaneous supersymmetry breaking in 
Section~3.5.

The results we have just derived are exact consequences of the commutation
relations of supersymmetry.  It must then be true that the vacuum energy of a 
supersymmetric theory must vanish in perturbation theory. This is already
nontrivial for the free theory \leqn{freem}.  But it is correct.  The 
positive zero point energy of the boson field exactly cancels the negative
zero point energy of the fermion field. With some effort, one can show the
cancellation also for the leading-order diagrams in an interacting theory.
Zumino proved that this cancellation is completely general~\cite{Zuminovac}.

\begin{figure}
\begin{center}
\includegraphics[height=2.0in]{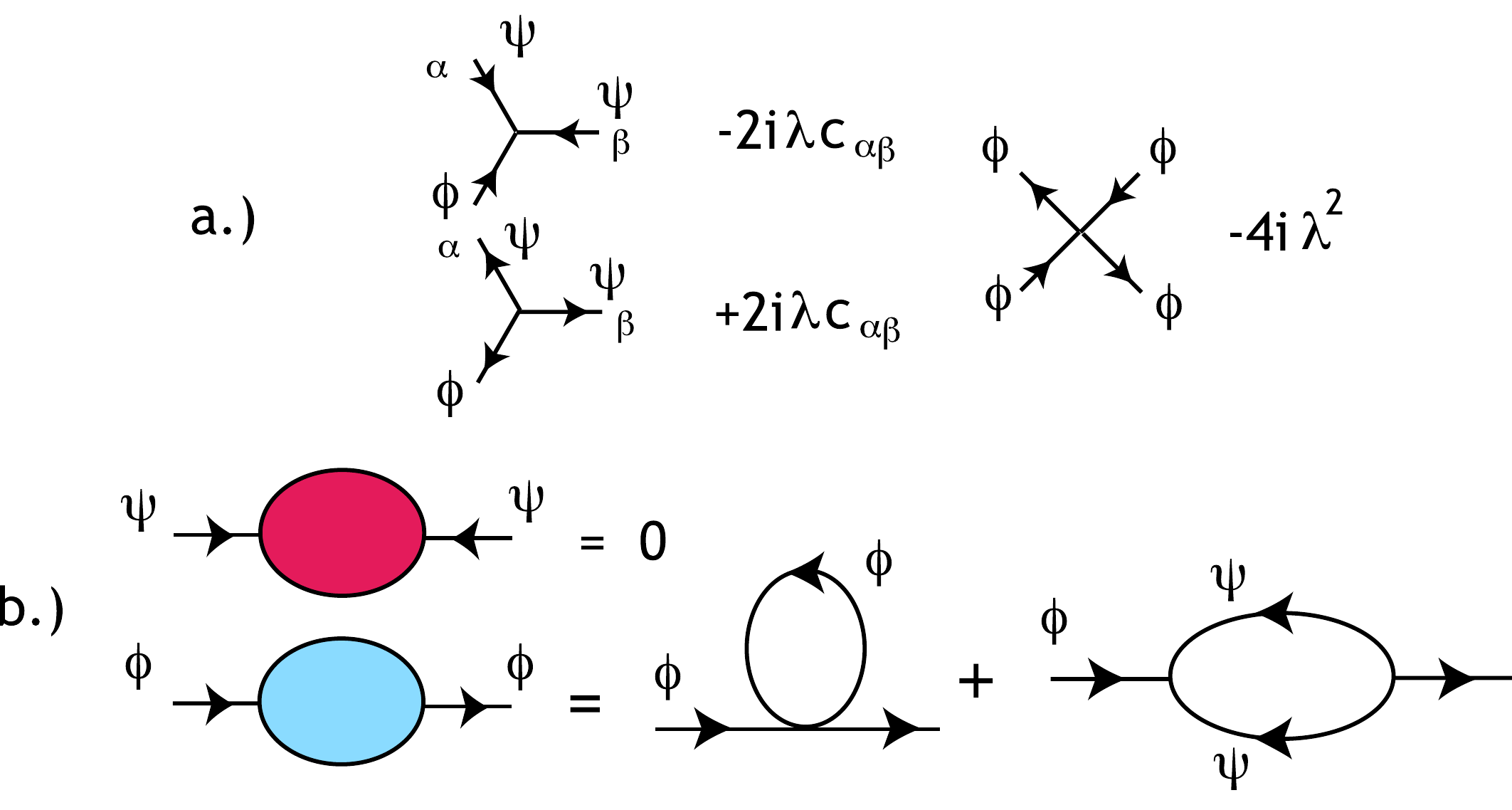}
\caption{Perturbation theory for the supersymmetric model \leqn{intchiral}:
  (a) vertices of the model; (b) corrections to the fermion and scalar 
   masses.}
\label{fig:chiralmodel}
\end{center}
\end{figure}

I would like to show you another type of cancellation that is also seen
in perturbation theory in models with chiral fields.  Consider the model
with one chiral field and superpotential
\beq
        W = {\lambda\over 3} \phi^3 \ .
\eeq{interactingW}
After eliminating $F$, the Lagrangian becomes
\beq
   \L = \del\phi^* \del_m \phi + \psi^\dagger i 
 \bar\sigma \cdot \del \psi - \lambda (\phi \psi^T c \psi 
 - \phi^* \psi^\dagger c 
     \psi^*) - \lambda^2 |\phi|^4 \ .
\eeq{intchiral}
The vertices of this theory are shown in Fig.~\ref{fig:chiralmodel}(a).

From our experience in \leqn{Higgsadd}, we might expect to find an 
addditive radiative correction to the scalar mass.  The corrections to the
fermion and scalar mass terms are given by the diagrams 
in Fig.~\ref{fig:chiralmodel}(b).  Actually, there are no diagrams that
correct the fermion mass; you can check that there it is not possible to 
match the arrows appropriately.  For the scalar mass correction, the 
two diagrams shown contribute
\beq
   - 4i \lambda^2 \int {d^4p\over (2\pi)^4} {i \over p^2} + 
   \half (-2i\lambda)(+2i\lambda) \int {d^4 p\over (2\pi)^4}
 \tr \left[ {i \sigma\cdot p\over p^2 } c { i\sigma^T\cdot (-p)\over p^2 } c 
 \right] 
\eeq{scalarmass}
Using $\sigma\cdot p \bar\sigma \cdot p = p^2$ in the second term and then
taking the trace, we see that these two contributions cancel precisely.
In this way, supersymmetry really does control radiative corrections to the
Higgs mass, following the logic that we presented in Section~1.2. 

\begin{figure}
\begin{center}
\includegraphics[height=1.0in]{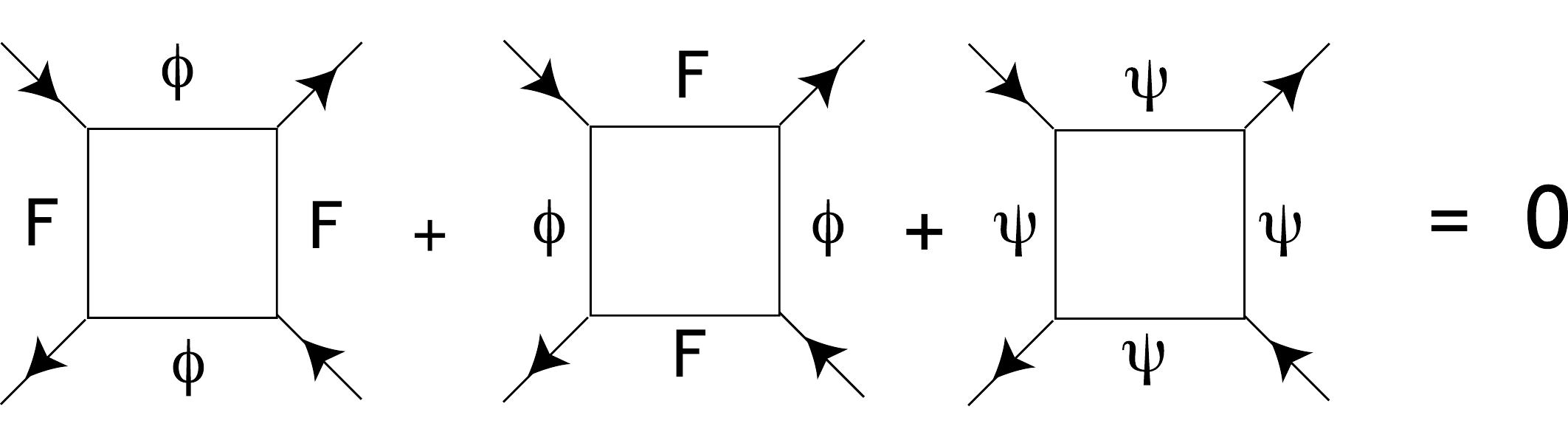}
\caption{Scheme of cancellations of one-loop 
 corrections to the F-term potential.}
\label{fig:oneLoop}
\end{center}
\end{figure}

In fact, it can be shown quite generally that not only the mass term but the
whole superpotential $W$ receives no additive 
radiative corrections in any order 
of perturbation theory~\cite{Supergraphs}.  
For example, the one-loop corrections to quartic 
terms in the Lagrangian cancel in a simple way that is 
indicated in Fig.~\ref{fig:oneLoop}.
The field strength renormalization of chiral fields can be nonzero,
so the form of $W$ can be changed by radiative corrections
by the rescaling of fields.   Examples are known in which $W$ receives
additive radiative corrections from nonperturbative effects~\cite{ADS}.

\subsection{Superspace}

Because the commutation relations of supersymmetry include the generators
of translations, supersymmetry is a space-time symmetry.  It is an 
attractive idea that supersymmetry is the natural set of translations on a 
generalized space-time with commuting and anticommuting coordinates.
In this section, I will introduce the appropriate generalization of 
space-time and use it to re-derive some of the results of 
Section~2.2.

Consider, then, a space with four ordinary space-time coordinates 
$x^\mu$ and four 
anticommuting coordinates  $(\theta_\alpha, \bar\theta_\alpha)$.
I will take the  coordinates $\theta_\alpha$ to transform as  2-component
Weyl spinors; the
$\bar\theta_\alpha$ are the complex conjugates of the $\theta_\alpha$.
This is {\it superspace}.
A {\it superfield} is a function of these superspace coordinates:
$\Phi(x, \theta, \bar \theta)$.

It is tempting to define supersymmetry transformations as translations
$\theta \to \theta + \xi$.  However, this does not work.  These 
transformations commute, $[\delta_\xi, \delta_\eta] = 0$, and we have 
seen in Section~1.2 that this implies that the S-matrix of the resulting
field theory must be
trivial.  To construct a
 set of transformations with the correct commutation relations, we must 
write
\beq
        \delta_\xi \Phi = {\cal Q}_\xi \Phi \ ,
\eeq{superspacetrans}
where
\beq
  {\cal Q}_\xi = \left( - {\del\over \del \theta} - i \bar\theta \bar \sigma^m
  \del_m \right)\xi + \xi^\dagger \left( {\del\over \del\bar\theta} + i 
 \bar \sigma^m \theta \del_m \right)\ .
\eeq{Qdefin}
This is a translation of the fermionic coordinates ($\theta$, $\bar\theta$)
plus 
 a translation of the ordinary space-time
coordinates proportional to $\theta$, $\bar \theta$.  
It is straightforward to show that these operators 
satisfy
\beq
      [ {\cal Q}_\xi, {\cal Q}_\eta ] = -  2i 
\left( \xi^\dagger \bar \sigma^m \eta
          - \eta^\dagger \bar \sigma^m \xi \right) \ \del_m \ .
\eeq{QACR}
Despite the fact that this equation has an extra minus sign on the right-hand
side with respect to \leqn{deltaACR}, it is the relation that we want.
(The difference is similar to that between active and passive transformations.)
Combined with the decomposition of the superfield that I will introduce
below, this relation 
will allow us to derive the chiral supermultiplet transformation 
laws \leqn{chiralrep}.

Toward this goal, we need one more ingredient.  Define the superspace 
derivatives
\beq
   D_\alpha =  {\del\over \del \theta_\alpha} - i (\bar \theta \sigma^m)_\alpha
 \del_m \qquad   
   \bar D_\alpha = -  {\del\over \del \bar \theta_\alpha} +
        i ( \sigma^m \theta )_\alpha   \del_m \ ,
\eeq{DbarDdefin}
such that $(D_\alpha \Phi)^\dagger = \bar D_\alpha \Phi^\dagger$.
These operators commute with ${\cal Q}_\xi$:
\beq
    [ D_\alpha, {\cal Q}_\xi ] = 0 \qquad  
   [ \bar D_\alpha, {\cal Q}_\xi ] = 0 \ .
\eeq{Dcommutes}
Thus, we can constrain $\Phi$ by the equation
\beq
      D_\alpha \Phi = 0  \qquad   \mbox{or} \qquad \bar D_\alpha \Phi = 0 \ ,
\eeq{Dconstraints}
and these constraints are consistent with supersymmetry.  What we have 
just shown is that the general superfield $\Phi(x,\theta, \bar \theta)$ is 
a {\it reducible } representation of supersymmetry.  It can be decomposed
into a direct sum of three smaller representations, one constrained by the
first of the relations \leqn{Dconstraints}, one constrained by the second
of these relations, and the third containing whatever is left over in $\Phi$ 
when these pieces are removed.

Let's begin with the constraint $\bar D_\alpha \Phi = 0$.  The solution of
this equation can be written
\beq
     \Phi(x, \theta, \bar \theta ) = \Phi(x + i \bar\theta \bar \sigma^m 
 \theta, \theta) \ ,
\eeq{chiralPhi}
that is, this solution is parametrized by a general function of $x$ and 
$\theta$.  Since $\theta$ is a two-component anticommuting object, this
general function of $x$ and $\theta$ can be represented as
\beq
  \Phi(x,\theta) =  \phi(x) + \sqrt{2} \theta^T c \psi(x) + \theta^T c \theta
     F(x) \ .
\eeq{Phicontent}
The field content of this expression is exactly that of the chiral 
supermultiplet. The supersymmetry transformation of this field should
be 
\beq
          \delta_\xi \Phi =  {\cal Q}_\xi \Phi(x + i \bar\theta \bar \sigma^m 
 \theta, \theta) \ .
\eeq{transPhi}
 It is straightforward to compute the right-hand side of \leqn{transPhi} in 
terms of $\theta$, $\bar\theta$, and the component fields of \leqn{Phicontent}.
  The coefficients of 
powers of $\theta$ are precisely the supersymmetry variations 
given in \leqn{chiralrep}.  Thus a superfield satisfying
\beq
            \bar D_\alpha \Phi = 0
\eeq{Chiraldefin}
is equivalent to a chiral supermultiplet, and the transformation 
\leqn{transPhi} gives the supersymmetry transformation of this multiplet.
A superfield satisfying \leqn{Chiraldefin} is called a {\it chiral 
superfield}.  Similarly, a superfield satisfying
\beq
            D_\alpha \Phi = 0
\eeq{antiChiraldefin}
is called an {\it antichiral superfield}.  This superfield has a 
component field decomposition $(\phi^*, \psi^*, F^*)$, on which $\cal Q_\xi$
induces the transformation \leqn{antichiralrep}.
  I will describe the remaining content of the general
superfield $\Phi$ in Section~2.5.

A Lagrangian on Minkowski space is integrated over $d^4 x$. A superspace
Lagrangian should be also be integrated over the $\theta$ coordinates.
  Integration over fermionic coordinates is defined to be proportional 
to the coefficient of the highest power of $\theta$.  I will define 
integration over superspace coordinates by the formulae
\beq
     \int d^2 \theta \, 1 = \int d^2 \theta \, \theta_\alpha = 0 \qquad
      \int d^2\theta (\theta^T c \theta) = 1
\eeq{thetaint}
and their conjugates.
To use these  formulae, expand the
superfields in powers of $\theta$ and pick out the  terms proportional to
$(\theta^T c \theta)$.  Then, if $\Phi$ is a chiral superfield constrained
by \leqn{Chiraldefin} and $W(\Phi)$ is an analytic function of $\Phi$,
\beqa
            \int d^2 \theta\ \Phi(x,\theta) &=& F(x) \CR
           \int d^2 \theta\ W(\Phi) &=& F(x){\del W\over \del\phi} - 
    \half \psi^T c \psi {\del^2 W\over \del^2 \phi} \ , 
\eeqa{chiralints}
where, in the second line, $W$ on the right-hand side is evaluated with 
$\Phi = \phi(x)$.  With somewhat more effort, one can show
\beq
   \int d^2 \theta \int d^2 \bar \theta \ \Phi^\dagger \Phi = 
 \del^m \phi^* \del_m\phi  + \psi^\dagger i \bar\sigma\cdot \del \psi
       + F^* F \ .
\eeq{doublethetaint}

These formulae produce the invariant Lagrangians of chiral supermultiplets
from a superspace point of view.  The most general Lagrangian of chiral
superfields $\Phi_k$ takes the form
\beq
    \L   =   \int d^4\theta \, K(\Phi, \Phi^\dagger) + 
         \int d^2 \theta \, W(\Phi) + \int d^2\bar \theta \, 
            (W(\Phi))^\dagger \ , 
\eeq{mostgeneralL}
where $W(\Phi)$ is an analytic function of complex superfields and $K(\Phi, 
\Phi^\dagger)$ is a general real-valued function of the superfields.  The
Lagrangian \leqn{Lchiralgen} is generated from this expression by taking
$ K(\Phi, \Phi^\dagger) = \Phi^\dagger_k \Phi_k$.  The most
general {\it renormalizable} Lagrangian of chiral supermultiplets is
obtained by taking $K$ to be of this simple form and taking $W$ to be
a polynomial of degree at most~3.

Because the integral $d^2\theta$ exposes the Lagrange multiplier $F$ in 
\leqn{Phicontent}, I will refer to a term with this superspace integral as
an {\it F-term}.  For similar reasons that will become concrete in the 
next section, I will call a term with a $d^4\theta$ integral a {\it D-term}.

In the remainder of these lectures, I will restrict myself to discussing
renormalizable supersymmetric 
theories. But, still, it is interesting to ask what theories
we obtain when we take more general forms for $K$.  The Lagrangian 
for $\phi$ turns out to be a nonlinear sigma model for which the target
space is a complex manifold with the metric~\cite{Kahler}
\beq
           g_{m\bar n} = {\del^2 \over \del\Phi^m \del\Phi^{\dagger \bar n}}
                     K(\Phi, \Phi^\dagger)
\eeq{metricofK}
A complex manifold whose metric is derived from a potential in this way
is called a {\it K\"ahler manifold}.  The function $K$ is the {\it
K\"ahler potential}.  It is remarkable that, wherever in ordinary 
quantum field theory we find a general structure from real analysis, 
the supersymmetric version of the theory has a corresponding complex
analytic structure.

Now that we have a Lagrangian in superspace, it is possible to derive
Feynman rules and compute Feynman diagrams in superspace.  I do not have
space here to discuss this formalism; it is discussed, for example, 
in~\cite{WessBagger} and~\cite{Supergraphs}.  I would like to 
state one important consequence
of this formalism.  It turns out that, barring some special circumstances
related to perturbation theory anomalies, these Feynman diagrams
always generate 
corrections to the effective Lagrangian that are D-terms,
\beq
         \int d^4 \theta \,  X(\Phi, \Phi^\dagger)\ .
\eeq{Xform}
The perturbation theory does not produce terms that are integrals
$\int d^2\theta$.
This leads to an elegant proof of the result cited at the end of the previous
section that the superpotential is not renormalized at any order in 
perturbation theory~\cite{Supergraphs}.

\subsection{Supersymmetric Lagrangians with Vector Fields}

To construct a supersymmetric model that can include the Standard Model, 
we need to be able to write supersymmetric Lagrangians that include
Yang-Mills vector fields.  In this section, I will discuss how to do 
that.

To prepare for this discussion, let me present my notation for 
gauge fields in a general quantum field theory.  The couplings of 
gauge bosons to matter are based on the covariant derivative, which
I will write as
\beq
      \D_m \phi = (\del_m - i g A_m^a t^a_R) \phi
\eeq{covariantD}
for a field $\phi$ that belongs to the representation $R$ of the gauge
group $G$.  In this formula, $t^a_R$ are the representation matrices of
the generators of $G$ in the representation $R$.  These obey
\beq
      [t^a_R, t^b_R] = i f^{abc} t^c_R
\eeq{Gcomm}
The coefficients $f^{abc}$ are the {\it structure constants} of $G$. They are 
independent of $R$; essentially, their values define the multiplication
laws of 
$G$.  They can be taken to be totally antisymmetric.

The generators of $G$ transform under $G$ according to a representation 
called the {\it adjoint representation}.  I will denote this representation
by $R = G$.  Its representation matrices are
\beq
       (t^a_G)_{bc} = i f^{bac}
\eeq{adjstr}
These matrices satisfy \leqn{Gcomm} by virtue of the Jacobi identity.  The 
covariant derivative on a field in the adjoint representation takes the 
form
\beq
     \D_m \Phi^a =  \del_m\Phi^a + g f^{abc} A_m^b \Phi^c
\eeq{covariantDinG}
The field strengths $F^a_{mn}$ are defined from the covariant derivative
(in any representation) by
\beq
      [ \D_m, \D_n ] = -ig F^a_{mn} t^a_R \ .
\eeq{Fadef}
This gives the familiar expression
\beq
      F^a_{mn} = \del_m A^a_n - \del_n A^a_m + g f^{abc} A^b_m A^c_n \ .
\eeq{Fdefin}

Now we would like to construct a supersymmetry multiplet that contains the
gauge field $A^a_m$.  The fermion in the multiplet should differ in spin
by $\half$ unit.  To write a renormalizable theory, we must take this to 
be a spin-$\half$ Weyl fermion.  I will then define the {\it vector 
supermultiplet}
\beq
        ( A^a_m,  \lambda^a_\alpha, D^a)
\eeq{vectorsup}
including the gauge field, a Weyl fermion in the adjoint representation of
the gauge group, and an auxililary real scalar field, also in 
the adjoint representation,  that will have no 
independent particle content.  The particle content of this multiplet is
one massless vector boson, with two transverse polarization states, and 
one massless fermion and antifermion, for each generator of the gauge group.
The fermion is often called a {\it gaugino}.
The number of physical states is again equal between bosons and fermions.

The supersymmetry transformations for this multiplet are
\beqa
    \delta_\xi A^{am} &=& [ \xi^\dagger \bar\sigma^m \lambda^a + 
          \lambda^{\dagger a} \bar \sigma^m \xi ] \CR
   \delta_\xi \lambda^a &=& [ i \sigma^{mn} F^a_{mn} + D^a ] \xi \CR
   \delta_\xi \lambda^{\dagger a} &=& \xi^\dagger 
      [ i \bar \sigma^{mn} F^a_{mn} + D^a ]  \CR
   \delta_\xi D^a &=& -i [ \xi^\dagger \bar \sigma^m \D_m \lambda^a - 
         \D_m \lambda^{\dagger a} \bar \sigma^m \xi ] 
\eeqa{vectorrep}
where 
\beq
        \sigma^{mn} = {1\over 4}(\sigma^m \bar \sigma^n - \sigma^n 
            \bar\sigma^m ) \ .
\eeq{sigmadef}
I encourage you to verify that these tranformations obey the algebra
\beq
    [\delta_\xi, \delta_\eta ] = 2 i \left(\xi^\dagger \bar\sigma^m \eta
      - \eta^\dagger \bar\sigma^m \xi \right) \, \del_m + \delta_\alpha \ ,
\eeq{vectortrans}
where $\delta_\alpha$ is a gauge tranformation with the gauge parameter
\beq
         \alpha = -2i (\xi^\dagger\bar\sigma^m\eta - \eta^\dagger \bar\sigma^m
                  \xi) A^a_m \ .
\eeq{thegaugetrans}
Acting on $\lambda^a$, the  extra term $\delta_\alpha$ in \leqn{vectortrans} 
can be combined with 
the translation to produce  the commutation relation
\beq
    [\delta_\xi, \delta_\eta ]\, \lambda^a 
       = 2 i \left(\xi^\dagger \bar\sigma^m \eta
      - \eta^\dagger \bar\sigma^m \xi \right) \, (\D_m \lambda)^a\ .
\eeq{vectortranstwo}
 This rearrangement applies also for the auxiliary field
$D^a$ and for any matter field
that tranforms linearly under $G$.  The gauge field $A^{a m}$ 
does not satisfy this last criterion; instead, we find
\beqa
 [ \delta_\xi, \delta_\eta] A^a_m & =& 2i (\xi^\dagger \bar \sigma^n \eta - 
         \eta^\dagger \bar \sigma^n \xi) (\del_n A_m^a - {\cal D}_m A_n) \CR
      &=&  2i (\xi^\dagger \bar \sigma^n \eta - 
         \eta^\dagger \bar \sigma^n \xi)   \, F_{nm}^a
\eeqa{deltaforA}
The proof that \leqn{vectorrep} satisfies the supersymmetry algebra
is more tedious than for \leqn{phiACR}, but it is not 
actually difficult.  For the transformation of $\lambda^a$ we need 
both the Fierz identity \leqn{firstFierz} and the relation
\beq
           \eta_\alpha \xi_\beta - (\xi \leftrightarrow \eta) = 
       - ( \xi^T c \sigma_{pq} \eta)     (\sigma^{pq}c)_{\alpha\beta}\ .   
\eeq{secondFierz}
The matrices $\sigma^{pq}c$ and $c\bar\sigma^{pq}$ are symmetric in their 
spinor indices.

Again, the transformation laws leave a simple Lagrangian invariant. For 
the vector supermultiplet, this Lagrangian is that of the renormalizable 
Yang-Mills theory including the gaugino:
\beq
         \L_F = - {1\over 4} (F^a_{mn})^2 + \lambda^{\dagger a} i \bar \sigma
       \cdot \D \lambda^a + \half (D^a)^2
\eeq{vectorL}
The kinetic term for $D^a$ contains no derivatives, so this field will be a 
Lagrange multiplier.

The vector supermultiplet can be coupled to matter particles in chiral 
supermultiplets.  To do this, we must first modify the transformation
laws of the chiral supermultiplet so that the commutators of supersymmetry
transformations obey \leqn{vectortrans} or \leqn{vectortranstwo}.
The modified transformation laws are:
\beqa
    \delta_\xi \phi &=& \sqrt{2} \xi^T c \psi  \CR
    \delta_\xi \psi & =& \sqrt{2} i \sigma^n c \xi^* \D_n \phi + 
        \sqrt{2} F \xi \CR
    \delta_\xi F &=&  - \sqrt{2} i \xi^\dagger \bar \sigma^m \D_m \psi 
             - 2 g \xi^\dagger c \lambda^{a*} t^a \phi
\eeqa{chiralreptwo}
In this formula, the chiral fields $\phi$, $\psi$, $F$ must belong to the
same representation of $G$, with $t^a$ a representation matrix in that 
representation.  From the transformation laws, we can construct the 
Lagrangian.  Start from \leqn{Lchiral}, replace the derivatives by
covariant derivatives, add terms to the Lagrangian involving 
the $\lambda^a$ to cancel the supersymmetry variation of these terms,
and then add terms involving $D^a$ to cancel the remaining supersymmetry
variation of the $\lambda^a$ terms.  The result is
\beqa
  \L_D &=& \D^m \phi^* \D_m\phi  + \psi^\dagger i \bar\sigma\cdot \D \psi
       + F^* F \CR
       & & \qquad - \sqrt{2} g (\phi^* \lambda^{aT} t^a c \psi - \psi^\dagger
          c   \lambda^{a *} t^a \phi) + g D^a \phi^a t^a \phi \ .
\eeqa{Lchiralgauge}
The proof that this Lagrangian is supersymmetric, $\delta_\xi \L = 0$,
 is completely straightforward,
but it requires a very large sheet of paper.

 The gauge invariance of the theory requires the 
superpotential Lagrangian $\L_W$ to be invariant under $G$ as a global 
symmetry.  Under this condition, $\L_W$,  which contains no derivatives,
is invariant under \leqn{chiralreptwo} without modification.   The 
combination of $\L_F$, $\L_D$, and $\L_W$, with $W$ a polynomial of 
degree at most 3, gives the most general renormalizable supersymmetric
gauge theory.

As we did with the $F$ field of the chiral multiplet, it is interesting to
eliminate the Lagrange multiplier $D^a$.  For the Lagrangian which is 
the sum of \leqn{vectorL} and \leqn{Lchiralgauge}, the equation of motion for
$D^a$ is 
\beq
                 D^a = - g \phi^* t^a \phi\ .
\eeq{Dtermval}
Eliminating $D^a$ gives a second potential energy term proportional to 
$(D^a)^2$.  This is the {\it D-term potential} promised below \leqn{VF}. 
I will write the result for a theory with several chiral 
multiplets:
\beq
        V_D = \half g^2 \left( \sum_k \phi^*_k t^a \phi_k \right)^2 \  .
\eeq{Dpotential}
As with the F-term potential,
 $V_D \geq 0$ and vanishes if and only if $D^a = 0$. It can be 
shown by an argument similar to \leqn{psitoF} that 
\beq
            \bra{0} D^a \ket{0}  = 0 
\eeq{Dvacval}
unless supersymmetry is spontaneously broken.

It makes a nice illustration of this formalism to show how the Higgs
mechanism works in supersymmetry.  For definiteness, consider a 
supersymmetric gauge theory with the gauge group $U(1)$. 

Introduce
chiral supermultiplets $\phi_+$, $\phi_-$, and $X$, with charges
+1, $-1$, and 0, respectively, and the superpotential
\beq
        W = \lambda (\phi_+ \phi_- - v^2) X\ .
\eeq{Wforhiggs}
The $F = 0$ equations are
\beq
     F_X^* =  (\phi_+ \phi_- - v^2) = 0 \qquad   F_\pm^* = \phi_\pm X = 0 \ .
\eeq{Fequations}
To solve these equations, set 
\beq
    X = 0 \qquad   \phi_+ = v/y  \qquad   \phi_- = vy \ ,
\eeq{Fsolutions}
where $y$ is a complex-valued parameter.  The $D = 0$ equation is
\beq
         \phi^\dagger_+ \phi_+ -    \phi^\dagger_- \phi_- = 0 \ .
\eeq{Dequations}
This implies $|y| = 1$.  So $y$ is a pure phase and can be removed by a 
$U(1)$ gauge transformation.

Now look at the pieces of the Lagrangian that give mass to gauge bosons, 
fermions, and scalars.  The gauge field receives mass from the Higgs
mechanism.  To compute the mass, we can look at the scalar kinetic 
terms
\beq
  \phi_+^\dagger (-\D^2) \phi_+ +  \phi_-^\dagger (-\D^2) \phi_- =  \cdots
 +   \phi_+^\dagger ( g^2 A^2) \phi_+ +  \phi_-^\dagger (g^2 A^2) \phi_- \ .
\eeq{gaugekin}
Putting in the vacuum expectation values $\phi_+ = \phi_- = v$, we find
\beq
         m^2 =  4 g^2 v^2 
\eeq{theHiggsmass}
for the vector fields. The mode of the scalar field
\beq
       \delta \phi_+ = \eta/\sqrt{2} \qquad \delta\phi_- = - \eta/\sqrt{2}\ ,
\eeq{goodmode}
with $\eta$ real, receives a mass from the $D$-term potential energy
\beq
          {g^2\over 2} (\phi_+^\dagger \phi_+ - \phi^\dagger_- \phi_-)^2
\eeq{scalartomass}
Expanding to quadratic order in $\eta$, we see that  $\eta$ also receives
the mass $m^2 = 4 g^2 v^2$.  The corresponding mode for $\eta$ imaginary is
the infinitesimal version of the phase rotation of $y$ that we have already
gauged away below \leqn{Dequations}.
  The mode of the fermion fields
\beq
       \delta \psi_+ = \chi/\sqrt{2} \qquad \delta\psi_- = - \chi/\sqrt{2}
\eeq{goodfmode}
mixes with the gaugino through the term
\beq
    - \sqrt{2} g ( \phi^\dagger_+ \lambda^T c \psi_+ - \phi^\dagger_- 
    \lambda^T c \psi_-) + h.c. 
\eeq{fermiontomass}
Putting in the vacuum expectation values  $\phi_+ = \phi_- = v$, we find
a Dirac mass with the value
\beq
         m =  2 g v
\eeq{diracmassval}
In all, we find a massive vector boson, a massive real scalar, and a 
massive Dirac fermion, all with the mass $m = 2 g v$.  The system has
four physical bosons and four physical 
fermions, all with the same mass, as supersymmetry
requires. 

\subsection{The Vector Supermultiplet in Superspace}

The vector supermultiplet has a quite simple representation
in superspace.  This multiplet turns out to be the answer to the question
that we posed in our discussion of superspace in the previous section:  When
the chiral and antichiral components of a general superfield 
 are removed, what is left over?  To analyze this issue, I will
write a Lagrangian containing a local symmetry that allows us to gauge 
away the chiral and antichiral components of this superfield.  Let 
$V(x, \theta, \bar\theta)$ be a real-valued superfield, acted on by a 
local gauge transformation  in superspace
\beq
      \delta V = - {i\over g} (\Lambda - \Lambda^\dagger)
\eeq{Vgaugetrans}
where $\Lambda$ is a chiral superfield and $\Lambda^\dagger$ is its 
conjugate.  Since $\Lambda$ satisfies \leqn{Chiraldefin}, its expansion 
in powers of $\theta$ contains
\beq
   \Lambda(x,\theta,\bar\theta) = \Lambda(x+ i \bar\theta \bar \sigma \theta, 
\theta) 
  =  \alpha(x) + \cdots + i \bar\theta \bar \sigma^m \theta \del_m \alpha(x)
 + \cdots
\eeq{Lambdadecomp}
The general superfield $V$ contains a term\footnote{The factor 2 in this 
equation is convenient but disagrees with some standard treatments, \eg,
\cite{WessBagger}.}
\beq
         V(x, \theta, \bar\theta) = \cdots + 2  \bar\theta \bar \sigma^m \theta
  \,  A_m(x)  + \cdots
\eeq{Adecomp}
So the superfield $V$ contains a space-time vector field $A_m(x)$, and under
\leqn{Vgaugetrans}, $A_m$ transforms as
\beq
          \delta A_m = {1\over g} \del_m ( \Re\, \alpha) \ .
\eeq{Agaugetrans}
This is just what we would like for an Abelian gauge field.  So we should
accept \leqn{Vgaugetrans} as the generalization of the Abelian gauge
 transformation to 
superspace.

The real-valued superfield transforming under \leqn{Vgaugetrans} is called 
a {\it vector superfield}.  To understand its structure, use the gauge 
transformation to remove all components with powers of $\theta$ or 
$\bar\theta$ only. This choice is called 
{\it Wess-Zumino gauge}~\cite{WZgauge}.  What remains after this gauge choice
is
\beq
   V(x,\theta, \bar\theta) = 2  \bar\theta \bar \sigma^m \theta
  \,  A_m(x)  +  2 \bar \theta^2 \theta^T c \lambda - 2 \theta^2 \bar \theta^T
 c \lambda^* + \theta^2 \bar\theta^2 D \ .
\eeq{Vdecomp}
This expression has exactly the field content of the Abelian 
vector supermultiplet 
$(A_m, \lambda, D)$. 

This gauge multiplet can be coupled to matter described by chiral 
superfields.  For the moment, I will continue to discuss the Abelian gauge
theory.  For a chiral superfield $\Phi$ with charge $Q$, 
the gauge transformation
\beq
    \delta \Phi = i Q \Lambda \Phi
\eeq{Phigaugetrans}
contains a standard Abelian gauge transformation with gauge parameter
$\Re\, \alpha(x)$ and also preserves the chiral nature of $\Phi$.  Then
the superspace Lagrangian
\beq
    \int d^2\theta d^2\bar\theta \  \Phi^\dagger e^{g Q V} \Phi
\eeq{gaugematterL}
is gauge-invariant.  Using the representation \leqn{Vdecomp} and the rules
\leqn{thetaint}, it is 
straightforward to carry out the integrals explicitly and show that 
\leqn{gaugematterL} reduces to \leqn{Lchiralgauge}, with $t^a = Q$ for 
this Abelian theory.

We still need to construct the pure gauge part of the Lagrangian.  To 
do this, first note that, because a quantity antisymmetrized on three 
Weyl fermion indices vanishes,
\beq
       \bar D_\alpha  {\bar D}^2 X = 0
\eeq{DDXid}
for any superfield $X$.  Thus, acting with ${\bar D}^2$ makes any 
superfield a chiral superfield.  The following is a chiral superfield
that also has the property that its leading component is the gaugino
field $\lambda(x)$:
\beq
     W_\alpha = - {1\over 8} {\bar D}^2 (D c)_\alpha V \ .
\eeq{Wdefin}
Indeed, working this out in full detail, we find that $W_\alpha = 
W_\alpha(x + i \bar \theta \sigma\theta, \theta)$, with 
\beq
  W_\alpha(x,\theta) = \lambda_\alpha + \left[ ( i \sigma^{mn} F_{mn} + D)
  \theta\right]_\alpha + \theta^Tc\theta
 \left[ \del_m\lambda^* i \bar\sigma^m c
  \right]_\alpha\ .
\eeq{Wcontent}

The chiral superfield $W_\alpha$ is the superspace analogue of the 
electromagnetic field strength.  The Lagrangian
\beq
      \int d^2 \theta \, \half W^T c W  
\eeq{WWLagrangian}
reduces precisely to the Abelian version of \leqn{vectorL}.  It is odd
that the kinetic term for gauge fields is an F-term rather than a 
D-term.  It turns out that this term can be renormalized by loop corrections
as a consequence of the trace anomaly~\cite{GrisaruWW}.  However, 
the restricted form of the correction has implications, both some 
simple ones that I will discuss later in Section~4.3 and 
and more profound implications discussed, for example, 
in~\cite{NSWZ,AHM}.

I will simply quote the generalizations of these results to the 
non-Abelian case.  The gauge transformation of a chiral superfield 
in the representation $R$ of the gauge group is
\beq
     \Phi \to e^{i \Lambda^a t^a } \Phi \qquad 
     \Phi^\dagger \to \Phi^\dagger e^{-i \Lambda^{\dagger a} t^a }\ ,
\eeq{PhiNAtrans}
where $\Lambda^a$ is a chiral superfield in the adjoint representation 
of $G$ and $t^a$ is is the representation of the generators of $G$ in the 
representation $R$.  The gauge transformation of the vector superfield
is
\beq
      e^{g V^a t^a} \to  e^{i \Lambda^{\dagger a} t^a }  e^{g V^a t^a}
     e^{-i \Lambda^a t^a }
\eeq{VNAtrans}
Then the Lagrangian 
\beq
      \int d^2\theta d^2 \bar \theta \, \Phi^\dagger e^{g V^a t^a} \Phi
\eeq{PhiNALag}
is locally gauge-invariant.  Carrying out the integrals in the gauge
\leqn{Vdecomp} reduces this Lagrangian to \leqn{Lchiralgauge}.

The form of the field strength superfield is rather more complicated
than in the Abelian case,
\beq
     W_\alpha^a t^a = - {1\over 8 g} {\bar D}^2 e^{-gV^a t^a} (Dc)_\alpha
           e^{g V^a t^a}
\eeq{NAWdefin}
In Wess-Zumino gauge, 
this formula does reduce to the non-Abelian version of \leqn{Wcontent},
\beq
  W_\alpha^a(x,\theta) = \lambda_\alpha^a
 + \left[ ( i \sigma^{mn} F_{mn}^a + D^a)
  \theta\right]_\alpha + \theta^T c\theta
 \left[ \D_m\lambda^{* a} i \bar\sigma^m c
  \right]_\alpha\ .
\eeq{NAWcontent}
Then the Lagrangian
\beq
       \int d^2\theta \, \tr[W^T c W] 
\eeq{NAWWLag}
reduces neatly to \leqn{vectorL}.  

The most general renormalizable supersymmetric Lagrangian can be 
built out of these ingredients.  We need to put together
the Lagrangian \leqn{NAWWLag}, plus a term \leqn{PhiNALag} for each 
matter chiral superfield, plus a superpotential Lagrangian to 
represent the scalar field potential energy.  These formulae can be 
generalized to the case of a nonlinear sigma model on  a K\"ahler manifold,
with the gauge symmetry associated with an isometry of this target 
space.  For the details, see~\cite{WessBagger}.

\subsection{R-Symmetry}

The structure of the general superspace action for a renormalizable
theory of scalar and fermion fields suggests that this theory has a
natural continuous symmetry.  

The superspace Lagrangian is 
\beq
    \L   =   \int d^2\theta \, \tr[W^T c W] + 
 \int d^4\theta \, \Phi^\dagger e^{gV\cdot t} \Phi + 
         \int d^2 \theta \, W(\Phi) + \int d^2\bar \theta \, 
            (W(\Phi))^\dagger \ . 
\eeq{mostgeneralrenL}
Consider first the case in which $W(\phi)$ contains only dimensionless
parameters and is therefore a cubic polynomial in the scalar fields.
Then $\L$ is invariant under the $U(1)$ symmetry
\beq
   \Phi_k(x,\theta) \to e^{-i2\alpha/3} \Phi_k(x, e^{i\alpha}\theta) \ ,
    \quad   V^a(x,\theta, \bar\theta) \to V^a(x, e^{i\alpha}\theta,
                 e^{-i\alpha}\bar\theta)
\eeq{Rsymmetry}
or, in components,  
\beq
   \phi_k \to e^{-i2\alpha/3} \phi_k \ , \quad 
                      \psi_k \to  e^{i\alpha/3} \psi_k \ , \quad 
           \lambda^a \to e^{-i \alpha} \lambda^a\ ,
\eeq{Rsymmetrycomps}
and the gauge fields are invariant.
This transformation is called {\it R-symmetry}.  Under R-symmetry, the 
charges of bosons and fermions differ by 1 unit, in such a way that that
the gaugino and superpotential vertices have zero net charge.

Since all left-handed fermions have the same charge 
under \leqn{Rsymmetrycomps}, the R-symmetry will have an axial vector 
anomaly.  It can be shown that the R-symmetry current (of dimension 3, spin 1)
forms a supersymmetry multiplet together with the supersymmetry
current (dimension ${7\over 2}$, spin ${3\over 2}$) and the energy-momentum 
tensor (dimension 4, spin 2)~\cite{FerraraZumino}.  All three currents
have perturbation-theory anomalies; the anomaly of the energy-momentum 
tensor is the trace anomaly, associated with 
the breaking of scalar invariance by coupling constant 
renormalization.  The R-current anomaly is thus connected to the running
of coupling constants and gives a useful formal approach to study  
this effect in supersymmetric models. 

It is often possible to combine the transformation \leqn{Rsymmetry} with 
other apparent $U(1)$ symmetries of the theory to define a non-anomalous
$U(1)$ R-symmetry.  Under such a symmetry, we will have 
\beq
   \Phi_k(x,\theta) \to e^{-i\beta_k} \Phi_k(x, e^{i\alpha}\theta) \ ,
 \quad \mbox{such\ that} \quad 
   W(x,\theta) \to e^{2i\alpha} W(x,e^{i\alpha}\theta) \ . 
\eeq{generalR}
Such symmetries also often arise in models in which the superpotential 
has dimensionful coefficients.

In models with extended, $N > 1$, supersymmetry, the R-symmetry group is also 
extended, to $SU(2)$ for $N = 2$ and to $SU(4)$ for $N=4$ supersymmetry.

\section{The Minimal Supersymmetric Standard Model}

\subsection{Particle Content of the Model}

Now we have all of the ingredients to construct a supersymmetric
generalization of the Standard Model.  To begin, let us construct a 
version of the Standard Model with exact supersymmetry.  To do this,
we assign the vector fields in the Standard Model to vector supermultiplets
and the matter fields of the Standard Model to chiral supermultiplets.

The vector supermultiplets correspond to the generators of $SU(3) \times
SU(2) \times U(1)$.  In these lectures, I will refer to the gauge bosons
of these groups as $A^a_m$, $W^a_m$, and $B_m$, respectively.   I will 
represent the Weyl fermion partners of these fields 
as $\s g^a$, $\s w^a$, $\s b$.  I will
call these fields the {\it gluino}, {\it wino}, and {\it bino}, or,
collectively, {\it gauginos}. In the later parts of these lectures, I 
will drop the tildes over the gaugino fields when they are not needed
for clarity.

I will assign the  quarks and leptons to be fermions in chiral superfields.
I will use the convention presented in Section~1.3 of considering
left-handed Weyl fermions as the basic particles and right-handed Weyl
fermions as their antiparticles.  In the Standard Model, the left-handed
fields in a fermion generation have the quantum numbers
\beq
    L = \pmatrix{ \nu\cr e \cr} \qquad \bar e \qquad
    Q =  \pmatrix{ u\cr d \cr } \qquad \bar u \qquad 
        \bar d
\eeq{Lgeneration}
The field $\bar e$ is the left-handed positron; the fields $\bar u$, 
$\bar d$ are the left-handed antiquarks.  The right-handed Standard Model
fermion fields are the conjugates of these fields.
To make a generalization to supersymmetry, we will extend each of the
fields in \leqn{Lgeneration}---for each of the three generations---to 
a chiral supermultiplet.  I will use the symbols
\beq
     \s L \qquad   \s{\bar e} \qquad \s Q \qquad \s{\bar u} \qquad
            \s{\bar d}
\eeq{sfermions}
to represent both the supermultiplets and the scalar fields in these
multiplets.  Again, I will drop the tilde if it is unambiguous that I am 
referring to the scalar partner rather than the fermion.  The 
scalar particles in these supermultiplets are called {\it sleptons} and 
{\it squarks}, collectively, {\it sfermions}.

What about the Higgs field?  The Higgs field of the Standard Model should
be identified with a complex scalar component of a chiral supermultiplet.
But it is ambiguous what the quantum numbers of this multiplet should be.
In the Standard Model, the Higgs field is a color singlet with $I = \half$, 
but we can take the hypercharge of this field to be either $Y = +\half$ or
$Y = -\half$, depending on whether we take the positive hypercharge field
or its conjugate to be primary.  In a supersymmetric model, the choice 
matters.  The superpotential is an analytic function of superfields, so it
can only contain the field, not the conjugate.  Then
different Higgs couplings will be 
 allowed depending on the choice that we make.

The correct solution to this problem is to include {\it both} possibilities,
That is, we include a Higgs supermultiplet with $Y= +\half$
and a second Higgs supermultiplet with $Y = -\half$. I will call
the scalar components of these multiplets $H_u$ and $H_d$, respectively:
\beq
      H_u =\pmatrix{ H^+_u \cr H^0_u \cr } \qquad
      H_d =\pmatrix{ H^0_d \cr H^-_d \cr } 
\eeq{Higgsfields}
I will refer to the Weyl fermion components with these quantum numbers
as $\s h_u$, $\s h_d$.  These fields or particles are called {\it Higgsinos}. 

\begin{figure}
\begin{center}
\includegraphics[height=0.8in]{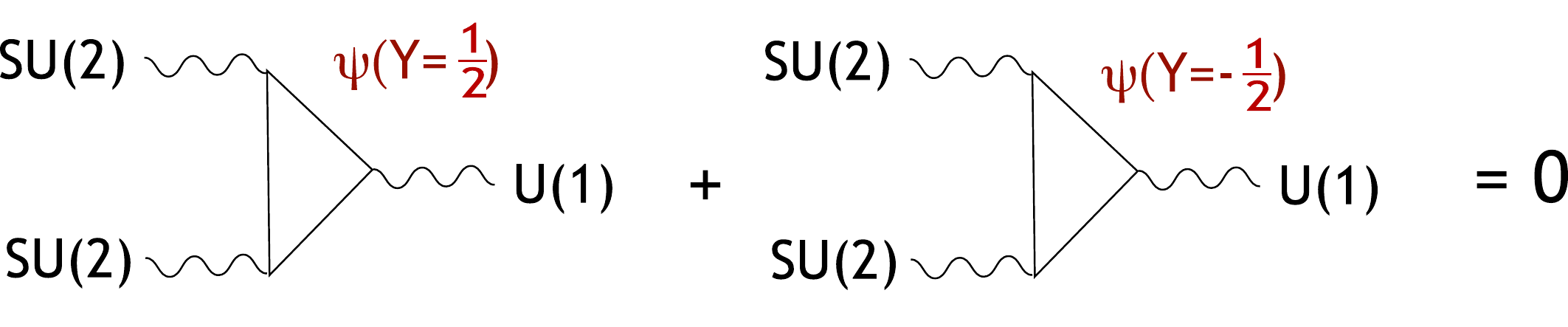}
\caption{The anomaly cancellation that requires two doublets of Higgs 
   fields in the MSSM.}
\label{fig:anomaly}
\end{center}
\end{figure}

I will argue below that it is necessary to 
include both Higgs fields in order to obtain all of the needed 
couplings in the superpotential.  However, there is another argument.
The axial vector anomaly of one $U(1)$ and two 
$SU(2)$ currents (Fig.~\ref{fig:anomaly}) must vanish to maintain the 
gauge invariance of the model.  In the Standard Model, the anomaly
cancels nontrivially between the quarks and the leptons.  In the 
supersymmetric generalization of the Standard Model, each Higgsino makes
a nonzero contribution to this anomaly.  These contributions cancel if we
include a pair of Higgsinos with opposite hypercharge.

\subsection{Grand Unification}

Before writing the Lagrangian in detail, I would like to point out that
there is an interesting conclusion that follows from the 
quantum number assignments of the new particles that we have introduced
to make the Standard Model supersymmetric.

An attractive feature of the Standard Model is that the quarks and 
leptons of each generation fill out multiplets of the simple 
gauge group $SU(5)$.  This suggests a very beautiful picture,
called {\it grand unification}, in which $SU(5)$, or a group such as 
$SO(10)$ or $E_6$ for which this is a subgroup,
is the  fundamental gauge symmetry at very short
distances. This unified symmetry will be  spontaneously broken to 
the Standard Model gauge group
$SU(3) \times SU(2) \times U(1)$.   

For definiteness, I will examine the  model in which  the grand unified
symmetry group is $SU(5)$.  The generators of $SU(5)$ can be represented 
as $5\times 5$ Hermitian matrices acting on the 5-dimensional vectors
in the fundamental representation.  To see how the Standard Model 
is embedded in $SU(5)$, it is convenient to write these matrices as blocks
with 3 and 2 rows and columns.  Then the Standard Model generators 
can be identified as
\beq
SU(3)\ : \quad \pmatrix{ t^a &  \cr & 0 \cr } \ ; \quad
SU(2)\ : \quad \pmatrix{ 0 &  \cr & \sigma^a/2 \cr }\ ; \quad
U(1)\ : \quad \sqrt{{3\over 5}}
        \pmatrix{ -{1\over 3} \mbox{\bf 1} &  \cr &
                \half \mbox{\bf 1} \cr }  \ .
\eeq{SUfivegens}
In these expressions, $t^a$ is an $SU(3)$ generator, $\sigma^a/2$ is an $SU(2)$
generator, and all of these matrices are normalized to $\tr[T^A T^B] = 
\half \delta^{AB}$.  We should identify the last of these matrices with
$\sqrt{3/5}\ Y$. 

The symmetry-breaking can be caused by the vacuum
expectation value of a Higgs field in the adjoint representation of 
$SU(5)$. The expectation value
\beq
     \VEV{\Phi} =  V \cdot  \pmatrix{ -{1\over 3} \mbox{\bf 1} &  \cr &
                \half \mbox{\bf 1} \cr } 
\eeq{VEVofadj}
commutes with the generators in \leqn{SUfivegens} and fails to commute with 
the off-diagonal generators.  So this vacuum expectation value gives mass
 to the off-diagonal generators and breaks the gauge group to $SU(3) \times
SU(2) \times U(1)$.  

Matter fermions can be organized as left-handed Weyl fermions in the 
$SU(5)$ representations $\bar 5$ and $10$.  The $\bar 5$ is the conjugate
of the fundamental representation of $SU(5)$; the $10$ is the antisymmetric
matrix with two $5$ indices.
\beq
       \bar 5\ : \quad \pmatrix{
    \bar d \cr \bar d \cr \bar d \cr e \cr \nu \cr }_L \ ; \qquad 
10 \ :\quad \pmatrix{
  0 & \bar u & \bar u & u & d \cr     & 0 & \bar u & u & d \cr
  & & 0  & u & d \cr  & &    & 0 & \bar e \cr  & & & & 0 \cr 
        }_L
\eeq{fiveandten}
It is straightforward to check that each entry listed has the quantum 
numbers assigned to that field in the Standard Model. To compute the 
hypercharges, we act on the $\bar 5$ with $(-1)$ times the hypercharge
generator in \leqn{SUfivegens}, and we act on the $10$ with the 
hypercharge generator on each index.  This gives the standard results,
for example, $Y = + {1\over 3}$ for the $\bar d$ and $Y = -\third + \half 
= {1\over 6}$ for $u$ and $d$. 

The $SU(5)$ covariant derivative is 
\beq
   \D_m = (\del_m - i g_U A^A_m T^A) \ , 
\eeq{SUfivecovderiv}
where $g_U$ is the $SU(5)$ gauge coupling.  There is only room for one 
value here.  So this model predicts that the three Standard Model
gauge couplings are related by 
\beq
          g_3 =  g_2 = g_1 = g_U \ ,
\eeq{threecouplings}
where
\beq 
     g_3 = g_s \qquad  g_2 = g \qquad g_1 = \sqrt{5\over 3} g' \ .
\eeq{idcouplings}
Clearly, this prediction is not correct for the gauge couplings that we 
measure in particle physics.

However, there is a way to save this prediction.  In quantum field theory,
coupling constants are functions of length scale and change their values
significantly from one scale to another by renormalization group evolution.
It is possible that the values of $g'$, $g$, and $g_s$ that we measure
could evolve at very short distances into values that obey 
\leqn{threecouplings}.

I will now collect the formulae that we need to analyze this question.
Let 
\beq
     \alpha_i = {g_i^2\over 4 \pi}
\eeq{alphaidef}
for $i = 1,2,3$. 
The one-loop renormalization group equations for gauge couplings are
\beq
  {d g_i \over d \log Q} = - {b_i\over (4\pi)^2} g_i^3 \qquad \mbox{or}
\qquad  {d \alpha_i \over d \log Q} = - {b_i\over (2\pi)} \alpha_i^2  \ .
\eeq{RGEsforalpha}
For $U(1)$, the coefficient $b_1$ is 
\beq
         b_1 = - {2\over 3} \sum_f {3\over 5} Y_f^2 
         - {1\over 3} \sum_b  {3\over 5} Y_b^2 \ ,
\eeq{bonefromY}
where the two sums run over the multiplets of left-handed Weyl fermions
and complex-valued bosons. The factors ${3\over 5}Y^2$ are the squares
of the $U(1)$ charges defined by \leqn{SUfivegens}. 
 For non-Abelian groups, the expressions
for the $b$ coefficients are
\beq
   b = - {11\over 3} C_2(G) - {2\over 3} \sum_f C(r_f) - {1\over 3} 
\sum_b C(r_b) \ , 
\eeq{bcoeff}
where $C_2(G)$ and $C(r)$ are the standard group theory coefficients.
For $SU(N)$,
\beq
   C_2(G) = C(G) =  N \ , \quad C(N) = \half  \ .
\eeq{Cvals}
The solution of the renormalization group equation \leqn{RGEsforalpha} is 
\beq
    \alpha^{-1}(Q) = \alpha^{-1}(M) - {b_i\over 2\pi} \log {Q\over M} \ .
\eeq{RGsolution}

Now consider the situation in which the three couplings $g_i$ 
become equal at the mass scale $M_U$, the mass scale of $SU(5)$ 
symmetry breaking.  Let $\alpha_U$ be the value 
of the $\alpha_i$ at this scale.  Using \leqn{RGsolution}, we can then 
determine the Standard Model couplings at any lower mass scale.  The 
three $\alpha_i(Q)$ are determined by two parameters.  We can eliminate
those parameters and obtain the relation
\beq
   \alpha^{-1}_3 = (1 + B) \alpha^{-1}_2 - B \alpha^{-1}_1
\eeq{arelation}
where 
\beq
      B = {b_3 - b_2 \over b_2 - b_1 }  \ . 
\eeq{Bvalue}
  The values of the $\alpha_i$ are known very accurately
at $Q = \mz$~\cite{couplingvals}:
\beq
      \alpha_3^{-1} =  8.50\pm 0.14 \qquad    \alpha_2^{-1} = 29.57\pm 0.02
    \qquad  \alpha_1^{-1} = 59.00\pm 0.02 \ .
\eeq{alphavals}
Inserting these values into \leqn{arelation}, we find
\beq
         B = 0.716 \pm 0.005 \pm 0.03 \ .
\eeq{Bvaluenumber}
In this formula, the first error is that propagated from the errors
in \leqn{alphavals} and the second is my estimate of the systematic error
from neglecting the  two-loop renormalization group coefficients and other
 higher-order corrections. 

We can compare the value of $B$ in \leqn{Bvaluenumber} to the values of 
\leqn{Bvalue} from different models.  The hypothesis that the three
Standard Model couplings unify is acceptable only if the gauge theory
that describes physics between $\mz$ and $M_U$ gives a value of 
$B$ consistent with \leqn{Bvaluenumber}.  The minimal Standard Model 
fails this test.  The values of the $b_i$ are
\beqa
        b_3 &=& 11 - {4\over 3} n_g \CR
        b_2 &=& {22\over 3} - {4\over 3} n_g - {1\over 6} n_h \CR
        b_1 &=& \phantom{{22\over 3}}- {4\over 3} n_g  - {1\over 10} n_h 
\eeqa{MSMbvals}
where $n_g$ is the number of generations and $n_h$ is the number of 
Higgs doublets.  Notice that $n_g$ cancels out of \leqn{Bvalue}.  
This is to be expected.  The Standard Model fermions form complete 
representations of $SU(5)$, and so their renormalization effects cannot
lead to differences among the three couplings.  For the minimal case
$n_h  = 1$ we find  $B = 0.53$.  To obtain a value consistent with 
\leqn{Bvaluenumber}, we need $n_h = 6$.

We can redo this calculation in the minimal supersymmetric version of the
Standard Model.  First of all, we should rewrite \leqn{bcoeff} for a 
supersymmetric model with one vector supermultiplet, containing a 
vector and a Weyl fermion in the adjoint representation, and a set of 
chiral supermultiplets indexed by $k$, 
each with a Weyl fermion and a complex boson.
Then \leqn{bcoeff} becomes
\beqa
      b_i &=& {11\over 3} C_2(G) - {2\over 3} C_2(G) - \left(
  {2\over 3} + {1\over 3} \right) \sum_{k} C(r_k) \CR
 &=& 3  C_2(G) -  \sum_k C(r_k) 
\eeqa{bSUSYvals}
The formula \leqn{bonefromY} undergoes a similar rearrangement.
Inserting the values of the $C(r_k)$ for the fields of the Standard Model, 
we find
\beqa
        b_3 &=& 9  - 2 n_g \CR
        b_2 &=& 6 - 2 n_g - {1\over 2} n_h \CR
        b_1 &=& \phantom{6}- 2 n_g  - {3\over 10} n_h 
\eeqa{MSSMbvals}
For the minimal Higgs content $n_h = 2$, this gives
\beq
        B = {5\over 7} = 0.714
\eeq{BSUSYval}
in excellent agreement with \leqn{Bvaluenumber}.

\begin{figure}
\begin{center}
\includegraphics[height=3.0in]{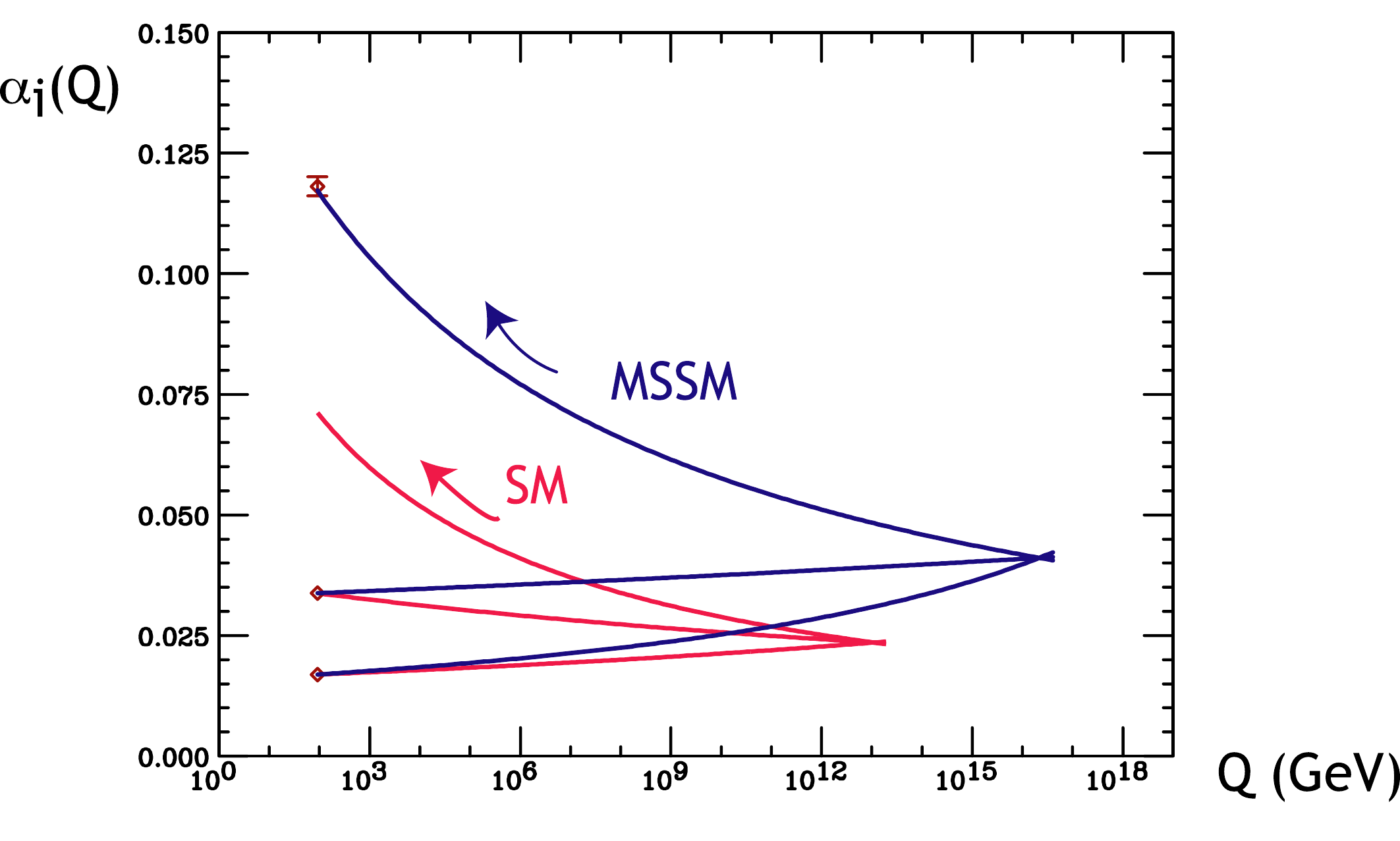}
\caption{Prediction of the $SU(3)$ gauge coupling $\alpha_s$ from the 
     electroweak coupling constants using grand unification, in the
        Standard Model and in  the MSSM.}
\label{fig:unification}
\end{center}
\end{figure}

In Fig.~\ref{fig:unification}, I show the unification relation 
pictorially.  The three data points on the the left of the figure
represent the measured values of the three couplings \leqn{alphavals}.
Starting from the values of $\alpha_1$ and $\alpha_2$, we can integrate
\leqn{RGEsforalpha} up to the scale at which these two couplings converge.
Then we can integrate the equation for $\alpha_3$ back down to $Q = \mz$
and see whether the result agrees with the measured value. The 
lower set of curves presents the result for the Standard Model with $n_h =1$.
The upper set of curves shows the result for the supersymmetric extension 
of the Standard Model with $n_h =2 $.  This choice gives excellent agreement
with the measured value of $\alpha_s$.  

Actually, I slightly overstate the case for supersymmetry by ignoring 
two-loop terms in the renormalization group equations, and also by integrating
these equations all the way down to $\mz$ even though, 
from searches at high-energy colliders, most of the squarks and gluinos must
be heavier than 300~GeV.  A more accurate prediction of $\alpha_s(\mz)$ from 
the electroweak coupling constants gives a 
 slightly higher value, 0.13 instead
of 0.12.  However, these corrections could easily be compensated by similar
corrections to the upper limit of the integration, following the 
details of the particle spectrum at the grand unification scale.
For a more detailed formal analysis of these corrections, see~\cite{Yamada}, 
and
for a recent evaluation of their effects, see~\cite{GUTinSUSY}. 
 It remains a remarkable fact that 
the minimal supersymmetric extension of the Standard Model is approximately
compatible with grand unification `out of the box', with no need for further
model-building.

\subsection{Construction of the Lagrangian}

Now I would like to write the full Lagrangian of the minimal supersymmetric
extension of the Standard Model, which I will henceforth call the MSSM.

The kinetic terms and gauge couplings of the MSSM Lagrangian are completely
determined by supersymmetry, the choice of the gauge group $SU(3) \times
SU(2) \times U(1)$, and the choice of the quantum numbers of the matter
fields.  The Lagrangian is a sum of terms of the forms
\leqn{vectorL}  and \leqn{Lchiralgauge}.  Up to this point, 
the only parameters that 
need to be introduced are the gauge couplings $g_1$, $g_2$, and $g_3$. 

Next, we need a superpotential $W$. The superpotential is the source of 
nonlinear fermion-scalar interactions, so we should include the appropriate
terms to generate the Higgs Yukawa couplings needed to give mass to the 
quarks and leptons.  The appropriate choice is
\beq
 W_Y = y_d^{ij} \bar d^i H_{d\alpha} \epsilon_{\alpha\beta} Q^j_\beta + 
 y_e^{ij} \bar e^i H_{d\alpha} \epsilon_{\alpha\beta} L^j_\beta 
- y_u^{ij} \bar u^i H_{u\alpha} \epsilon_{\alpha\beta} Q^j_\beta \ .
\eeq{WYform}
The notation for the quark and lepton multiplets is that in \leqn{Lgeneration};
the indices $i,j = 1,2,3$ run over the three generations. The indices
$\alpha, \beta = 1,2$ run over SU(2) isospin indices.  Notice that 
the first two terms require a Higgs field $H_d$ with $Y= -\half$, while
the third term requires a Higgs field $H_u$ with $Y = \half$. If we leave out
one of the Higgs multiplets, some quarks or leptons will be left
massless. This is the 
second argument that requires two Higgs fields in the MSSM.

I have written \leqn{WYform} including the most general mixing between 
left- and right-handed quarks and leptons of different generations.  However,
as in the minimal Standard Model, we can remove most of this flavor 
mixing by appropriate field redefinitions.  The coupling constants
$y_d$, $y_e$, $y_u$ are general  $3\times 3$ complex-valued matrices. Any
such matrix can be diagonalized using two unitary transformations.  Thus,
we can write
\beq
   y_d = W_d Y_d V^\dagger_d \qquad   y_e = W_e Y_e V^\dagger_e  \qquad
           y_u = W_u Y_u V^\dagger_u \ ,  
\eeq{Yukawadiag}
with $W_a$ and $V_a$ $3\times 3$ unitary matrices and $Y_a$ real, positive,
and diagonal.  The unitary transformations cancel out of the kinetic 
energy terms and gauge couplings in the Lagrangian, except that the 
$W$ boson coupling to quarks is transformed
\beq
     g u^\dagger \bar \sigma^m d W_m^+ \to g u^\dagger \bar\sigma^m 
                (V^\dagger_u V_d) d \, W_m^+ \ .
\eeq{CKMdef}
From this equation, we can identify $(V^\dagger_d V_u) = V_{CKM}$, the
Cabibbo-Kobayashi-Maskawa weak interaction mixing matrix. The Lagrangian
term \leqn{WYform} thus introduces the remaining parameters of the Standard 
Model, the 9 quark and lepton masses (ignoring neutrino masses) and 
the 4 CKM mixing angles.  The field redefinition \leqn{Yukawadiag} can also
induce or shift a QCD theta parameter, so the MSSM, like the Standard Model,
 has a strong CP 
problem that requires an axion or another model-building 
solution~\cite{DineTASI}.

There are several other terms that can be added to $W$.  One possible 
contribution is a pure Higgs term
\beq
  W_\mu = - \mu H_{d\alpha} \epsilon_{\alpha\beta} H_{u\beta} \ .
\eeq{Wmudef}
The parameter $\mu$ has the dimensions of mass, and consequently
this {\it mu term} provides a supersymmetric contribution to the 
masses of the Higgs bosons.  Because this term is in the superpotential,
it does not receive additive raditive corrections.  Even in a theory 
that includes grand unification and energies scale of the order of $10^{16}$
GeV, we can set the parameter $\mu$ to be of order 100 GeV without finding
this choice affected by large quantum corrections.
We will see in Section~4.2 that the mu term is needed for 
phenomenological reasons.  If $\mu = 0$, a Higgsino state will be massless
and should have been detected already in experiments.  It is odd
that a theory whose fundamental mass scale is the grand unification
scale should require a parameter containing a weak interaction mass scale.
I will present some models for the origin of this term in 
Section~3.5.

At this point, we have introduced two new parameters beyond those in the 
Standard Model.  One is the value of $\mu$.  The other is the result of 
the fact that we have two Higgs doublets in the model.  The ratio of the
Higgs vacuum expectation values
\beq
            \VEV{H_u} / \VEV{H_d} \equiv \tan \beta
\eeq{tanbetadef}
will appear in many of the detailed predictions of the MSSM.

There are still more superpotential terms that are consistent with 
the Standard Model gauge symmetry and quantum numbers.  These are
\beqa 
 W_{\not R} &=& \eta_1 \epsilon_{ijk}\bar u_i \bar d_j \bar d_k + 
   \eta_2 \bar d \epsilon_{\alpha\beta}L_\alpha Q_\beta \CR
& & +   \eta_3 \bar e \epsilon_{\alpha\beta}L_\alpha L_\beta 
    + \eta_4 \epsilon_{\alpha\beta} L_\alpha H_{u \beta}  \ . 
\eeqa{Rpvdef}
Here $i,j,k$ are color indices, $\alpha,\beta$ are isospin indices, and 
arbitrary generation mixing is also possible.  These terms 
violate baryon and lepton number through operators with dimensionless
coefficients.  In constructing supersymmetric models, it is necessary 
either to forbid these terms by imposing appropriate discrete symmetries or
to arrange by hand that some of the dangerous couplings are extremely 
small~\cite{DimopoulosHallRPV}.

If baryon number $B$ and lepton number $L$ are conserved in a supersymmetric
model, this model respects a discrete symmetry called R-parity,
\beq
           R = (-1)^{3B + L + 2J} \ . 
\eeq{Rparitydef}
Here $(3B)$ is quark number and $J$ is the spin of the particle.  This 
quantity is constructed so that $R = +1$ on the particles of the Standard 
Model (including the Higgs bosons) and $R = -1$ on their supersymmetry
partners. $R$ acts differently on particles of different spin in the 
same supermultiplet, so R-parity is a discrete subgroup of a continuous
R-symmetry. 

In a model with grand unification, there will be 
baryon number and lepton number violation, and so $B$ and $L$ cannot be 
used as fundamental symmetries.  However, we can easily forbid most of the 
superpotential 
terms \leqn{Rpvdef} by introducing a discrete symmetry that distinguishes
the field $H_d$ from the lepton doublets $L_i$.   A similar strategy can be
used to forbid the first, 3-quark, term.  With these additional
discrete symmetries,  the MSSM, including all 
other terms considered up to this point, will conserve R-parity.

\subsection{The Lightest Supersymmetric Particle}

If R-parity  is conserved, the lightest supersymmetric particle
will be absolutely stable.  This conclusion 
has an important implication for the
relation of supersymmetry to cosmology.  If  a supersymmetric particle
is  stable for a time longer than the
age of the universe, and if this particle is electrically neutral, that
particle is a good candidate for the cosmic dark matter. In Sections~6.3
and 6.4, I will discuss in some detail the properties of models in which
the lightest Standard Model superpartner is the dark matter particle.

However, this is not the only possibility.  Over times 
 much longer than those of particle physics experiments---minutes,
years, or billions of years---we need to consider the possibility that 
the lightest Standard Model superpartner will decay to a particle with 
only couplngs of gravitational strength.  Complete supersymmetric models
of Nature must include a superpartner of the graviton, a spin-${3\over 2}$
particle called the {\it gravitino}.  In a model with exact supersymmetry,
the gravitino will be massless, but in a model with spontaneously broken
supersymmetry, the gravitino acquires a mass through an analogue of the 
Higgs mechanism.  If the supersymmetry breaking is induced by one dominant
$F$-term, the value of this mass is~\cite{DZum}
\beq
           m_{3/2} =  {8\pi\over 3} {\VEV{F}\over m_\Pl}\ .
\eeq{gravitinomass}
This expression is of the same order of magnitude as the expressions for 
Standard Model superpartner masses that I will give in Section~3.6.
In string theory and other unified models, there may be additional Standard
Model singlet fields with couplings of gravitation strength, called 
{\it moduli}, that might also be light enough that long-lived Standard Model
superpartners could decay to them.

Supersymmetric models with R-parity conservation and 
 dark matter, then, divide into two classes, according
to the identity of the lightest supersymmetric particle---the LSP.
On one hand, the LSP could be a Standard Model
superpartner. Cosmology requires that this particle is neutral.
 Several candidates are available, including the fermionic
partners of the photon, $Z^0$, and neutral Higgs bosons and the scalar
partner of one of the neutrinos.  In all cases, these particles will be
weakly interacting; when they are produced at high-energy colliders, they
should not make signals in a particle detector.  On the other hand, the
LSP could be the gravitino or another particle with only gravitational
couplings.  In that case, the lightest Standard Model superpartner could
be a charged particle.  Whether this particle is visible or neutral and
weakly interacting, its decay should
be included in the phenomenology of the model.

\subsection{Models of Supersymmetry Breaking}

There is still one important effect that is missing in our construction
of the MSSM. The terms that we have written so far preserve 
exact supersymmetry.  A fully supersymmetric model would contain
a massless fermionic partner of the photon and a charged scalar particle 
with the mass of the electron. These particles manifestly do not exist.
So if we wish to build a  model of Nature with supersymmetry as a fundamental 
symmetry, we need to arrange that supersymmetry is spontaneously broken.

From the example of spontaneous symmetry breaking in the Standard Model, 
we would expect to do this by including in the MSSM a field whose 
vacuum expectation value leads to supersymmetry breaking.  This is not as 
easy as it might seem.  To explain why, I will first present some 
models of supersymmetry breaking.

The simplest model of supersymmetry breaking is the O'Raifeartaigh 
model~\cite{ORaif}, with three chiral supermultiplets $\phi_0$, $\phi_1$,
$\phi_2$ interacting through the superpotential
\beq
   W = \lambda \phi_0 + m \phi_1\phi_2 + g \phi_0 \phi_1^2 \ .
\eeq{ORAifW}
This superpotential implies the $F = 0$ conditions
\beqa
       0 =  F^*_0 &=& \lambda + g\phi^2_1 \CR
       0 =  F^*_1 &=& m \phi_2 + 2 g \phi_0 \phi_1\CR
       0 =  F^*_2 &=& m \phi_1 
\eeqa{ORaifF}
The first and third equations contradict one another. It is impossible
to satisfy both conditions, and so there is no supersymmetric vacuum state.
This fulfils the condition for spontaneous supersymmetry breaking that
I presented in Section~2.2. 

This mechanism of supersymmetry breaking has an unwanted corollary.  
Because one combination of the scalar fields appears in two 
different constraints in \leqn{ORaifF}, there must be an orthogonal
combination that does not appear at all. This means that the F-term
potential $V_F$ has a surface of degenerate vacuum states.  To see this 
explicitly, pick a particular vacuum solution
\beq
      \phi_0 = \phi_1 = \phi_2 = 0 \ .
\eeq{vacuumphis}
and expand the potential $V_F$ about this point.  There are 6 real-valued
boson fields with masses 
\beq
  0 \ , \quad 0 \ , \quad m \ , \quad m \ , \quad \sqrt{m^2- 2 \lambda g} \ ,
                 \quad  \sqrt{m^2+ 2 \lambda g}\ .
\eeq{oRaifmasses}
These six fields do not pair into complex-valued fields; that is already 
an indication that supersymmetry is broken.  The fermion mass term in 
\leqn{WkLagrangian} gives one Dirac fermion mass $m$ and leaves 
one Weyl fermion massless.  This massless fermion is the Goldstone
particle associate with spontaneous supersymmetry breaking. 

A property of these masses is that the sum rule for fermion and boson masses
\beq
       \mbox{str}[m^2] =   \sum m_f^2 - \sum m_b^2 = 0 
\eeq{sumrule}
remains valid even when supersymmetry is broken. This  sum rule is  the 
coefficient of the one-loop quadratic divergence in the vacuum energy.  Since 
supersymmetry breaking does not affect the ultraviolet structure of the 
theory, this coefficient must cancel even if  supersymmetry is 
spontaneously broken~\cite{LutyTASI}.  In fact, if $Q$ is a conserved 
charge in the model, the sum rule is valid in each charge sector $Q = q$:
\beq
               \mbox{str}_q[m^2] = 0  \ .
\eeq{sumruleq}

In the O'Raifeartaigh model, supersymmetry is spontaneously broken by a
nonzero expectation value of an $F$ term.  It is also possible to break 
supersymmetry with a nonzero expectation value of a $D$ term.  The 
$D$-term potential $V_D$ typically has zeros. For example, in an $SU(3)$
supersymmetric Yang-Mills theory,
\beq
       V_D = \half \left( \sum_{3} \phi^\dagger t^a \phi - 
            \sum_{\bar 3}  \bar \phi t^a \bar\phi^\dagger \right)^2
\eeq{VDexample}
and it is easy to find solutions in which the terms in parentheses sum to zero.
However, it is not difficult to arrange a $V_F$ such that the solutions
of the $F=0$ conditions do not coincide with the solutions of the $D= 0$
conditions.  This leads to spontaneous symmetry breaking, again with the 
sum rule \leqn{sumruleq} valid at tree level.

Unfortunately, the sum rule \leqn{sumruleq} is a disaster for the prospect of 
finding a simple model of spontaneously broken supersymmetry that extends the 
Standard Model. For the charge sector of the $d$ squarks, we would need all 
down-type squarks to have masses less than 5 GeV.  For the charge sector of 
the  charged leptons, we would need all sleptons to have masses less 
than 2 GeV.  

\subsection{Soft Supersymmetry Breaking}

The solution to this problem is to construct models of spontaneously broken 
supersymmetry using a different strategy from the one that we use for 
electroweak symmetry breaking in the Standard Model.  To break electroweak
symmetry, we introduce a Higgs sector whose mass scale is the same as the 
scale of the fermion and gauge boson masses induced by the symmetry breaking.
To break supersymmetry, however, we could introduce a new sector at a much 
higher mass scale, relying on a weak coupling of the new sector to the 
Standard Model particles to communicate the supersymmetry breaking terms.
In principle, a weak gauge interaction could supply this coupling.  However,
the default connection is through gravity.  Gravity and supergravity couple
to all fields.  It can be shown that supersymmetry breaking anywhere in 
Nature is communicated to all other sectors through supergravity 
couplings.

We are thus led to the following picture, which produces a phenomenologically
reasonable supersymmetric extension of the Standard Model:  We extend
the Standard Model fields to supersymmetry multiplets in the manner 
described in Section~3.1.   We also introduce a 
 {\it hidden sector} with no direct coupling to 
quark, leptons, and Standard Model gauge bosons. Supersymmetry is 
spontaneously broken in this hidden sector.  A weak interaction coupling
the two sectors then induces a supersymmetry-breaking effective interaction 
for the Standard Model particles and their superpartners.   If $\Lambda$ is
the mass scale of the hidden sector, the supersymmetry breaking mass 
terms induced for the Standard Model sector are of the order of
\beq
       m \sim {\VEV{F}\over M} \sim {\Lambda^2 \over M} \ ;
\eeq{estimateM}
where $M$ is the mass of the particle responsible for the weak connection
between the two sectors.  $M$ is called the {\it messenger scale}.
By default, the messenger is supergravity.  Then $M = m_\Pl$ and 
$\Lambda \sim 10^{11}$ GeV.  In this scenario, the superpartners acquire
masses of the order of the parameter $m$ in \leqn{estimateM}.

It remains true
that the quarks, leptons, and gauge 
bosons cannot obtain mass until $SU(2)\times U(1)$ is broken. It is 
attractive to think that the symmetry-breaking terms that give mass to 
the superpartners cause $SU(2)\times U(1)$ to be spontaneously broken, at
more or less the same scale.  I will discuss a mechanism by which this 
can happen in Section~6.1. 
The weak interaction scale would then  not be a fundamental scale in 
Nature, but rather one that arises dynamically from the hidden sector 
and its couplings.

The effective
interaction that are generated by messenger exchange generally involve
simple operators of low mass dimensions, to require the minimal number of 
powers of $M$ in the denominator.  These operators
are {\it soft} perturbations of
the theory, and so we say that the MSSM is completed by including
{\it soft supersymmetry-breaking interactions}.

However, the supersymmetry-breaking terms induced in this model will 
not include all possible low-dimension operators.  Since
these interactions arise by coupling into a supersymmetry theory, they
are formed by starting with a supersymmetric effective action and turning 
on $F$ and $D$ expectation values as spurions.  Only a subset of the 
possible supersymmetry-breaking terms can be formed 
in this way~\cite{GrisaruG}.  
By replacing a superfield
$\Phi$ by $ \theta^T c \theta \VEV{F}$, we can convert
\beqa
      \int d^4 \theta \,  K(\Phi,\phi) & \to&   m^2 \phi^\dagger \phi \CR
      \int d^2 \theta \,  f(\Phi) W^T c W  & \to&   m \lambda^T c \lambda \CR
      \int d^2 \,  W(\Phi, \phi) & \to&   B \phi^2 + A\phi^3
\eeqa{sampleeffectiveL}
However, as long as the $\phi$ theory is renormalizable, we cannot generate
the terms
\beq
             m \psi^T c \psi \ , \quad    C \phi^*\phi^2 \ , 
\eeq{notgenerated}
by turning on expectation values for $F$ and $D$ fields.  Thus, we cannot
generate super\-sym\-me\-try-break\-ing interactions that are
mass terms for the fermion field of a chiral multiplet or 
non-holomorphic cubic terms for the scalar fields.  

There is another difficulty with terms of the form \leqn{notgenerated}.
In models with Standard Model singlet scalar fields, which typically
occur in concrete models, these two interactions
can generate new quadratic divergences when they appear in loop 
diagrams~\cite{GrisaruG}.

Here is the most general supersymmetry-breaking effective Lagrangian 
that can be constructed following the rule just given that is consistent
with the gauge symmetries of the Standard Model:
\beqa
  \L_{soft} &=&  - M_f^2 |\s f|^2 - \half m_i \lambda^{Ta}_i c \lambda^a_i \CR
        & & - (A_d y_d \s{\bar d} H_{d \alpha} \epsilon_{\alpha\beta} 
    \s Q_\beta + A_e y_e \s{\bar e} H_{d \alpha} \epsilon_{\alpha\beta} 
    \s L_\beta   \CR
    & & \hskip 0.1in - A_u y_u \s{\bar u} H_{u \alpha} \epsilon_{\alpha\beta} 
    \s Q_\beta   -  B\mu  H_{d\alpha} \epsilon_{\alpha\beta} H_{u\beta} )
        - h.c.
\eeqa{generalsoft}
I have made the convention of scaling the $A$ terms with the corresponding 
Yukawa couplings and scaling the $B$ terms with $\mu$.  The parameters
$A$ and $B$ then have the dimensions of mass and are expected to be of the 
order of $m$ in \leqn{estimateM}.

For most of the rest of these lectures,
I will represent the effects of the hidden sector and supersymmetry breaking
simply by adding \leqn{generalsoft} to the supersymmetric Standard Model. 
I will then consider the MSSM to be defined by 
\beq
    \L = \L_F + \L_D + \L_W + \L_{soft}
\eeq{fullL}
combining the pieces from \leqn{vectorL}, \leqn{Lchiralgauge}, \leqn{WYform}, 
\leqn{Wmudef}, and \leqn{generalsoft}.

There are two problems with this story.  The first is the $\mu$ term in 
the MSSM superpotential.  This a supersymmetric term, and so $\mu$ can 
be arbitrarily large.  To build a successful phenomenology of the MSSM,
however, we need to have $\mu$ of the order of the weak scale.  Ideally,
$\mu$ should be parametrically equal to \leqn{estimateM}.

There are simple mechanisms that can solve this problem.
A fundamental theory that leads to the renormalizable Standard Model at 
low energies can also contain higher-dimension operators suppressed by the
high-energy mass scale.  Associate this scale with the messenger scale.
Then a supersymmetric higher-dimension operator in the superpotential
\beq
             \int d^2\theta \,  {1\over M} S^2 H_d H_u
\eeq{NKmu}
leads to a $\mu$ term if $S$ acquires a vacuum expectation value. If $S$ is 
a hidden sector field, we could find~\cite{NillesKim}
\beq
                 \mu = {\VEV{S^2}\over M} \sim {\Lambda^2 \over M} \ ,
\eeq{NKmuresult}
A supersymmetric higher dimension contribution to the K\"ahler potential
\beq
             \int d^4\theta \,  {1\over M} \Phi^\dagger H_d H_u
\eeq{GMmu}
leads to a $\mu$ term if $\Phi$ acquires a vacuum expectation value in 
its $F$ term. If $\Phi$ is 
a hidden sector field, we could find~\cite{GiudiceMas}
\beq
                 \mu = {\VEV{F_\Phi}\over M} \sim {\Lambda^2 \over M} \ ,
\eeq{GMmuresult}
In models with weak-coupling dynamics, 
higher-dimension operators are associated with the string
or Planck scale; then, these mechanisms work most naturally if supergravity
is the mediator.   However, it is also possible to apply these strategies
in  models with strong-coupling dynamics in the 
hidden sector at an intermediate scale.

Generating the $\mu$ term typically requires breaking all continuous 
R-symmetries of the model.   This is unfortunate, because an R-symmetry
might be helpful phenomenologically, for example, to keep gaugino masses
small while allowing sfermion masses to become large, or because it might
be difficult to break an R-symmetry using a particular  
explicit mechanism of supersymmetry breaking.  In this case, it is necessary
to add Standard Model singlet fields to the MSSM to allow all gaugino
and Higgsino fields to acquire nonzero masses.  Models of this type 
are presented in~\cite{preserveR,preserveRtoo}. 

\begin{figure}
\begin{center}
\includegraphics[height=1.0in]{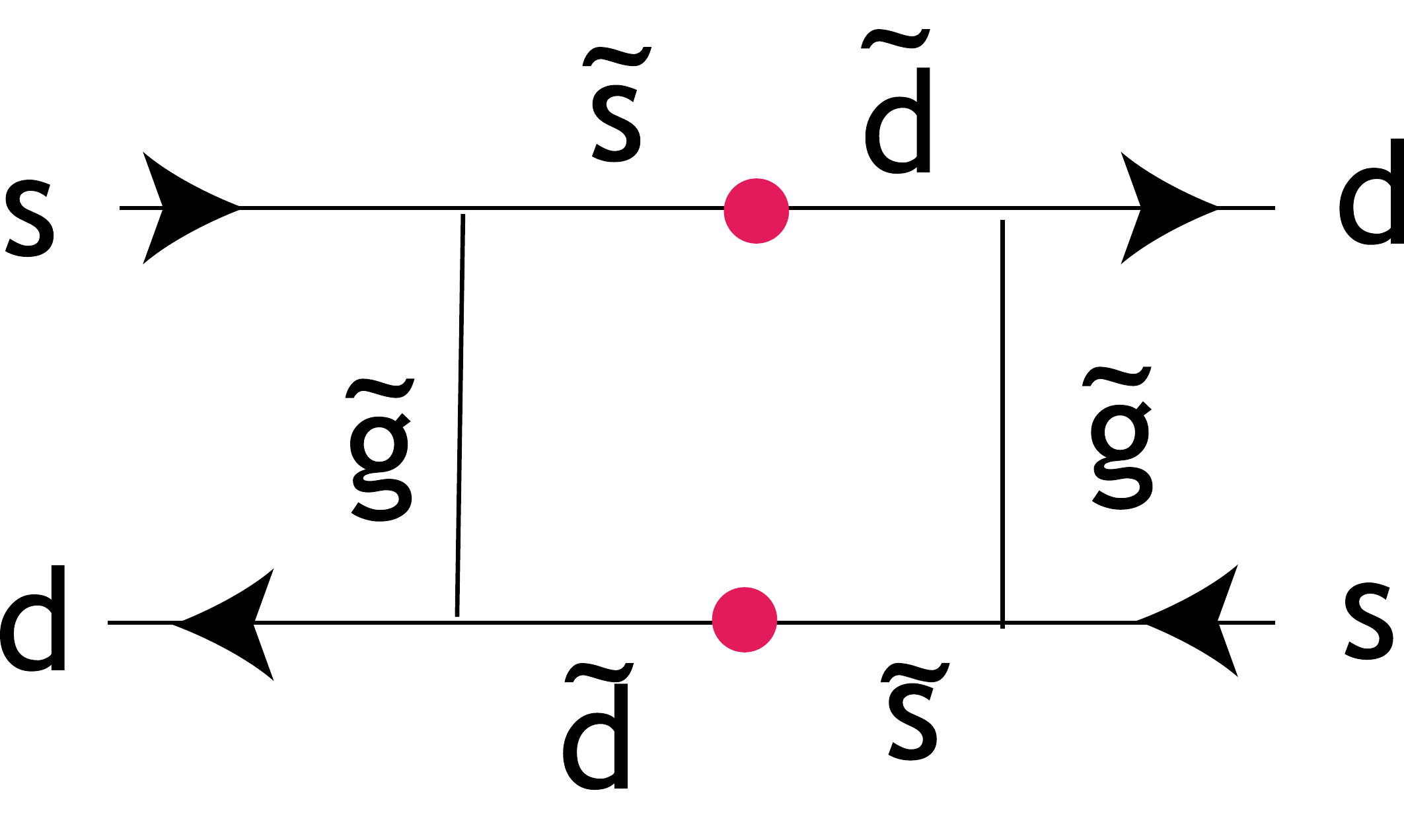}
\caption{A dangerous contribution to $K$-$\bar K$ mixing involving gluino
     exchange and flavor mixing in the squark mass matrix.}
\label{fig:KKmix}
\end{center}
\end{figure}

The second problem involves the flavor structure of the soft supersymmetry
breaking terms.  In writing \leqn{generalsoft}, I did not write flavor 
indices.  In principle, these terms could have flavor-mixing that is
arbitrary in structure and different from that in \leqn{WYform}. 
Then the flavor-mixing
would not be transformed away when \leqn{WYform} is put into canonical form.
However, flavor-mixing from the soft supersymmetry breaking terms is 
highly constrained by experiment.  Contributions such as the one shown in 
Fig.~\ref{fig:KKmix} give contributions to $K^0$, $D^0$, and $B^0$ 
mixing, and to $\tau\to \mu \gamma$ and $\mu\to e\gamma$, that can be 
large compared to the measured values or limits.   Theories of the origin
of the soft terms in models of supersymmetry breaking should address
this problem.  For example,  the models of 
{\it gauge-mediated}~\cite{DineNelson} and 
{\it anomaly-mediated}~\cite{GMLR,RSanomaly} 
supersymmetry breaking induce soft terms that depend only on the $SU(2)\times
U(1)$ quantum number and are therefore automatically diagonal in flavor.
A quite different solution, based on a extension of the MSSM with
 a continuous R-symmetry, is presented 
in \cite{Weiner}.

If I assume that the soft supersymmetry-breaking Lagrangian is 
diagonal in flavor but is otherwise arbitrary, it introduces 22 new parameters.
With arbitrary flavor and CP violation, it introduces over 100 new parameters.
This seems a large amount of parameter freedom.  I feel that it is not 
correct, though, to think of these as new fundamental parameters in physics.
The soft Lagrangian is computed from the physics of the hidden sector, and so
we might expect that these parameters are related to one another as a part of
a theory of supersymmetry breaking.  Indeed, 
the values of these parameters are the essential data from which we will
infer the properties of the hidden sector and its new high energy 
interactions.  

If supersymmetry is discovered at the weak interaction 
scale, it will be a key problem to measure the coefficients in the soft
Lagrangian and to understand their pattern and implications.  Most of 
my discussion in the next two sections will be devoted to the question of
how the soft parameters can be determined from data at the LHC and ILC.

\section{The Mass Spectrum of the MSSM}

\subsection{Sfermion Masses}

Our first task in this program is to ask how the parameters of the MSSM
Lagrangian are reflected in the mass spectrum of the superparticles.
The relation between the MSSM parameters and the particle masses is
surprisingly complicated, even at the tree level.   For each particle,
we will need to collect
all of the pieces of the Lagrangian \leqn{fullL} that can contribute to the
mass term.  Some of these will be direct mass contributions; others 
will contain Higgs fields and contribute to the masses when these fields
obtain their vacuum expectation values.  In this discussion, and in the 
remainder of these lectures, I will ignore all flavor-mixing.

Begin with the squark and slepton masses. For light quarks and leptons, 
we can ignore the fermion masses and Higgs couplings. Even with this 
simplification, though, there are two sources
for the scalar masses.  One is the soft mass term
\beq
     \L_{soft} =   - M_f^2 |\s f|^2 \ .
\eeq{softmsquared}
The other comes from the $D$-term potential.  The $SU(2)$ and $U(1)$ potentials
contain the cross terms between the Higgs field and sfermion field 
contributions 
\beqa
   V_D &=& {g^2\over 2}\cdot 2 \cdot (H^\dagger_d {\sigma^3\over 2} H_d
          + H_u^\dagger {\sigma^3\over 2} H_u ) \cdot (\s f^* t^3 \s f)\CR
       &  & + {g^{\prime 2}\over 2}\cdot  2 \cdot
 ( -\half H^\dagger_d H_d + \half
         H_u^\dagger  H_u ) \cdot (\s f^* Y \s f)\ .
\eeqa{VDforsf}
To evalute this expression, we must insert the vacuum expectation 
values of the two Higgs
fields.  In terms of the angle $\beta$ defined in \leqn{tanbetadef}, 
these are
\beq
  \VEV{H_u} = \pmatrix{ 0 \cr {1\over \sqrt{2}}v \sin\beta \cr
                     } \qquad
  \VEV{H_d} = \pmatrix{  {1\over \sqrt{2}}v \cos\beta \cr  0\cr
                     } \ ,
\eeq{Higgsvevs}
where $v = 246$ GeV so that $m_W = gv/2$.

Inserting the Higgs vevs into the potential  \leqn{VDforsf}, we find
\beqa
  V_D & = & \s f^* \bigl[ {v^2\over 4} (\cos^2\beta - \sin^2\beta)(g^2 I^3 
          - g^{\prime 2} Y) \bigr] \s f \CR
&= & \s f^* \bigl[ {(g^2 + g^{\prime 2})v^2\over 4} \,\cos 2\beta\, 
          (I^3 - s_w^2(I^3+Y))\bigr] \s f \CR
 &= & \s f^* \bigl[ \mz^2 \cos 2 \beta (I^3 - s_w^2 Q)\bigr] \s f \ .
\eeqa{sfmassfromD}
Then, if we define 
\beq
          \Delta_f = (I^3 - s_w^2 Q) \cos 2\beta \, \mz^2 \ ,
\eeq{Deltafdef}
the mass of a first- or second-generation sfermion takes the form
\beq
       m^2_f = M^2_f + \Delta_f \ 
\eeq{mfmass}
when contributions proportional to fermion masses can be neglected.
The D-term contribution can have interesting effects.  For example, $SU(2)$
invariance of $M_f^2$ implies that
\beq
      m^2(\s e) - m^2(\s \nu) = \left|\cos 2\beta \right|\, \mz^2 > 0 \ .
\eeq{sesnumass}
For some choices of parameters, the measurement of this mass difference is a 
good way to determine $\tan \beta$~\cite{FengMoroi}.

For third-generation fermions, the contributions to the mass term from Yukawa
couplings and from $A$ terms can be important. For the $\s b$ and $\s{\bar b}$,
these contributions come from the terms in the effective Lagrangian
\beqa
  |F_b|^2 + |F_{\bar b}|^2 &=& |y_b\VEV{H^0_d} \s b|^2 
      + |y_b \s{\bar b}\VEV{H^0_d}|^2  =  m_b^2 (|\s b|^2 + |\s {\bar b}|^2)\CR
 |F_{Hd}|^2 &=& (-\mu \VEV{H^0_d})^* (y_b \s{\bar b}\, \s b) + h.c. = 
      -\mu m_b \tan\beta \,  \s{\bar b}\, \s b + h.c.\CR
 -\L_{soft} &=& A_b y_b \VEV{H^0_d} \s{\bar b}\,\s b 
                        = A_b m_b \s{\bar b}\, \s b\ .
\eeqa{FAtermsformf}
In all, we find a mass matrix with mixing between the two scalar partners 
of the $b$ quark,
\beq
   \pmatrix{  \s b ^*& \s {\bar b}^* \cr } \M_b^2
  \pmatrix{  \s b \cr \s {\bar b} \cr } \ ,
\eeq{bmatrixM}
with
\beq
  \M_b^2 = 
   \pmatrix{ M_b^2 +\Delta_b + m_b^2  &   m_b (A_b - \mu\tan\beta)\cr
   m_b (A_b - \mu\tan\beta)&  \M_{\bar b}^2  +\Delta_{\bar b} + m_b^2\cr
 }
\eeq{Mbmatrix}
The mass matrix for $\s \tau$, $\s{\bar \tau}$ has the same structure.  For 
$\s t$, $\s {\bar t}$, replace $\tan\beta$  by  $\cot\beta$.

The mixing terms in the mass matrices of the third-generation sfermions often
play an important role in the qualitative physics of the whole SUSY model.
Because of the mixing, one sfermion eigenstate is pushed down in mass.  
This state is often the lightest squark or even the lightest superparticle 
in the theory.

\subsection{Gaugino and Higgsino Masses}

In a similar way, we can compute the mass terms for the gauginos and 
Higgsinos.  Since the gauginos and Higgsino have the same quantum numbers 
after $SU(2)\times U(1)$ breaking, they will mix.  We have seen in 
Section~2.4 that this mixing plays an essential role in 
the working of the 
Higgs mechanism in the limit where soft supersymmetry breaking terms are 
turned off.

The charged gauginos and Higgsinos receive mass from three sources.  First,
there is a soft SUSY breaking term
\beq
           - \L_{soft} = m_2  \s w^{-T} c \s w^+  \ .
\eeq{winomass}
The $\mu$ superpotential term contributes
\beq
           -\L_W = \mu \s h^{-T}_d c \s h^+_u \ .
\eeq{higgsinomass}
The  gauge kinetic terms contribute
\beq
   -\L = \sqrt{2} {g\over \sqrt{2}} \bigl( \VEV{H^0_d} \s w^{+T} c \s h^-_d
         +  \VEV{H^0_u} \s w^{-T} c \s h^+_u \bigr)
\eeq{winokinetic}
Inserting the Higgs field vevs from \leqn{Higgsvevs}, we find the mass term
\beq
 \pmatrix{  \s w^{-T} & \s h^{-T}_d \cr } c \ 
       m_C 
 \pmatrix{  \s w^{+} \cr \s h^{+}_u \cr } \ ,
\eeq{charginomass}
with
\beq
  m_C = \pmatrix{   m_2    &    \sqrt{2} \mw \sin\beta \cr
          \sqrt{2} \mw \cos\beta &   \mu \cr } \ .
\eeq{mCmassmatrix}

The mass matrix for neutral gauginos and Higgsinos also receives contributions
from these three sources.  In this case, all four of the states
\beq
       ( \s b, \s w^0, \s h^0_d, \s h^0_u ) 
\eeq{nmixing}
have the same quantum numbers after $SU(2)\times U(1)$ breaking and 
can mix together.  The mass matrix is
\beq  m_N = \pmatrix{   m_1   & 0 & -\mz c_\beta s_w & \mz s_\beta s_w\cr
       0 &     m_2   & \mz c_\beta c_w & -\mz s_\beta c_w\cr
  -\mz c_\beta s_w & \mz c_\beta c_w &  0 &  -\mu\cr
  \mz s_\beta s_w & -\mz s_\beta c_w &  -\mu & 0 \cr    }\ .
\eeq{neutralinomatrix}

The mass eigenstates in these systems are referred to collectively as 
{\it charginos} and {\it neutralinos}.  The matrix \leqn{neutralinomatrix}
is complex symmetric, so it can be diagonalized by a unitary matrix 
$V_0$,\footnote{Note that this formula is different from that which 
diagonalizes a Hermitian matrix.  A detailed discussion of the diagonalization
of mass matrices appearing in SUSY can be found in the Appendix of 
\cite{ChoiHaber}.}
\beq  
             m_N = V_0^* D_N V_0^\dagger  \ .
\eeq{mNdiag}
I will denote the neutralinos as $\s N^0_i$, $i = 1, \ldots, 4$, in order
of mass with $\s N_1^0$ the lightest. Elsewhere in the literature, you will
see these states called $\s \chi^0_i$ or $\s Z_i^0$.
 The mass eigenstates are related to the
weak eigenstates by the transformation
\beq
    \pmatrix{ \s b^0 \cr \s w^0 \cr \s h^0_d \cr \s h^0_u \cr
        }       
  =   V_0    \pmatrix{ \s N_1 \cr \s N_2 \cr \s N_3 \cr \s N_4 \cr 
        }   \ .
\eeq{wmtransforN}
Note that the diagonal matrix $D_N$ in \leqn{mNdiag} may have 
negative or  complex-valued
elements.  If that is true, the physical fermion masses of the 
$\s N_i$ are the 
absolute values of the corresponding elements of $D_N$.  The phases will
appear in the three-point couplings of the $\s N_i$ and can lead to observable
interference effects.  Complex phases in $D_N$ would provide a new 
source of CP violation.

The chargino mass matrix \leqn{mCmassmatrix} is not symmetric, so in general
it is diagonalized by two unitary matrices
\beq
       m_C = V_-^* D_C V_+^\dagger \ .
\eeq{mCdiag}
I will denote the charginos as $\s C^\pm_i$, $i = 1, 2$, in order
of mass with $\s C_1^\pm$ the lighter. Elsewhere in the literature, you will
see these states called $\s \chi^\pm_i$ or $\s W_i^\pm$.
 The mass eigenstates are related to the
weak eigenstates by the transformation
\beq
    \pmatrix{ \s w^+\cr \s h^+_u \cr
        }       
  =   V_+   \pmatrix{ \s C_1^+ \cr \s C_2^+\cr 
         }   \ ,\qquad 
    \pmatrix{ \s w^-\cr \s h^-_u \cr
        }       
  =   V_-   \pmatrix{ \s C_1^- \cr \s C_2^-\cr 
         }   \ .
\eeq{wmtransforC}

It should be noted that $\mu$ are must be nonzero. 
If  $\mu = 0$, the determinant of \leqn{neutralinomatrix}
vanishes and so the lightest neutralino must be massless. This
neutralino will also have a large Higgsino content and thus an order-1
coupling to the $Z^0$.  It is excluded by searches for an excess of 
invisible $Z^0$ decays and for $Z^0 \to \s N_1 \s N_2$.  The condition
$\mu = 0 $ also implies that the lightest chargino has a mass below the
current limit of about 100~GeV. 

\begin{figure}
\begin{center}
\includegraphics[height=3.0in]{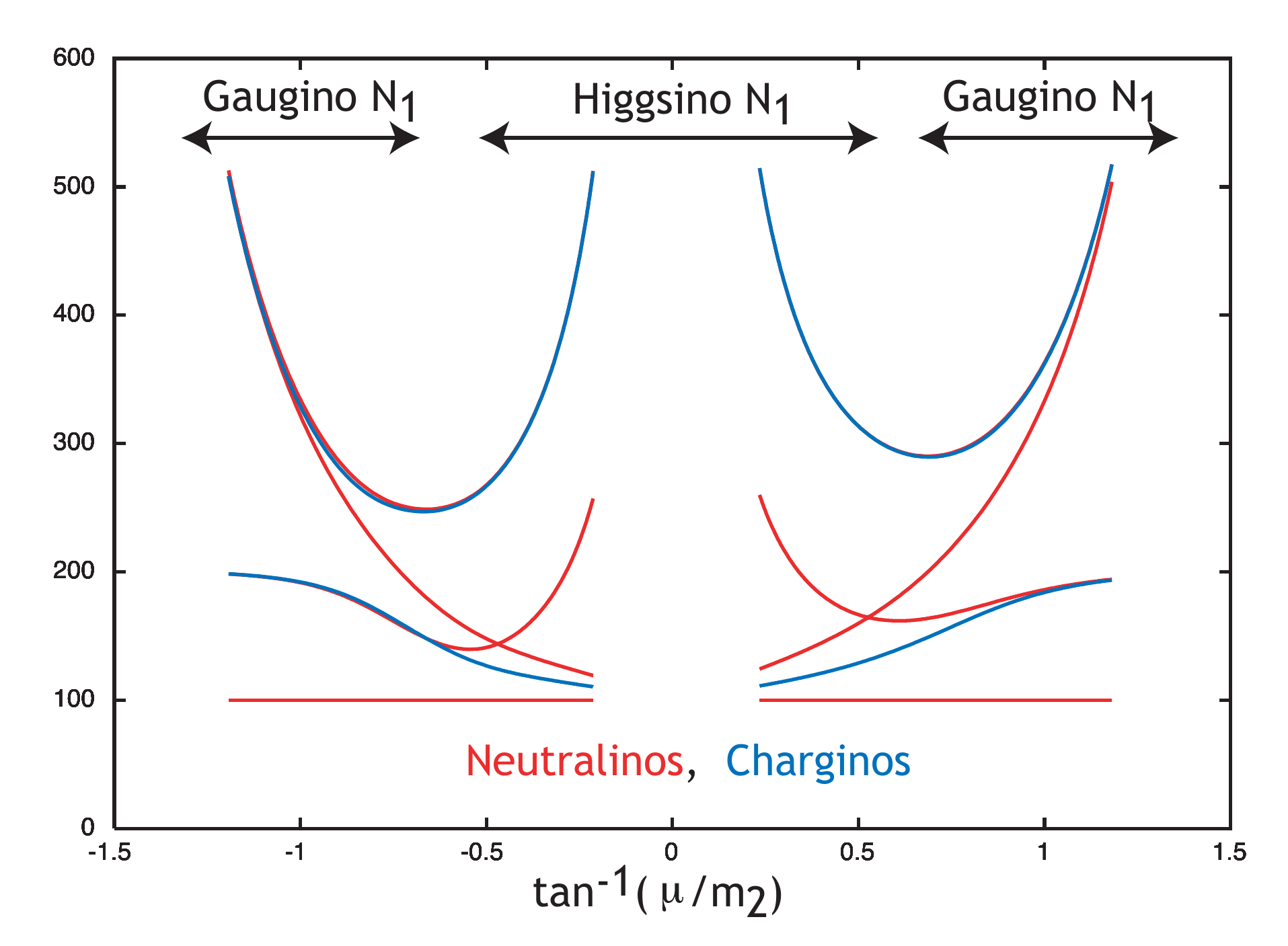}
\caption{Masses of the four neutralinos and two charginos along a line
in the SUSY 
parameter space on which $m(\s N^0_1) = 100$~GeV while the parameter
$\mu$ moves from large negative to large positive values.  The parameter 
 $m_1$ is set to $m_1 = 0.5 m_2$.  Note the approximate degeneracies in
the extreme limits of the gaugino and Higgsino regions.}
\label{fig:NCmixing}
\end{center}
\end{figure}

Often, one studies models for which   $m_1$, $m_2$, and $\mu$
are all large compared to $\mw$ and $\mz$.
The off-diagonal elements that mix the gaugino and Higgsino states are 
of the order of $\mw$ and $\mz$.  Thus, if the scale of masses generated 
by the SUSY breaking terms is large, the mixing is small and the 
individual eigenstates
are mainly gaugino or mainly Higgsino. However, there are two distinct
cases. The first is the {\it gaugino region}, where $m_1, m_2 < |\mu|$.
In this region of parameter space, the lightest states $\s N_1$, $\s C_1$ are 
mainly gaugino, while the heavy neutralinos and charginos are mainly 
Higgsino.  In the {\it Higgsino region}, 
 $m_1, m_2 > |\mu|$, the situation is reversed and $\s N_1$, $\s C_1$ are 
mainly Higgsino.  In this case, the two lightest neutralinos are almost
degenerate.   In Fig.~\ref{fig:NCmixing}, I show the mass eigenvalues
as a function of the mass matrix parameters along a line in the parameter
space on which the $\s N_1$ has a fixed mass of 100 GeV.  As we will see in 
Section~6.4, the exact makeup of the lightest neutralino as a mixture of
gaugino and Higgsino components is important to the study of
supersymmetric dark matter.

\begin{figure}
\begin{center}
\includegraphics[height=3.0in]{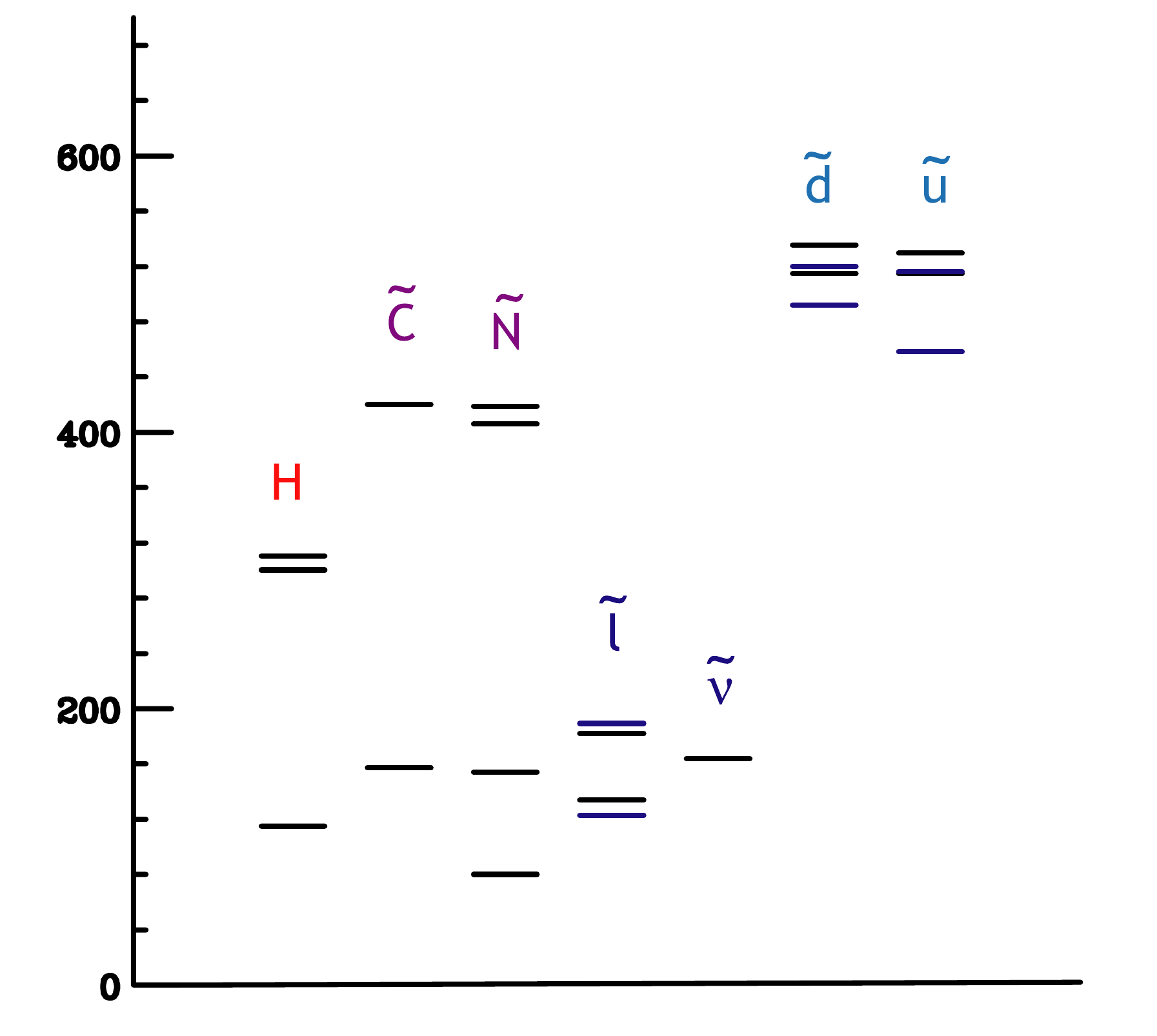}
\caption{Illustrative spectrum of supersymmetric particles.  The columns
contain, from the left, the Higgs bosons, the four neutralinos, the two 
charginos, the charged sleptons, the sneutrinos, the down squarks, and the
up squarks.  The gluino, not shown, is at about 800 GeV.}
\label{fig:onespectrum}
\end{center}
\end{figure}

To summarize this discussion, I present in Fig.~\ref{fig:onespectrum} the
complete spectrum of new particles in the MSSM at a representative point in 
its parameter space.  Notice that the third-generation sfermions are split 
off from the others in each group.  Note also that the parameter point 
chosen is in the gaugino region.  The lightest superparticle is the 
$\s N_1$.  I will discuss the spectrum of Higgs bosons in Section~6.2.

\subsection{Renormalization Group Evolution of MSSM Parameters}

The spectrum shown in Fig.~\ref{fig:onespectrum} appears to have been 
generated by assigning random values to the soft SUSY breaking parameters.
But, actually, I generated this spectrum by making very simple assumptions 
about the relationships of the soft parameters, at a high energy scale.
Specifically, I assumed that the soft SUSY breaking 
gaugino masses and (separately) the sfermion
masses were equal at the scale of grand unification.  The structure that you 
see in the figure is generated by the renormalization group evolution 
of these parameters from the grand unification scale to the weak scale.

The renomalization group (RG) evolution of soft parameters is likely to play a
very important role in the interpretation of measurements of the SUSY
particle masses.  Essentially, after measuring these masses, it will be
necessary
to decode the results by running the effective mass parameters up to a
higher energy at which their symmetries might become more apparent.  The 
situation is very similar to that of the Standard Model coupling constants, 
where a renormalization group analysis told us that the apparently random
values \leqn{alphavals} for the coupling constants at the weak scale actually 
corresponds to a unification of couplings at a much higher scale.

In this section, I will write the most basic RG equations for the 
soft gaugino and sfermion masses.  One further effect, which involves the 
Yukawa couplings and is important for the third generation, will be discussed
later in Section~6.1.

The RG equation for the gaugino masses is especially simple. This is 
because both the gaugino masses and the gauge couplings arise from the 
superpotential term \leqn{NAWWLag}, with the supersymmetry breaking
terms arising as shown in  \leqn{sampleeffectiveL}.   
As I have already noted, this F-term
receives a radiative correction proportional to the $\beta$ function
as a consequence of the trace anomaly~\cite{GrisaruWW,AHM}.
The corrections are the same for the gauge boson field strength and the
gaugino mass.  Thus, if gaugino masses and couplings
are generated at the scale $M$, they have the relation after RG running to 
the scale $Q$:
\beq 
         {m_i(Q)\over m_i(M) } = {\alpha_i(Q)\over \alpha_i(M)} \ .
\eeq{moveralpha}
If the $F$ term that generates the soft gaugino masses is
an $SU(5)$ singlet, the soft gaugino masses will be grand-unified at $M$.
Then, running down to the weak scale, they will have the relation
\beq
    m_1 \ : \ m_2 \ : \ m_3 \ = \ \alpha_1 \ : \ \alpha_2\ : \ \alpha_3 \ = 
        \   0.5 \ :\ 1 \ : \ 3.5 \ .
\eeq{gauginounif}
This relation of soft gaugino masses is known as {\it gaugino unification}.

\begin{figure}
\begin{center}
\includegraphics[height=0.7in]{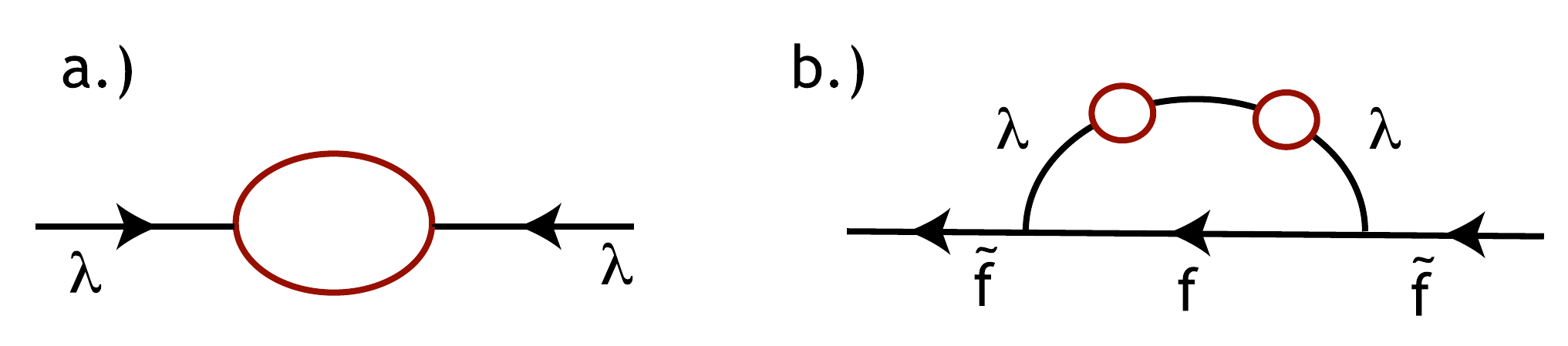}
\caption{Diagrams that generate the soft mass parameters in gauge mediated
     supersymmetry breaking: (a.) gaugino masses; (b.) sfermion masses.}
\label{fig:gaugemediation}
\end{center}
\end{figure}

There are other models of the soft gaugino masses that also lead to gaugino
unification.  In {\it gauge-mediated SUSY breaking}, the dynamics responsible
for SUSY breaking occurs at a scale much lower than the scale 
associated with mediation by supergravity.  At this lower scale $M_g$ (for 
example, 1000 TeV), some  heavy particles with nontrivial $SU(3)\times SU(2)
\times U(1)$ quantum numbers acquire masses from SUSY breaking.  These fields 
then couple to gauginos and generate SUSY breaking masses for those 
particles through the diagram shown in Fig.~\ref{fig:gaugemediation}(a).
The heavy particles must fall into complete $SU(5)$ representations; otherwise,
the coupling constant renormalization due to these particles between $M_g$
and the grand unification scale would spoil the grand unification of the gauge
couplings.  Then the diagram in Fig.~\ref{fig:gaugemediation}(a) generates
soft gaugino masses proportional to $\alpha(M_g)$.  Running these parameters
down to the weak scale, we derive the relation \leqn{gauginounif} from this
rather different mechanism.
   
\begin{figure}
\begin{center}
\includegraphics[height=0.7in]{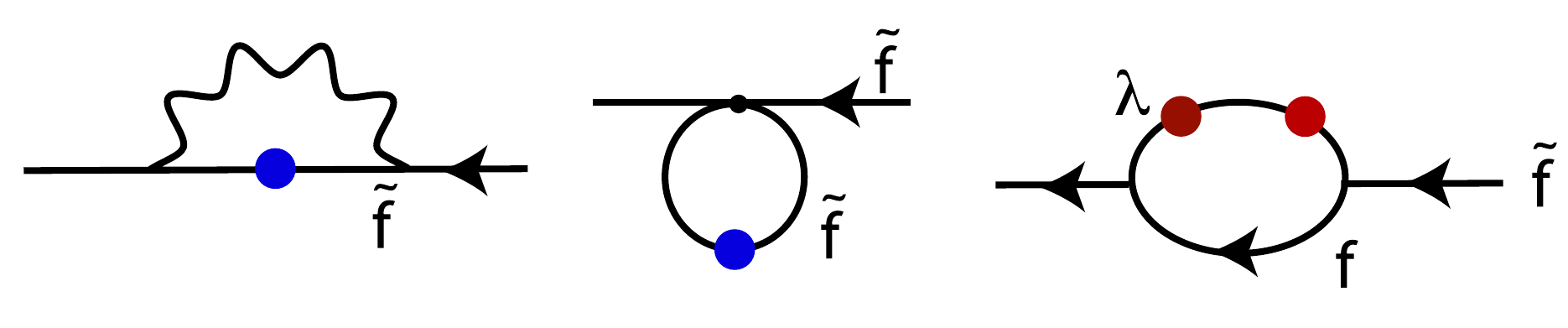}
\caption{Diagrams that generate the renormalization group evolution of the
    soft sfermion mass parameters $M_f^2$.}
\label{fig:MfRG}
\end{center}
\end{figure}

Now let us turn to the RG running of soft scalar masses.  In principle, 
there are two contributions, one from the RG rescaling of the soft mass
term $M_f^2$ and one from RG evolution generating $M_f^2$ from the gaugino
mass.  The Feynman diagrams that contribute to the RG coefficients are
shown in Fig.~\ref{fig:MfRG}.  The two one-loop 
diagrams proportional to $M_f^2$ cancel.  The third diagram, involving
 the gaugino 
mass, gives the RG equation
\beq
   {d M_f^2\over d\log Q} = 
         - {2\over \pi} \sum_i \alpha_i(Q) C_2(r_i) m_i^2(Q) \ ,
\eeq{RGforMf}
with $i = 1,2,3$ and $C_2(r_i)$ the squared charge in the fermion 
representation $r_i$ under the gauge group $i$.  This equation leads to a
positive contribution to $M_f^2$ as one runs the RG evolution from the 
messenger scale down 
to the weak scale.  The effect is largest for squarks, for 
which the SUSY breaking mass is induced from the gluino mass.

As an example of this mechanism of mass generation, assume gaugino unification
and assume that $M_f^2 = 0$ for all sfermions at the grand unification scale.
Then the weak scale sfermion masses will be in the ratio
\beqa
 & &    M(\s{\bar e}) \ : \ M(\s e) \ : \ M(\s{\bar d}) \ : \ M(\s{\bar u}) \ 
    : \ M(\s d, \s u)\ : \ m_2 \CR
  & & \hskip 0.4in = \ 0.5 \ : \ 0.9 \ : \ 3.09 \ : \ 3.10 \ : \ 3.24 \ : \ 
          1
\eeqa{noscaleMs}
This model of fermion mass generation is called {\it no-scale} SUSY breaking.
It has the danger that the lightest stau mass eigenstate could be lighter than
than the $\s N_1$, leading to problems for dark matter.  This problem can be 
avoided by RG running above the GUT scale~\cite{Skiba}.  Alternatively, 
it might actually be that the lightest Standard Model superpartner is a 
long-lived stau that eventually decays to a tau and a 
gravitino~\cite{FM,superWIMP}.

In gauge-mediated SUSY breaking, the diagram shown in 
Fig.~\ref{fig:gaugemediation}(b) leads to the qualitatively similar but
distinguishable formula
\beq
     M_f^2  =  2 \sum_i \alpha^3_i(M) C_2(r_i) \cdot\left( {m_2\over \alpha_2}
        \right)^2 \ .
\eeq{gaugemedMf}

\begin{figure}
\begin{center}
\includegraphics[height=4.0in]{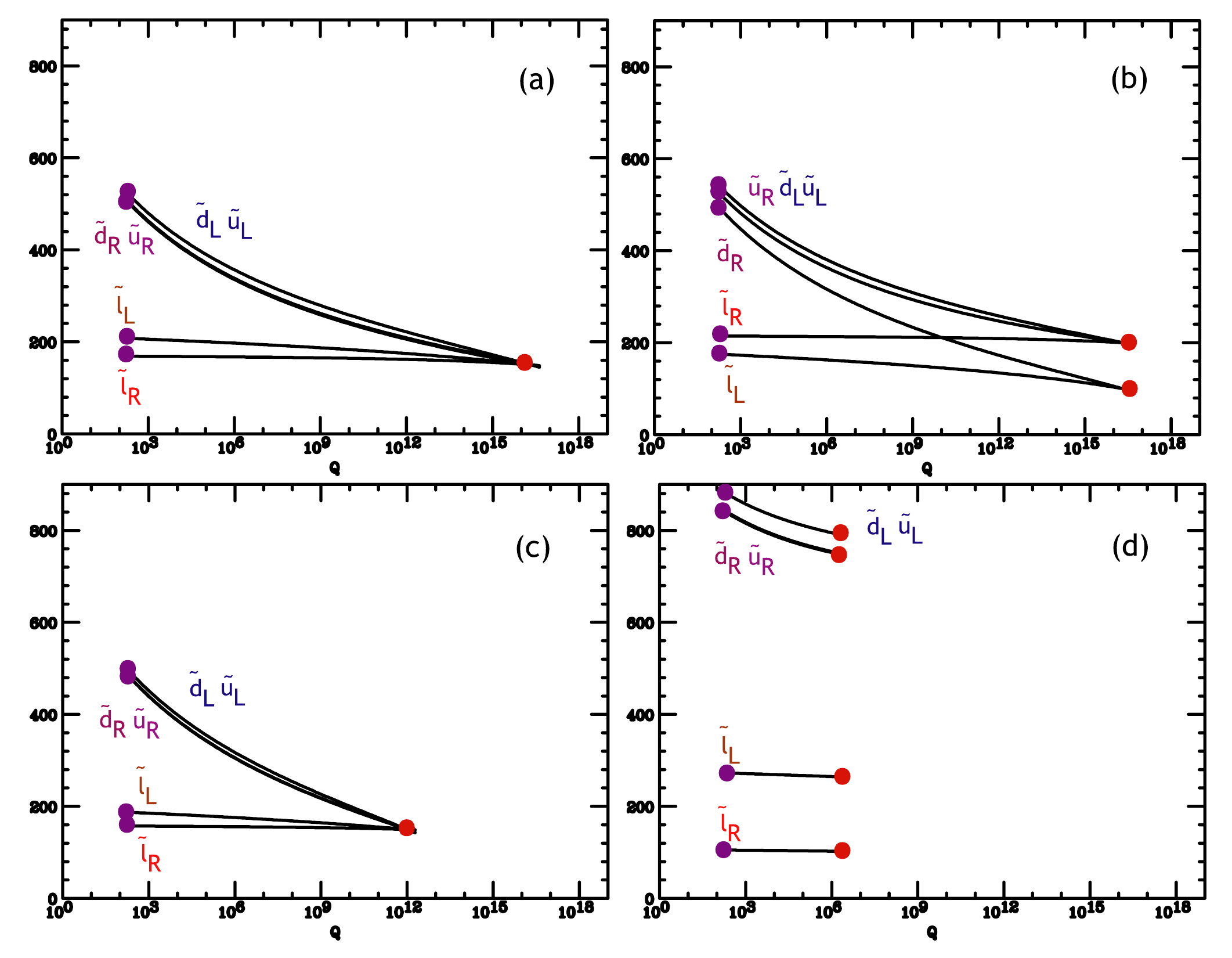}
\caption{Evolution of squark and slepton masses from the messenger scale
 down to the weak scale, for four different models of supersymmetry
breaking: (a.) universal sfermion masses at the grand unification 
scale $M_U$; (b.)  sfermion 
masses at $M_U$  that depend on the $SU(5)$ representation; (c.) 
 universal sfermion masses at an intermediate scale; (d.) gauge mediation
 from a sector of mass about 1000 TeV.}
\label{fig:fourschemes}
\end{center}
\end{figure}

Each model of SUSY breaking leads to its own set of relations among the 
various soft SUSY breaking parameters.  In general, the relations are 
predicted for the parameters defined at the messenger scale and must be
evolved to the weak scale by RG running to be compared with experiment.
Fig.~\ref{fig:fourschemes} shows four different sets of high-scale 
boundary conditions for the RG evolution, and the corresponding evolution
to the weak scale.  If we can measure the weak-scale values, we could
try to undo the evolution and recognize the pattern.  This will be a
very interesting study for the era in which superparticles are observed
at high energy colliders.

There are some features common to these spectra that are general features
of the RG evolution of soft parameters:
\begin{enumerate}
\item The pairs of sleptons $\s{\bar e}$ and $\s e$ 
can easily acquire a significant mass difference from RG evolution, and they 
might also have a different initial condition.  It is important to 
measure the mass ratio
$m(\s{\bar e})/m(\s e)$ as a diagnostic of the scheme of 
SUSY breaking. 
\item Gaugino unification is a quantitative prediction of 
certain schemes of SUSY breaking. It is important to find out whether
this relation is
 correct or not for the real spectrum of superparticles in Nature.
\item  When the RG effects on the squark
masses dominate the values of $M_f^2$ from the initial condition, the 
various species of 
squark have almost the same mass and are much heavier than the sleptons.
It is important to check whether most or all squarks appear at the same 
threshold.
\end{enumerate}

\section{The Measurement of Supersymmetry Parameters}

\subsection{Measurements of the SUSY Spectrum at the ILC}

Now that we have discussed the physics that determines the form of the 
spectrum of superparticles, we turn to the question of how we would
determine this spectrum experimentally.  This is not as easy as it might
seem. In this section, I will consider only
models in which the dark matter particle is the $\s N_1$, and  all other
SUSY particles decay to the $\s N_1$. This neutral and weakly interacting
particle would escape a collider detector unseen.  Nevertheless, methods have
been worked out not only to measure the masses of superparticles but also to
determine mixing angles and other information needed to convert these 
masses to values of the underlying parameters of the MSSM Lagrangian. 

Similar methods apply to other scenarios.  For example, in models in 
which the neutralino decays to a particle with gravitational
interactions, one would add that decay, if it is visible, to the analyses
that I will present.  It is possible in models of this type that the 
lightest Standard Model superpartner would be a charged slepton that is 
stable on the time scale of particle physics experiments.  That scenario
would produce very striking and characteristic events~\cite{FM}.

Most likely, this experimental study of the SUSY spectrum will begin in the 
next few years with the LHC experiments.  However, at a hadron collider like
the LHC, much of the kinematic information on superparticle production
is missing and so special tricks are needed even to measure the spectrum.  The 
study of supersymmetry should be much more straightforward at an $\ee$ collider
such as the planned International Linear Collider (ILC).  For this reason,
I would like to begin my discussion of the experiments in this section 
by discussing SUSY spectrum measurements at $\ee$ colliders.   More 
complete reviews of SUSY measurements at linear colliders can be 
found in~\cite{ResourceBook,FengNojiri}.

I first discuss slepton pair production, beginning with
the simplest process, $\ee\to \s \mu^+ \s\mu^-$ and considering 
successively the production of $\s\tau$ and $\s e$.  Each step will 
bring in new complexities and will allow new measurements of the SUSY 
parameters.

The process  $\ee\to \s \mu^+ \s\mu^-$, where $\s \mu$ is the partner of
either the left- or right-handed $\mu$, can be analyzed with the simple
formulae for scalar particle-antiparticle production.  The cross section 
for pair production from polarized initial electrons and positrons to 
final-state scalars with definite $SU(2)\times U(1)$ quantum numbers
is given by 
\beq
  {d\sigma\over d\cos\theta} = {\pi \alpha^2\over 2 s} \beta^3 \sin^2\theta
    \, |f_{ab}|^2  \ ,
\eeq{eetoscalarscalar}
where
\beq
  f_{ab} = 1 + {(I_e^3 + s_w^2)(I^3_\mu + s_w^2)\over c_w^2 s_w^2}
        {s\over s- \mz^2}
\eeq{fformula}
and, in this expression, $I^3 = -\half, 0$ for $a,b = L,R$.  For the
initial state, $a = L$ denotes the state $e^-_Le^+_R$ and $a=R$
denotes $e^-_Re^+_L$.  For the final state, $b = L$ denotes the 
$\s \mu$, $b = R$ the $\s{\bar \mu}$.  Notice that 
this cross section depends strongly on the polarization states:
\beqa
    |f_{ab}|^2 &=&   \quad  1.69  
               \quad e^-_Re^+_L \to \s{\bar \mu}^+\s {\bar\mu}^- \CR 
 &=&   \quad  0.42   \quad e^-_Le^+_R \to \s{\bar \mu}^+\s {\bar\mu}^- \CR 
 &=&   \quad  0.42   \quad e^-_Re^+_L \to \s \mu^+\s {\mu}^- \CR 
 &=&   \quad  1.98   \quad e^-_Le^+_R \to \s{\mu}^+\s {\mu}^- 
\eeqa{fforetosmu} 
The angular distribution is characteristic of pair-production of a spin 0 
particle; the normalization of the cross sections picks out the the correct
set of $SU(2)\times U(1)$ quantum numbers.

\begin{figure}
\begin{center}
\includegraphics[height=1.2in]{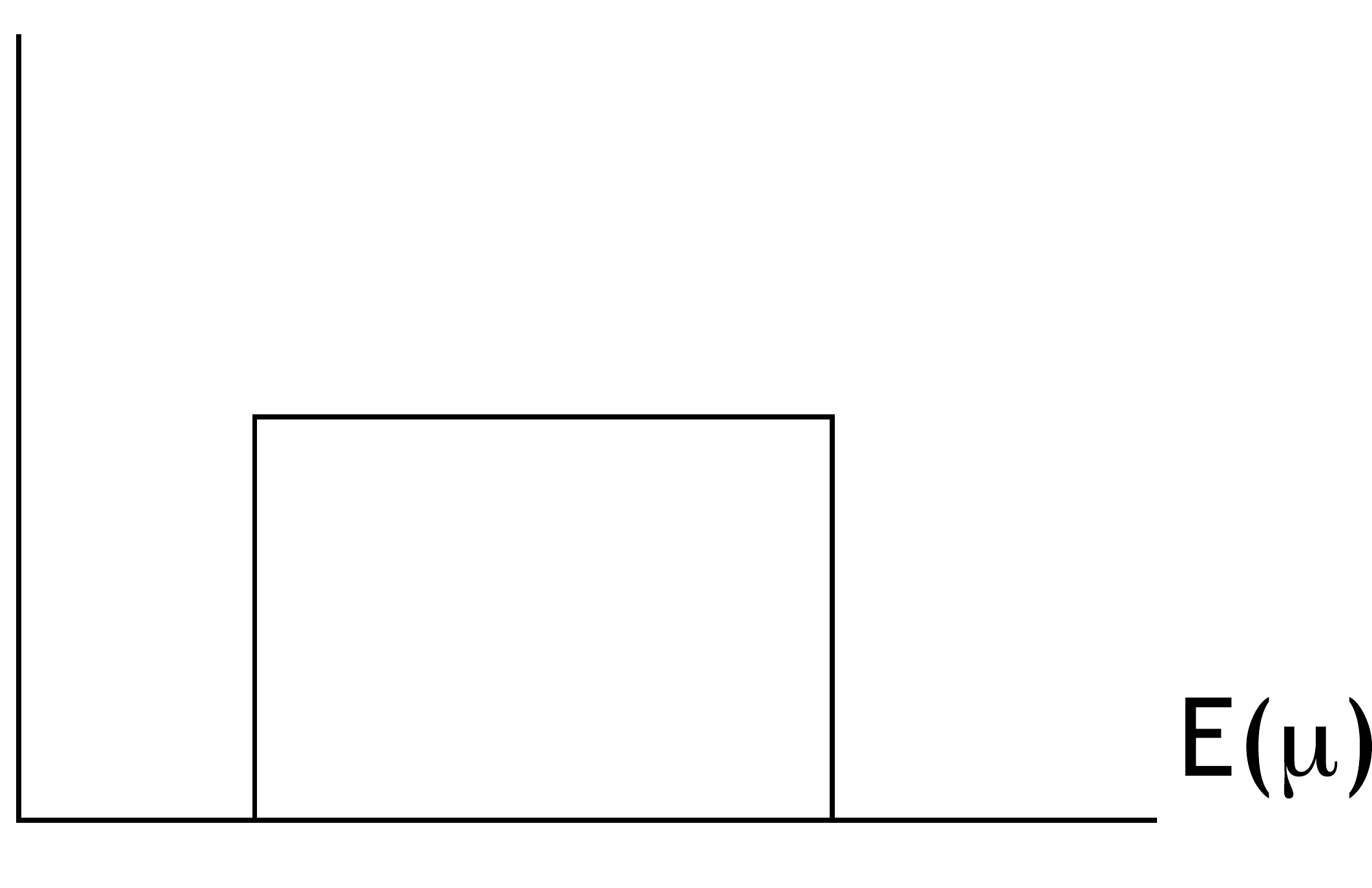}
\caption{Schematic energy distribution of final-state muons in 
     $\ee \to \s\mu^+ \s\mu^-$.}
\label{fig:smudecay}
\end{center}
\end{figure}

If the smuon is light, its only kinematically allowed decay might be 
$\s \mu \to \mu \s N_1^0$.  Even if the smuon is heavy, if the $\s N_1$ is 
mainly gaugino, this decay should be important.  As noted above, I am  assuming
that R-parity is conserved and that the $\s N_1$ is the lightest particle in 
the superparticle spectrum.  Then events with this decay on both sides 
will appear as
\beq
  \ee \to \mu^+\mu^- + \mbox{(missing\ $E$\ and\ $p$)}
\eeq{eetomumurxn}
The spectrum of the observed muons is very simple.  Since the $\s\mu$ has
spin 0, it decays isotropically in its own rest frame.  In $\ee$ 
production at a definite center of mass energy, the $\s \mu$ is produced
at a definite energy, and thus with a definite boost, in the lab.
The boost of an isotropic distribution is
a flat distribution in energy.  So, the muon energy distribution should be
flat, between endpoints determined by kinematics, as shown in 
the idealized Fig.~\ref{fig:smudecay}.

\begin{figure}
\begin{center}
\includegraphics[height=2.0in]{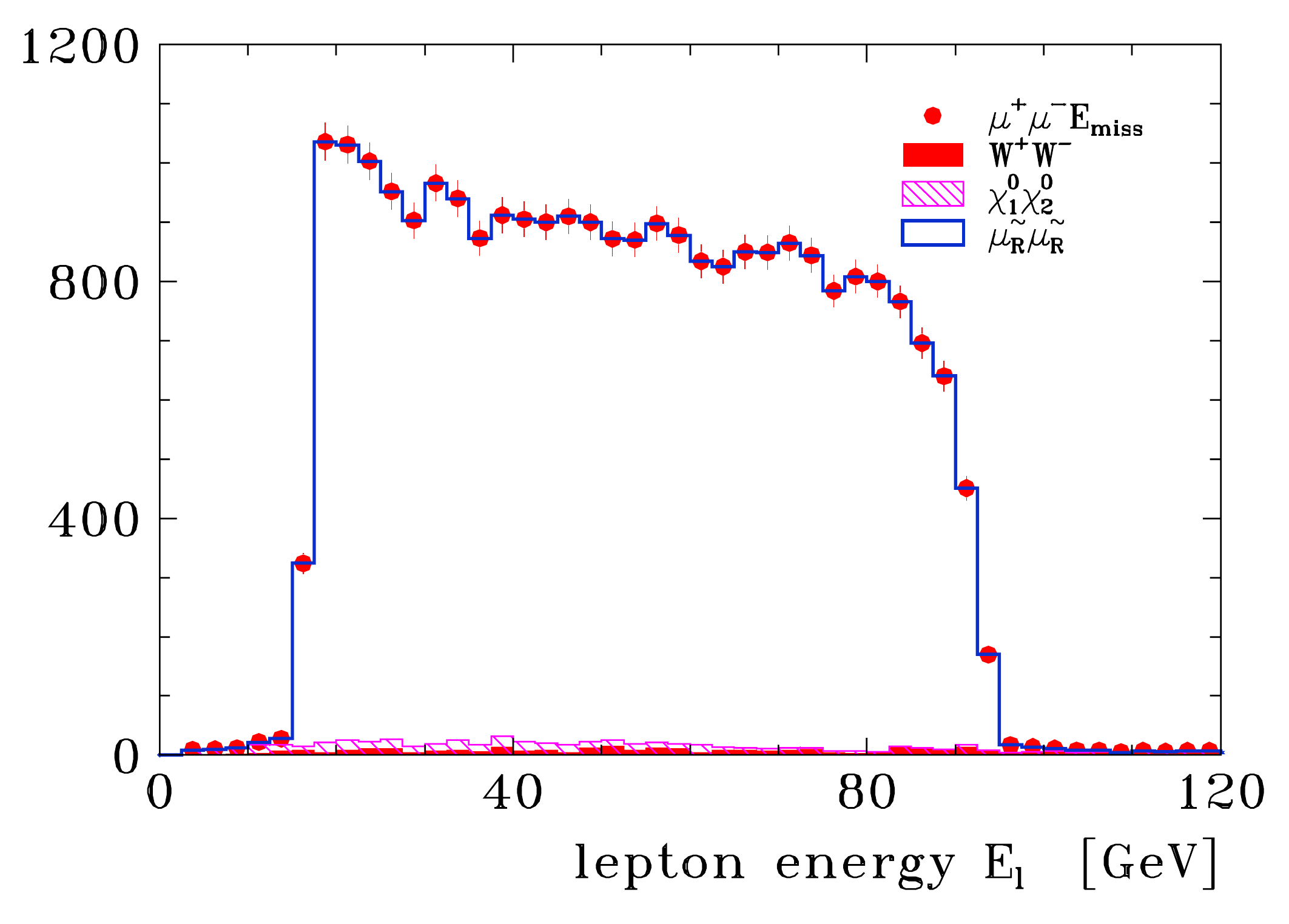}
\caption{Energy distribution of muons from $\ee\to \s{\bar \mu}^- 
\s{\bar\mu}^+$ at the ILC, in a simulation by Blair and Martyn
 that includes realistic momentum
resolution and beam effects~\cite{Weiglein}.}
\label{fig:BlairMartyn}
\end{center}
\end{figure}

The endpoint
positions are simple functions of the mass of the $\s\mu$ and the mass
of the $\s N_1$, 
\beq
       E_\pm = \gamma (1\pm \beta)\ {m^2(\s\mu) 
                        - m^2(\s N_1)\over 2 m(\s \mu)}\ ,
\eeq{smuendpoints}
where $\gamma = E_\CM/2m(\s\mu)$, $\beta = (1 - 4m^2(\s\mu)/E_\CM^2)^{1/2}$.
If we can identify both endpoint positions, we can solve for the two unknown
masses.    Figure~\ref{fig:BlairMartyn}
shows a simulation of the reconstructed smuon energy distribution  from
$\s{\bar \mu}$ pair production
 at the ILC~\cite{Weiglein}.
The high-energy edges of the distributions are rounded because of initial-state
radiation in the $\ee$ collision.  The experimenters expect to be
able to measure this effect and correct for it. Then they should obtain 
values of the smuon mass to an accuracy of about one hundred MeV, or
one part per mil.

A similar analysis applies to $\ee\to \s \tau^+\s\tau^-$, but there are 
several complications.  First, for the $\tau$ system, mixing between the 
$\s\tau$ and the $\s{\bar \tau}$ might be important, especially if $\tan\beta$
is large.  The production cross sections are affected directly by the
mixing.  For example, to compute the pair-production of the 
lighter $\s\tau$ mass eigenstate from a polarized initial state, 
 $e^-_Re^+_L\to \s\tau^-_1\s\tau^+_1$, 
we must generalize \leqn{eetoscalarscalar} to
\beq
  {d\sigma\over d\cos\theta} = {\pi \alpha^2\over 2 s} \beta^3 \sin^2\theta
    \, |f_{R1}|^2  \ ,
\eeq{eetotautau}
where
\beq
     f_{R1} = f_{RR} \cos^2\theta_\tau + f_{RL} \sin^2\theta_\tau
\eeq{eetotauf}
and $\theta_\tau$ is the mixing angle associated with the diagonalization
of the $\s\tau$ case of \leqn{Mbmatrix}.

Second, while the $\s{\bar \tau}^-$ can decay to $\tau^-_R \s b$ through gauge 
couplings, this weak eigenstate can also decay to $\tau^-_L \s h_d$ through 
terms proportional to the Yukawa coupling.  Both decay amplitudes contribute
to the observable decay $\s \tau_1 \to \tau \s N^0_1$. With the 
$\s\tau$ mixing angle fixed from the measurement of 
the cross section, the $\tau$ polarization
in $\s\tau$ decays can be used to determine the mixing angles in the 
diagonalization of the neutralino mass 
matrix \leqn{neutralinomatrix}~\cite{Nojiritau}.

\begin{figure}
\begin{center}
\includegraphics[height=2.0in]{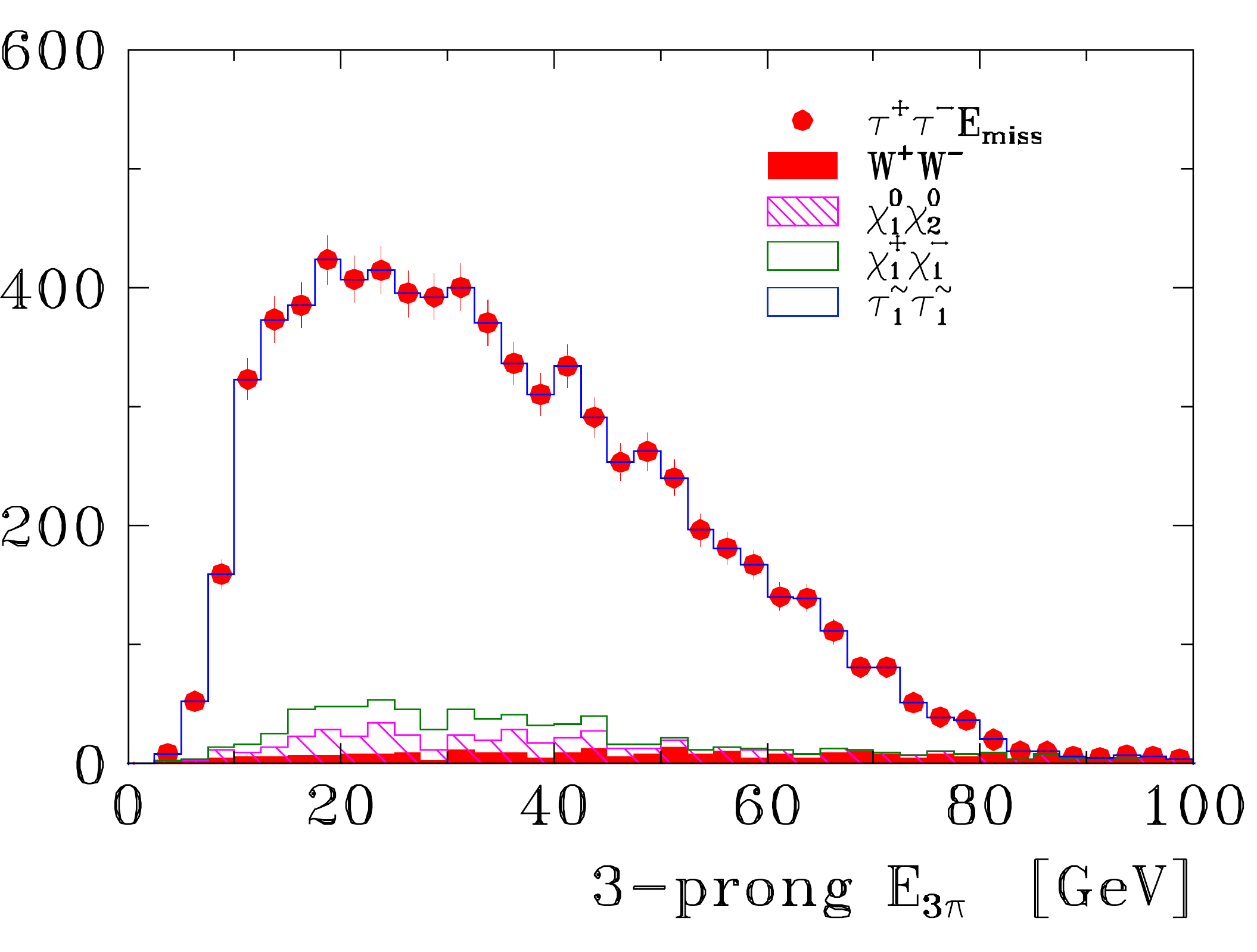}
\caption{Energy distribution of the three-pion system
 from $\ee\to \s{\tau}^-_1 
\s{\tau}^+_1$ at the ILC, with a $\tau$ decay to $3 \pi$,
 in a simulation by Blair and Martyn
 that includes realistic momentum
resolution and beam effects.~\cite{Weiglein}.}
\label{fig:taudists}
\end{center}
\end{figure}

In   Fig.~\ref{fig:taudists}, I  show the 
distribution of total visible energy in $\s\tau \to 3 \pi + \nu + \s N^0_1$
at the ILC.  Though there is no longer a sharp feature at the kinematic 
endpoint, it is still possible to accurately determine the $\s\tau$ mass
by fitting the shape of this distribution.

\begin{figure}
\begin{center}
\includegraphics[height=1.5in]{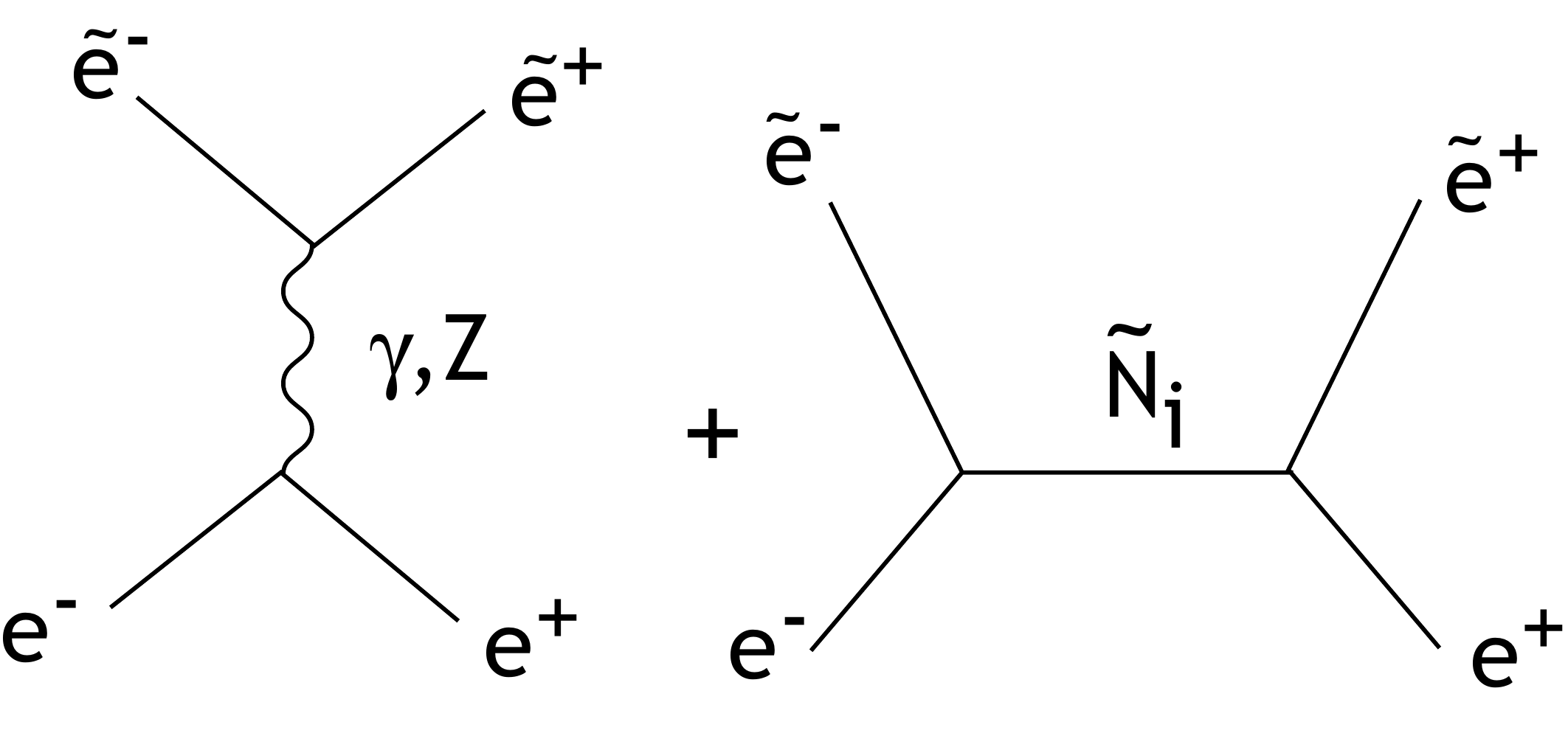}
\caption{Feynman diagrams contributing to $\ee\to \s e^-\s e^+$.}
\label{fig:eetosese}
\end{center}
\end{figure}

The physics of $\ee\to \s e^+ \s e^-$ brings in further new features.
In this case, there is a new Feyman diagram, involving $t$-channel neutralino
exchange.  The two diagrams contributing to the cross section for this 
process are shown in Fig.~\ref{fig:eetosese}.   The $t$-channel diagram
turns out to be the more important one,  dominating the $s$-channel gauge 
boson exchange and generating a large forward peak in selectron production.
The cross section for $e^-_Re^+_L\to \s{\bar e}^- \s{\bar e}^+$ is given
by another generalization of \leqn{eetoscalarscalar},
\beq
  {d\sigma\over d\cos\theta} = {\pi \alpha^2\over 2 s} \beta^3 \sin^2\theta
    \, |{\cal F}_{RR}|^2  \ ,
\eeq{eetosese}
where
\beq
 {\cal F}_{RR} = f_{RR} - \sum_i
\left| {V_{01i}\over c_w}\right|^2 {s\over m_i^2 - t} \ ,
\eeq{Fforse}
with the sum running over
 neutralino mass eigenstates.  The factor $V_{01i}$ is 
a matrix element of the unitary matrix introduced in \leqn{mNdiag}.  

The $t$-channel diagram also allows new processes such as $e^-_Le^+_L \to
\s e^- \s{\bar e}^+$.  Note the correlation of the initial-state electron
and position spins with the identities of the final-state selectrons.
A complete set of polarized 
cross sections for selectron pair production in $\ee$
and $e^-e^-$ collisions can be found in \cite{myselectrons}.

\begin{figure}
\begin{center}
\includegraphics[height=1.5in]{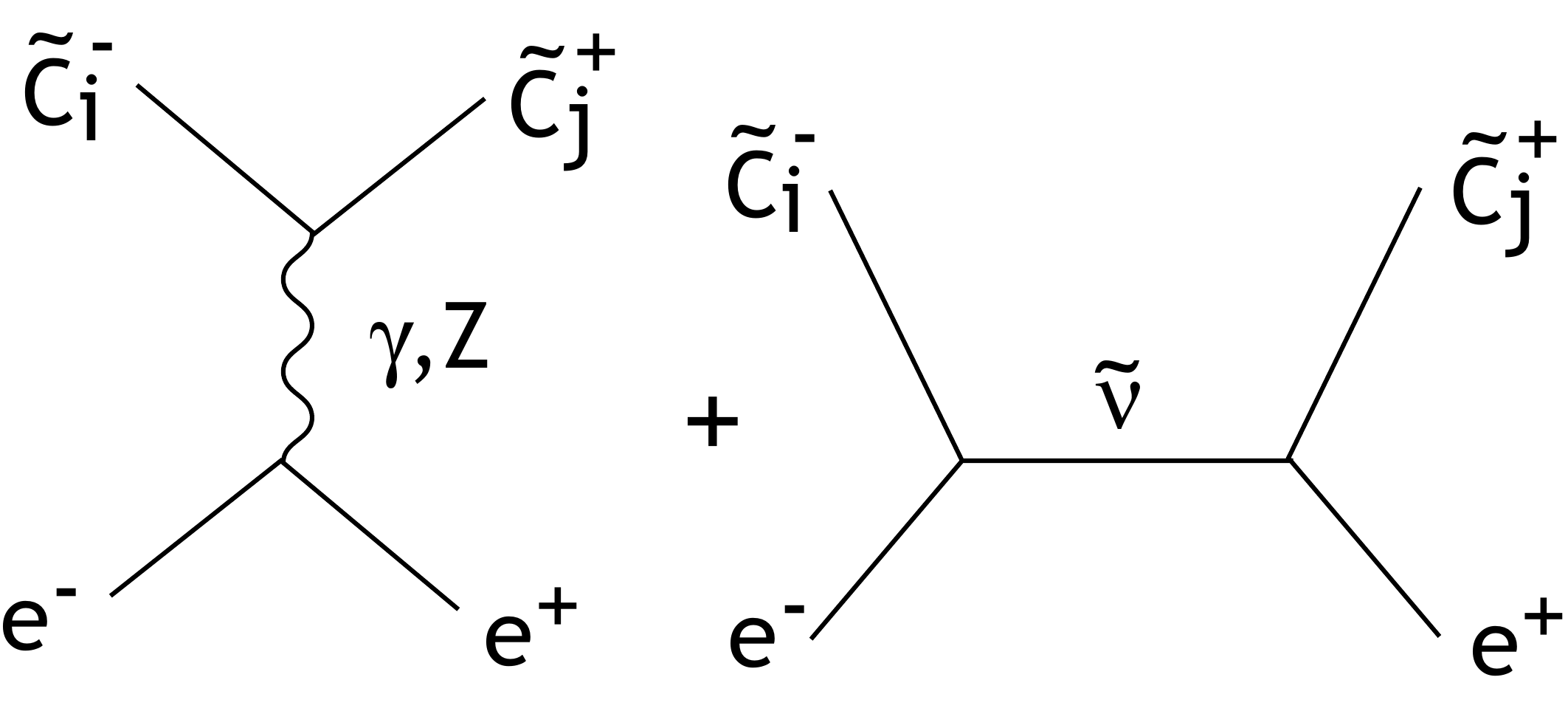}
\caption{Feynman diagrams contributing to $\ee\to \s C^-_i \s C^+_j$.}
\label{fig:charginopairs}
\end{center}
\end{figure}

\begin{figure}
\begin{center}
\includegraphics[height=3.0in]{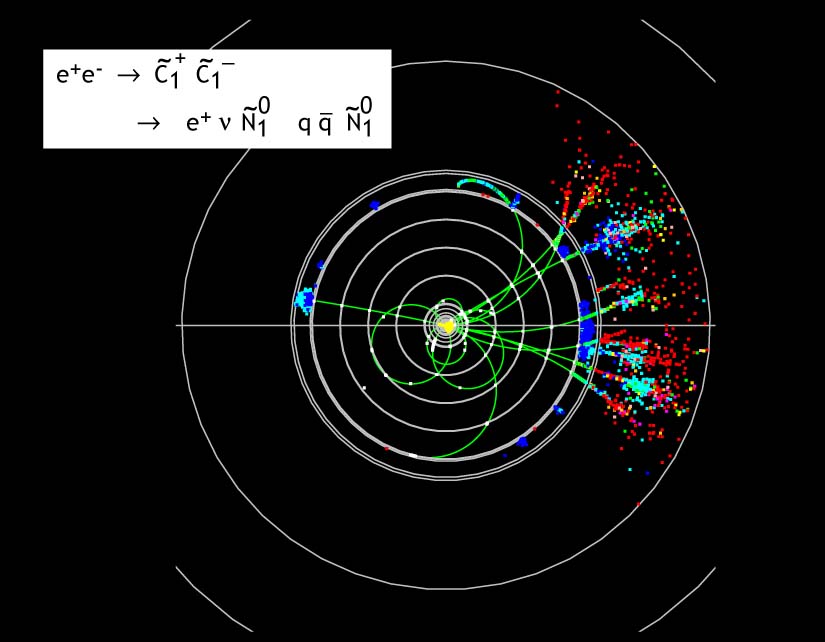}
\caption{A simulated chargino pair production event at the ILC~\cite{Graf}.}
\label{fig:charginoevent}
\end{center}
\end{figure}

The cross sections for chargino and neutralino pair production in $\ee$
collisions are somewhat more complicated, but still there are interesting 
things to say about these processes.  Chargino pair production is given by 
the Feynman diagrams shown in Fig.~\ref{fig:charginopairs}.  These 
diagrams are just the supersymmetric analogues of the diagrams for $\ee\to 
W^+W^-$.  As in that process, the most charcteristic final states are those
with a hadronic decay on one side of the event and a leptonic decay on the 
other side, for example,
\beq
     \s C^+_1 \to \ell^+ \nu \s N^0_1 \ ,
                  \quad \s C^-_1 \to d\bar u \s N^0_1 \ .
\eeq{Charginodecay}
A typical event of this kind is shown in Fig.~\ref{fig:charginoevent}.

The chargino and neutralino production cross sections have a strong dependence
on the  mixing angles in \leqn{mNdiag} and \leqn{mCdiag} and offer a number of 
strategies for the determination of these mixing angles.  Let me present one
such strategy here.  Consider the reaction from a polarized initial state
$e^-_Re^+_L\to \s C^-_1 \s C^+_1$.  Since we have an initial $e^-_R$, the 
$t$-channel diagram vanishes because the right-handed electron does not 
couple to the neutrino.  Now simplify the $s$-channel diagram by 
considering the limit of high energies,
$s \gg \mz^2$. In this limit, it
 is a good approximation to work with weak gauge 
eigenstates $(B^0, W^0)$ rather than the mass eigenstates $(\gamma, Z^0)$.
The weak eigenstate basis gives a nice simplification.  The initial $e^-_R$
couples only to $B^0$.  But $\s w^\pm$ couple only to $W^0$, so at high
energy the $s$-channel diagram gets contributions only from the Higgsino
components of the $\s C^-_1$ and $\s C^+_1$ eigenstates. 
If we go to still higher energies, $s \gg m(\s C_1)^2$, there is a further
simplification.  The cross section for $\s h_R^- \s h_L^+$ 
production is forward-peaked,
and the cross section for  $\s h_L^- \s h_R^+$ production is backward-peaked.
Then, the cross section for $e^-_Re^+_L\to \s C^-_1 \s C^+_1$ takes the 
form
\beq
  {d\sigma\over d\cos \theta} \sim {\pi \alpha^2\over 8 c_w^2 s}
 \bigl[ |V_{+21}|^4 (1 + \cos\theta)^2 + |V_{-21}|^2 (1-\cos\theta)^2 \bigr]\ .
\eeq{CCsigma}
In this limit, it is clear that we can read off both of the mixing angles
in \leqn{mCdiag} from the shape of this cross section.  

\begin{figure}
\begin{center}
\includegraphics[height=3.0in]{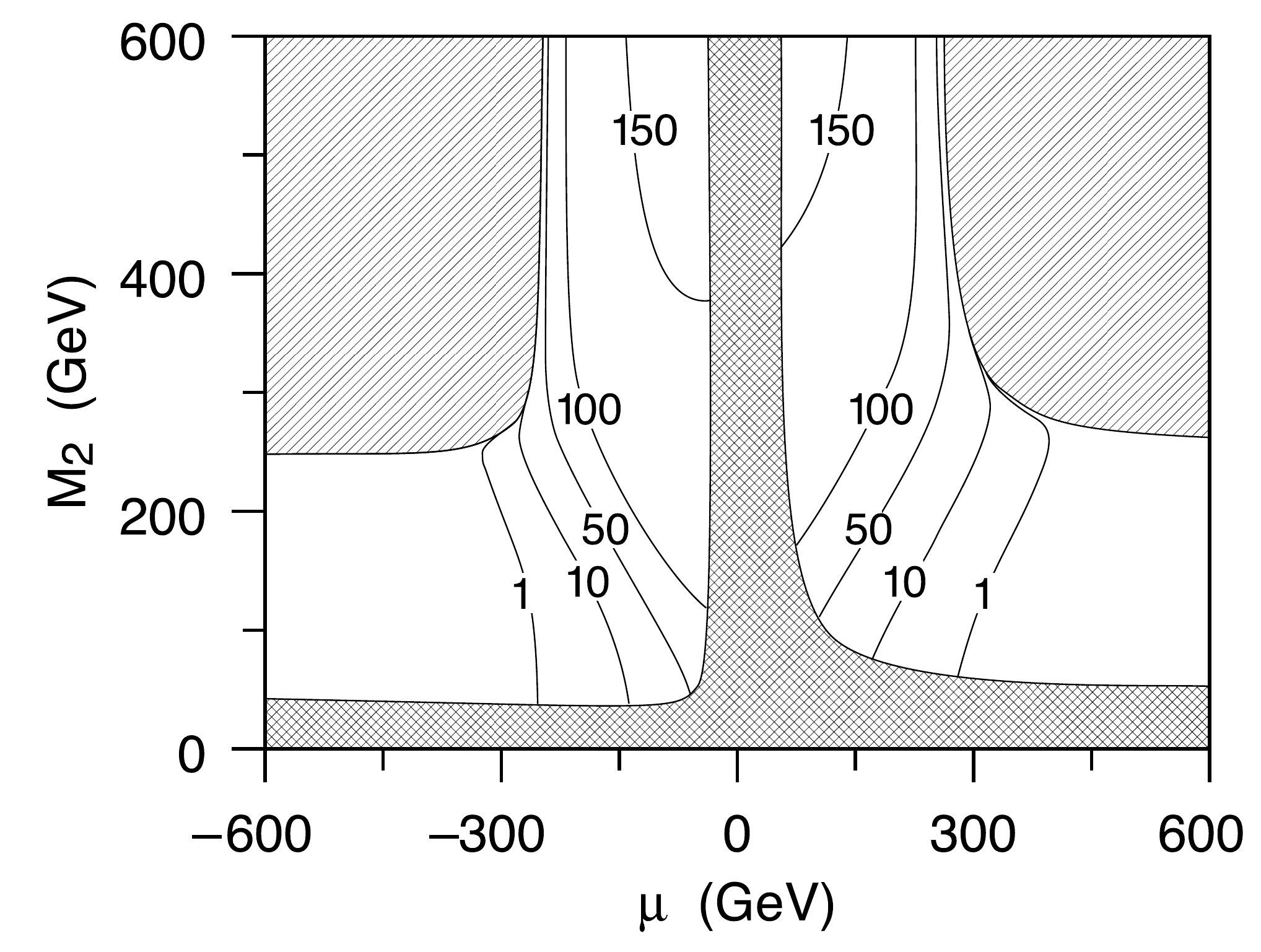}
\caption{Contours of constant cross section 
for the process $e^-_R e^+_L \to C^-_1 C^+_1$  (in fb, for $E_\CM = 500$ GeV),
as a function 
of the underlying SUSY parameters~\cite{FPMT}.
  The region shown is that in which 
the lightest chargino mass varies from 50 to 200 GeV. 
For fixed $\s C^+_1$ mass, the cross section increases from zero to about 
150 fb as we move from the gaugino region into the Higgsino region. }
\label{fig:CCmixing}
\end{center}
\end{figure}

The use of high-energy limits simplified this analysis, but the sentivity
of this cross section to the chargino mixing angles is
not limited to high energy.  Even relatively close to threshold,
the polarized cross sections for chargino production depend strongly on 
the chargino mixing angles and can be used to determine their values.
In Fig.~\ref{fig:CCmixing}, I show contours of constant cross section for 
 $e^-_Re^+_L\to \s C^-_1 \s C^+_1$ in the $(m_2, \mu)$ plane (for $\tan\beta
= 4$ and assuming gaugino unification)~\cite{FPMT}.  
The value of this cross section is
always a good measure of whether the SUSY parameters in Nature put us in the
gaugino or the Higgsino region of Fig.~\ref{fig:NCmixing}.

\subsection{Observation of SUSY at the LHC}

Now we turn to supersymmetry production processes at the LHC.  This 
subject, though more difficult, has immediate importance, since the 
LHC experiments are just about to begin.

The reactions 
that produce superparticles are typically much more complicated at hadron 
colliders than at lepton colliders.  This is true for several reasons.
High energy collisions of hadrons are intrinsically more complicated
because the final states include the fragments of the initial hadrons that 
do not participate in the hard reaction.  More importantly, 
the dominant reactions at
hadron colliders are those that involve strongly interacting superparticles.
This means that the primary particles are typically the heavier ones in the
spectrum, which then decay in several steps.  In addition, large backgrounds 
from QCD obscure the signatures of supersymmetric particle production in 
many channels.

Because of these difficulties, there is some question whether SUSY particle
production can be observed at the LHC.  However, as I will explain, the
signatures of supersymmetry are still expected to be striking and 
characteristic.   It is not so clear, though, to what extent it is possible
to measure the parameters of the SUSY Lagrangian, as I have described can 
be done from ILC experiments.  This is an important study that still offers
much room for new ideas.

The discovery of SUSY particles at the LHC and the measurement of SUSY 
parameters has been analyzed with simulations at a number of parameter
points.  Collections of interesting studies can be found 
in~\cite{Weiglein,ATLAS,CMS}.

\begin{figure}
\begin{center}
\includegraphics[height=3.0in]{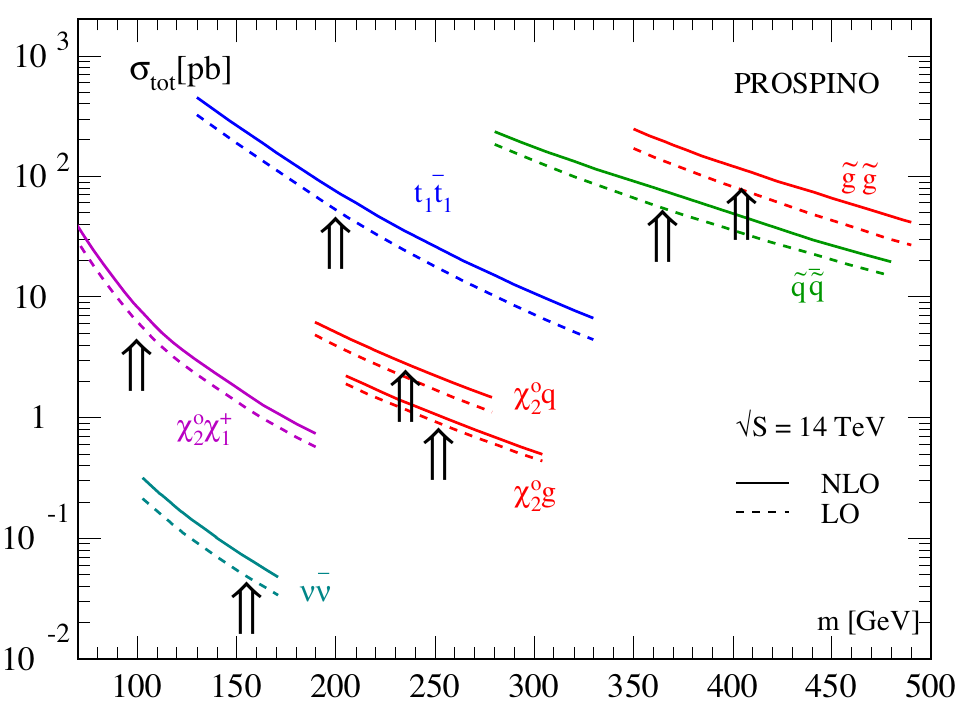}
\caption{Cross sections for the pair-production of supersymmetric
      particles at the LHC, from \cite{Spira}.}
\label{fig:SUSYcross}
\end{center}
\end{figure}

The dominant SUSY production processes at the LHC are
\beq
    gg \to \s g \s g \ , \   \s q \s q^* \qquad   g q \to \s g \s q
\eeq{ggrxns}
These cross sections are large---tens of pb in typical cases.  The 
values of numerous SUSY production cross sections at the LHC are 
shown in Fig.~\ref{fig:SUSYcross}~\cite{Spira}.

We have seen that the squarks and gluinos are typically the heaviest
particles in the supersymmetry spectrum.  The gluinos and squarks thus
will decay to lighter superparticles.  Some of these decays are 
simple, \eg, 
\beq 
             \s {\bar q} \to \bar q \s N^0_1\ .
\eeq{barqdecay}
However, other decays can lead to complex decay chains such as
\beq
      \s q \to q N^0_2 \to q (\ell^+\ell^-) \s N^0_1 \ , \qquad
     \s g \to u \bar d C^+_1 \to u\bar d W^+ \s N^0_1     \  .
\eeq{sqsgdecay}

\begin{figure}
\begin{center}
\includegraphics[height=3.0in]{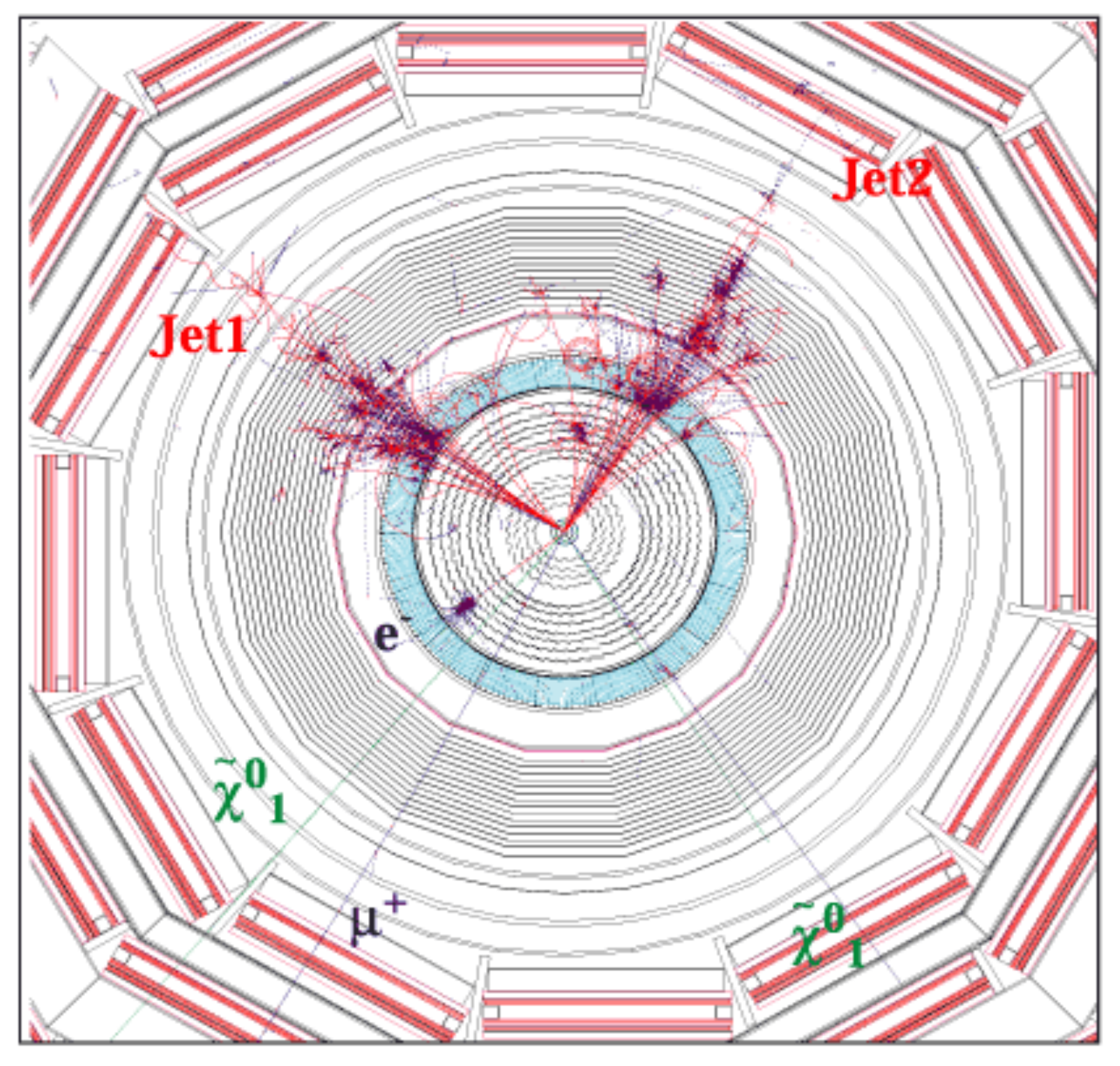}
\caption{Simulated SUSY particle production event in the CMS detector 
            at the LHC~\cite{CMSevent}.}
\label{fig:SUSYcartoon}
\end{center}
\end{figure}

With the assumptions that R-parity is conserved and that the $N^0_1$ is 
the LSP, all SUSY decay chains must end with the $N^0_1$, which is 
stable and very weakly interacting.  SUSY production processes at 
hadron colliders then have unbalanced visible momentum, accompanied by 
multiple jets and, possibility, isolated leptons or $W$ and $Z$ bosons.  
Momentum balance 
along the beam direction cannot be checked at hadron colliders, because
fragments of the initial hadrons exit along the beam directions, but 
an imbalance of transverse momentum will be visible and 
can be a characteristic signature of 
new physics.  SUSY events contain this signature and the general large
activity characteristic of heavy particle production.  A simulated event
of this type is shown in Fig.~\ref{fig:SUSYcartoon}.  

\begin{figure}
\begin{center}
\includegraphics[height=4.0in]{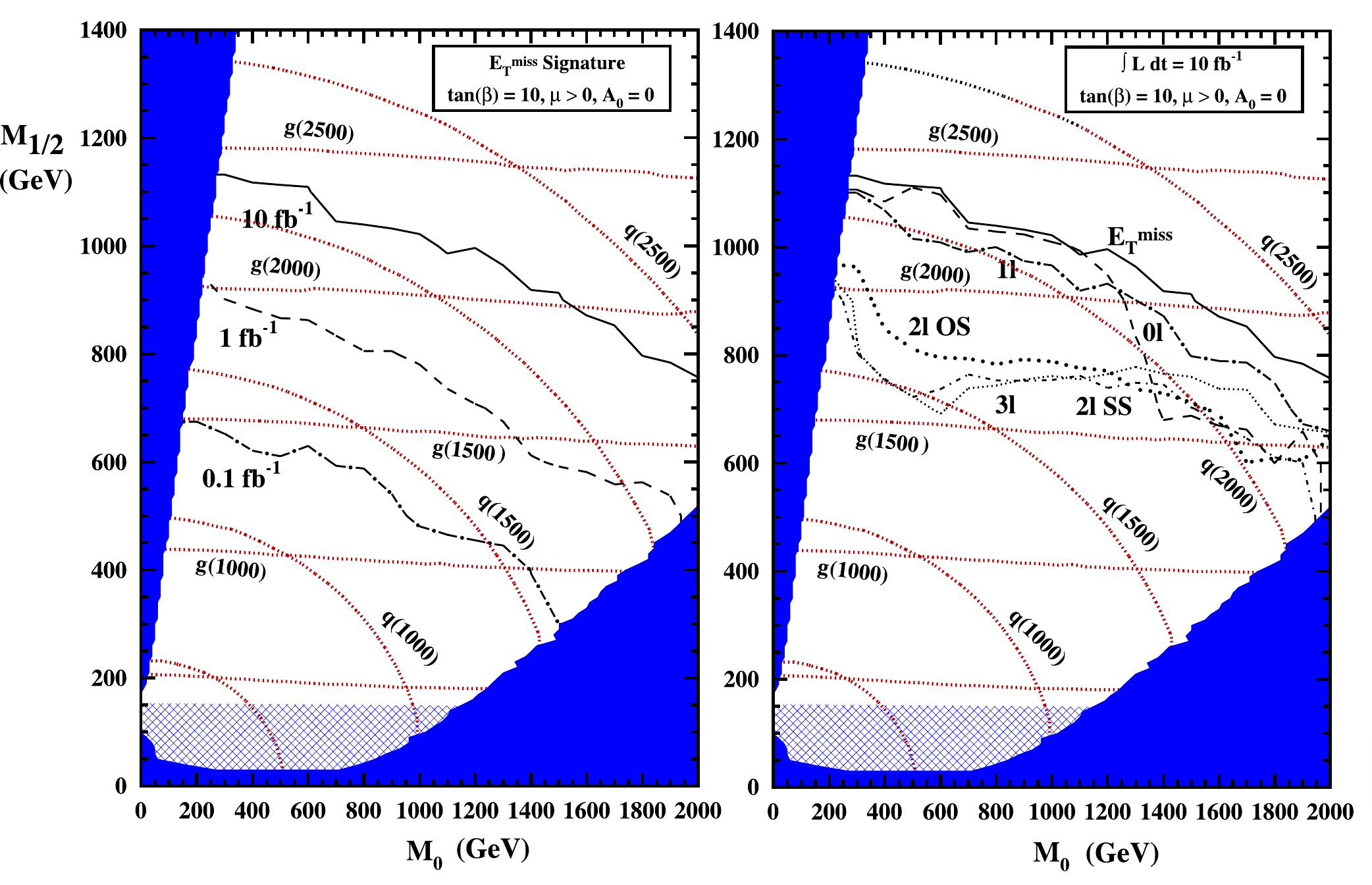}
\caption{Estimates by the ATLAS collaboration of the observability of 
       various signatures of SUSY at the LHC.  The plots refer to models
        with grand unification and universal sfermion and gaugino masses
        $M_0$ and $M_{1/2}$.  The left-hand plot shows the region of this
         parameter space in which it is possible to detect the 
      signature of missing $E_T$ plus multiple jets at various levels of
        integrated luminosity.  The right-hand plot shows the region of this
       parameter space in which it is possible to detect an excess of 
       events with one or more leptons in addition to jets and 
         missing $E_T$~\cite{Tovey}.}
\label{fig:Tovey}
\end{center}
\end{figure}

Figure~\ref{fig:Tovey} shows a set of estimates given by Tovey and the 
ATLAS collaboration of the discovery potential for SUSY as a function of
the LHC luminosity~\cite{Tovey}. The most important
backgrounds come from processes that are themselves relatively rare 
Standard Model reactions with heavy particle production,
\beq
     pp \to  (W, Z, t\bar t)\  + \ \mbox{jets}\ .
\eeq{ppbkgd}
With some effort, we can experimentally normalize and control 
these backgrounds and reliably discovery SUSY production as a new physics
process.    In the figure, the contours for 5$\sigma$ excesses of 
events above these backgrounds for
various signatures of SUSY events are plotted as a function of the 
so-called `mSUGRA' parameters.  The SUSY models considered are defined as 
follows:  Assume gaugino unification with a universal gaugino mass $m_{1/2}$
at the grand unification scale.  Assume also that all scalar masses, including
the Higgs boson mass parameters, are unified at the grand unification scale
at the value $m_0$.  Assume that the $A$ parameter is universal at the 
grand unification scale; in the figures, the value $A = 0$ is used.  Fix the
value of $\tan\beta$ at the weak scale.  Then it is possible to solve for
$\mu$ and $B$, up to a sign, from the condition that electroweak symmetry
is broken in such a way as to give the observed value of the $Z^0$ mass.
(I will describe this calculation in Section~6.1.)  This gives a
4-parameter subspace of the full 24-dimensional parameter space of the 
CP- and flavor-conserving MSSM, with the parameters
\beq
      m_0\ , \ m_{1/2}\ , \ A\ , \ \tan\beta\ , \ \mbox{sign}(\mu) \ .
\eeq{mSUGRA}
This subspace is often used to express the results of 
phenomenological analyses of supersymmetry.  In interpreting such results,
one should remember that this choice of parameters is used for simplicity
rather than being motivated by physics.

The figure shows contours below which the various signatures of 
supersymmetry significantly  modify the Standard Model expectations. 
  For clarity, the contours of constant squark and 
gluino mass are also plotted.  The left-hand plot shows Tovey's results 
for the missing transverse momentum plus multijets signature at various
levels of LHC integrated luminosity.    It is 
remarkable that, in the models in which the squark or gluino mass is below
1 TeV, SUSY should be discoverable with a data sample equivalent to a 
small fraction of a year of running.  The right-hand plot shows the 
contours for the discovery of a variety of SUSY signals, with up to 
three leptons plus jets plus missing transverse momentum, with roughly
one year of data at the initial design luminosity.  The signals 
are, as I have described, relatively robust with repect to 
uncertainties in the Standard Model backgrounds.  This makes it very likely
that, if SUSY is really present in Nature as the explanation of electroweak
symmetry breaking, we will discover it at the LHC.

\begin{figure}
\begin{center}
\includegraphics[height=3.0in]{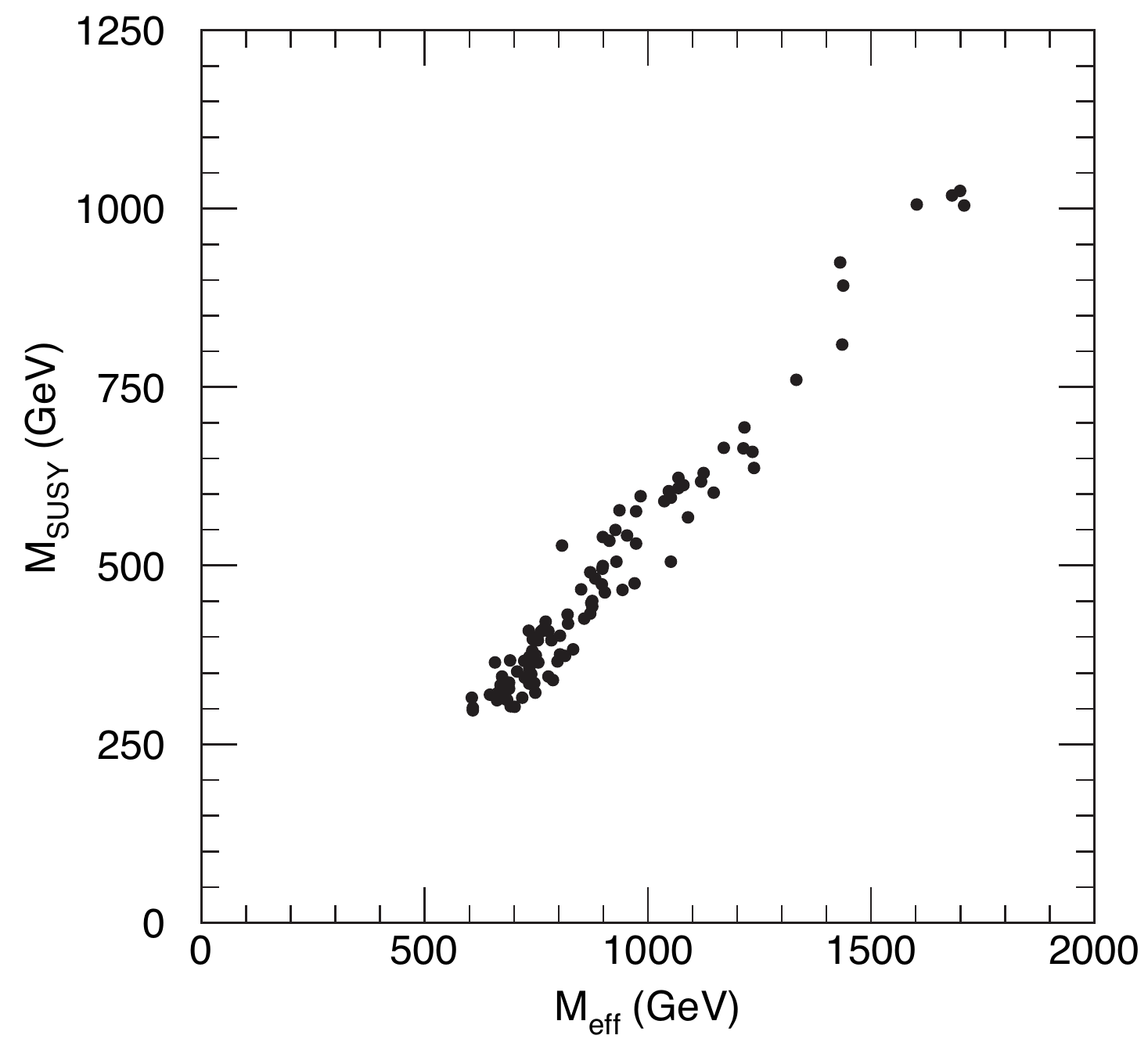}
\caption{Correlation between the value of the observable \leqn{Meff} and the
     lighter of the squark and gluino masses, from \cite{Hinch}.}
\label{fig:Meff}
\end{center}
\end{figure}

The general characteristics of SUSY events also allow us to estimate the 
SUSY mass scale in a relatively straightforward way.  In Fig.~\ref{fig:Meff},
I show a correlation pointed out by Hinchliffe and 
collaborators~\cite{Hinch}  between 
the lighter of the squark and gluino masses and the variable
\beq
       M_{eff} =   \Eslash_T + \sum_1^4 E_{Ti} 
\eeq{Meff}
given by the sum of
 the transverse momenta of the four highest $E_T$ jets together
with the value of the missing transverse momentum. The correlation applies
reasonably well to mSUGRA models.  In~other models with smaller mass 
gaps between the squarks and the lightest neutralino, this relation can 
break down, but $M_{eff}$ still measures the mass difference between the
squark or gluino and the $\s N^0_1$~\cite{KitanoNomura}.  Some more 
sophisticated techniques for determining mass scales in SUSY models 
from global kinematic variables are
described in~\cite{oset}.

\subsection{Measurements of the SUSY Spectrum at the LHC}

So far, I have only discussed the observation of the qualitative features 
of the SUSY model from global measures of the properties of 
events.  Now I would like to give some examples of analyses in which
specific details of the SUSY spectrum are measured with precision at the
LHC.  The examples that I will discuss involve the decay chain
\beq
       \s q \to q \s N^0_2 \ , \  \s N^0_2 \to \s N^0_1 \ell^+\ell^- \ ,
\eeq{sqdecaychain}
which is typically seen in models in which the gluino is heavier than 
the squarks and the LSP is gaugino-like.  

The decay of the $N^0_2$ can proceed by any of the mechanisms:
\beqa
    \s N^0_2 &\to& \ell^\pm + \s \ell^\mp \ , 
                \ \s \ell^\mp \to \ell^\mp \s N^0_1\CR
    \s N^0_2 &\to&  \s N^0_1 Z^0\ , \ Z^0 \to \ell^+\ell^-\CR
    \s N^0_2 &\to&  \s N^0_1 Z^{0*} \ , \ Z^{0*} \to  \ell^+\ell^-  \ .
\eeqa{Ntwodecay}
The last line indicates a virtual $Z^0$, decaying off-shell.  In a model
with gaugino unification and heavy Higgsinos, $\s N_2$ is mainly $\s w^0$
and $\s N_1$ is mainly $\s b^0$.  Then these modes are preferred in the 
order listed as long as they are kinematically allowed.  If the slepton
decay is allowed, this is the dominant model.  Otherwise, the decay to 
$\s N_1 Z^0$ or other open two-body decays dominate.  If no two-body
decays are open, the $\s N_2$ must decay through three-body processes such as
the last line of \leqn{Ntwodecay}.

\begin{figure}
\begin{center}
\includegraphics[height=3.0in]{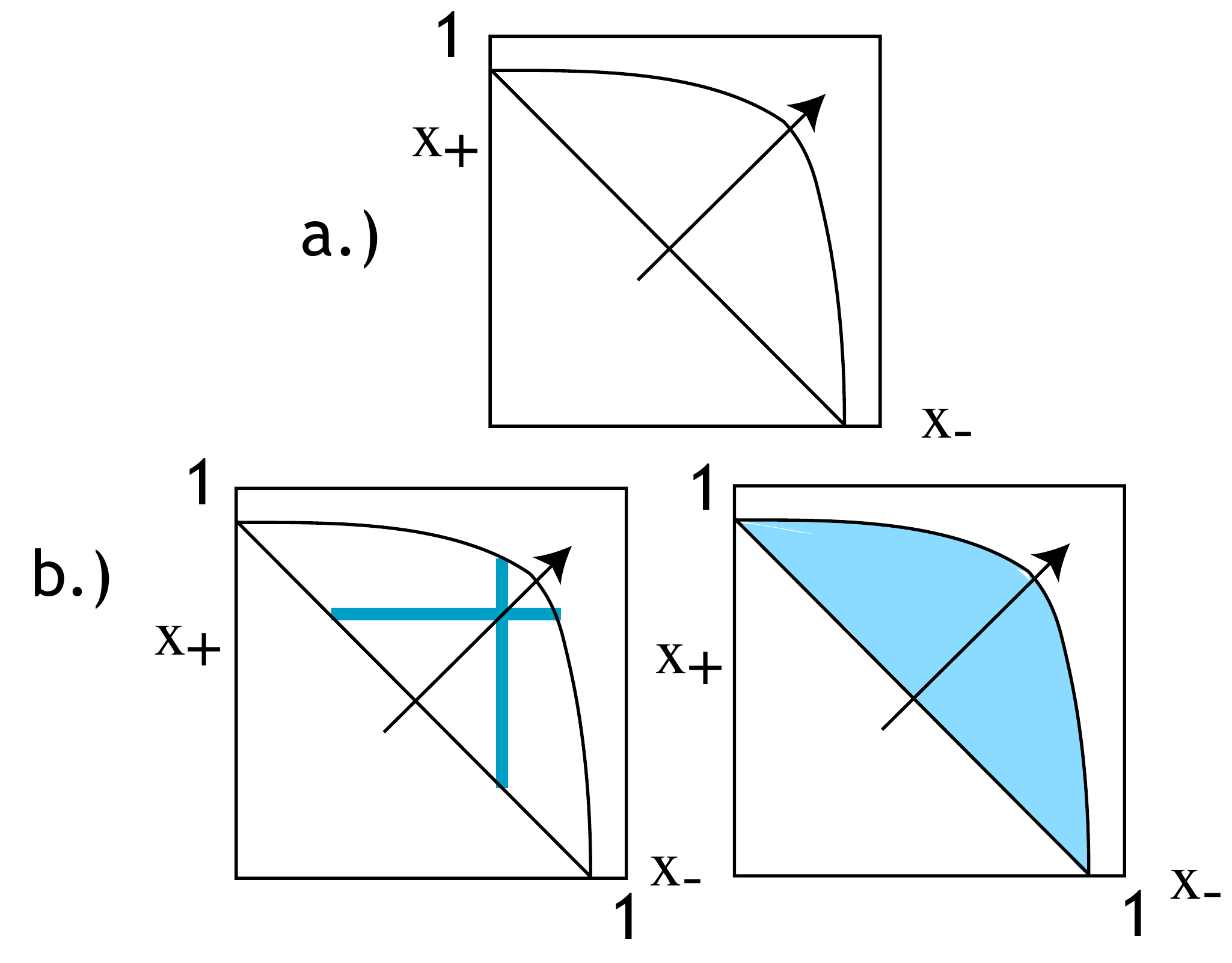}
\caption{The Dalitz plot describing 3-body neutralino decays, 
     $\s N_2^0 \to \s N^0_1 \ell^+\ell^-$.}
\label{fig:Dalitz}
\end{center}
\end{figure}

The decay to an on-shell $Z^0$ is hard to work with~\cite{EllisW}, but 
the other two cases
can be explored in depth.  It is useful to begin with the {\it Dalitz plot }
associated with the 3-body $(\s N_1, \ell^+, \ell^-)$ system.  Let
\beq
    x_0 = {2 E(\s N_1)\over m(\s N_2)}\ ,  \quad  
  x_+ = {2 E(\ell^+)\over m(\s N_2)}\ ,  \quad    
          x_- = {2 E(\ell^-)\over m(\s N_2)}
 \ ,
\eeq{xvariablesforN}
where the energies are measured in the rest frame of the $N_2$.  The three
variables are related by 
\beq
        x_0 + x_1 + x_2 = 2 \ .
\eeq{xidentity}
The three-body decay phase space is given by 
\beq
      \int d\Pi_3 =   {m^2(\s N_2)\over 128\pi^3} \int\, dx_+\, dx_-\ ;
\eeq{Pithree}
that is, phase space is flat in the variables \leqn{xvariablesforN}.
The basic kinematic identities involving the Dalitz plot
 variables are straightforward
to work out, especially if we ignore the masses of the leptons.
The kinematically allowed region is
a wedge of the $(x_+,x_-)$ plane bounded by the curves
\beqa
      x_+ + x_- &=&   1 - (m(\s N_1)/m(\s N_2))^2 \CR
    (1-x_+)(1-x_-) &=&   (m(\s N_1)/m(\s N_2))^2  \ ,
\eeqa{Dalitzregion}
as shown in Fig.~\ref{fig:Dalitz}(a).
The invariant masses of two-body combinations are given in terms of the $x_a$
by 
\beq
   {m^2(\s N_1 \ell^\pm) \over m^2(\s N_2)} = (1 - x_\mp) \ , \quad
 {m^2(\ell^+\ell^-)\over m^2(\s N_2)} = (1 - {m(\s N_1)^2\over m(\s N_2)^2})\ .
\eeq{massxrels}

I am assuming that the $\s N_1$ is stable and weakly interacting.  
In this case,
the $\s N_1$ will not be observed in the LHC experiments, 
and also the frame of 
the $\s N_2$ cannot be readily determined.  
The only property of this system
that is straightforward to measure is the two-body
invariant mass $m(\ell^+\ell^-)$.  So it is interesting to note that the 
distribution of this quantity distinguishes the first and third cases in 
\leqn{Ntwodecay}, in the manner shown in Fig.~\ref{fig:Dalitz}(b).
 In the case of a two-body decay to an intermediate 
slepton, the decays populate two lines on the Dalitz plot, leading to a 
sharp discontinuity at the kinematic endpoint.  In the case of a three-body
decay, the events fill the whole Dalitz plot, producing a distribution 
with a slope at the endpoint.  With a good understanding of the detector
resolution in the dilepton invariant mass, these cases can be distinguished
experimentally.

\begin{figure}
\begin{center}
\includegraphics[height=2.7in]{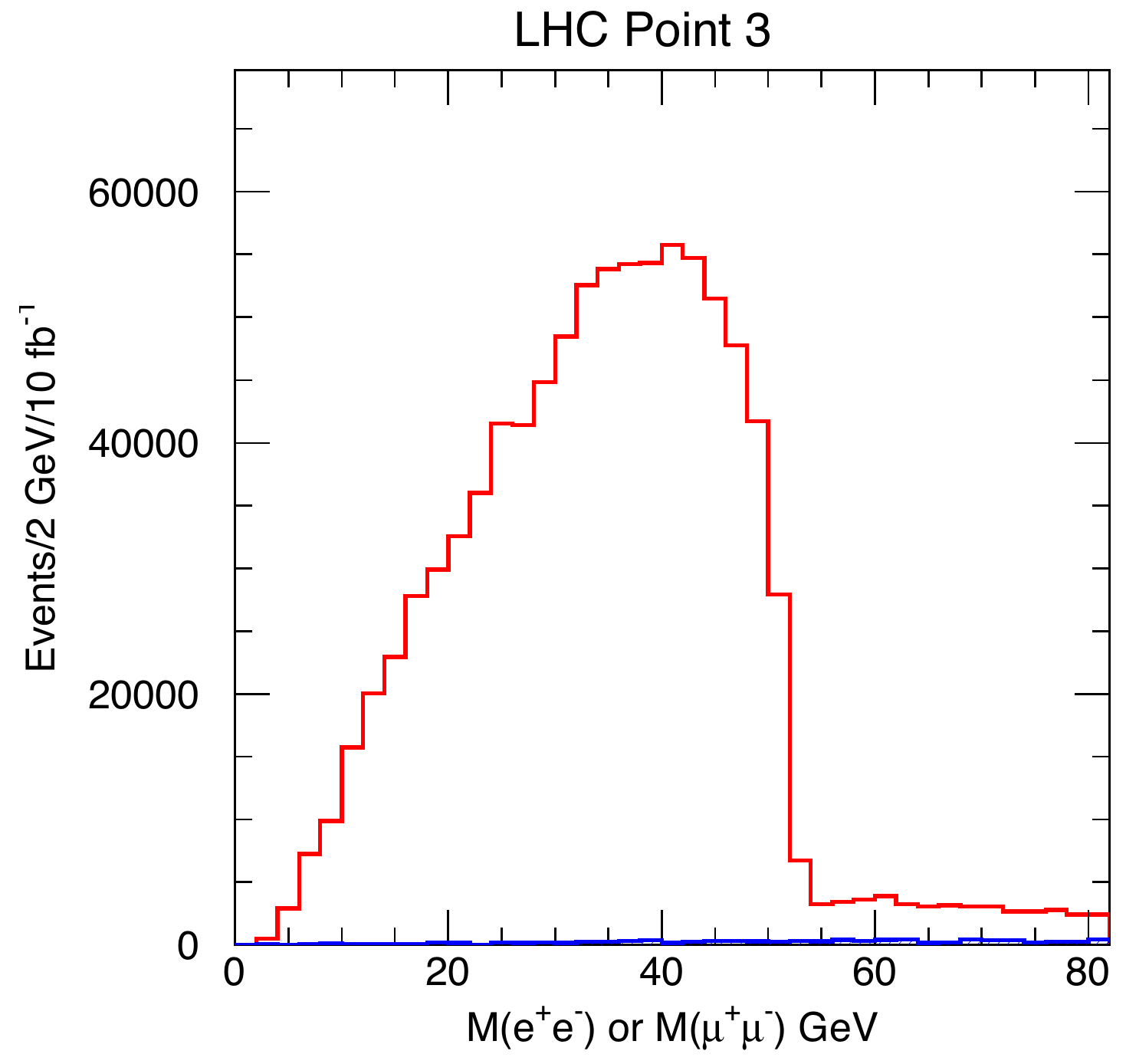}\quad
\includegraphics[height=2.7in]{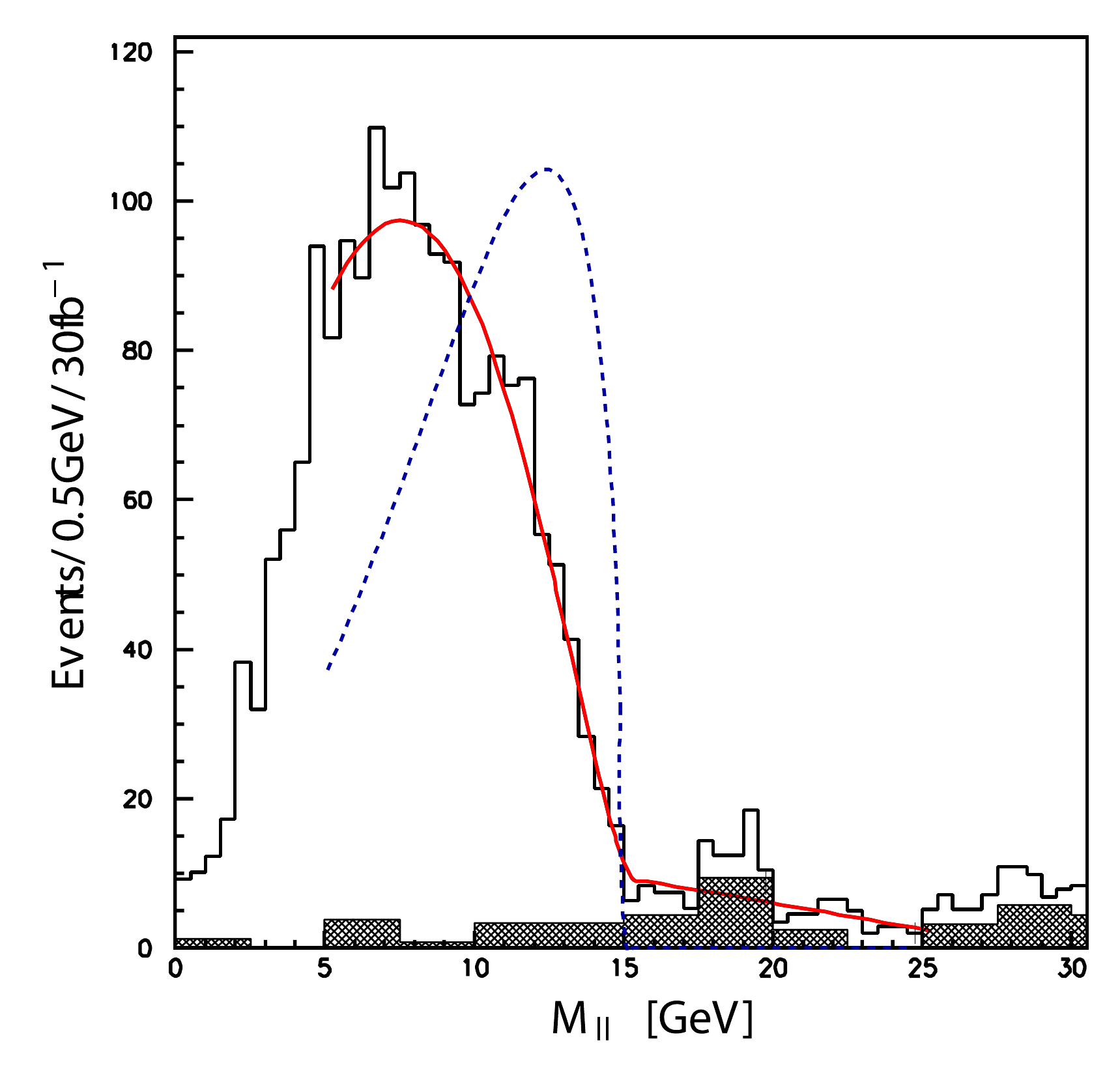}\quad
\caption{Distribution of the dilepton invariant mass in two supersymmetry
  models with 3-body neutralino decays:  (a.) a model with gaugino-like 
  neutralinos~\cite{Hinch}, (b.) a model with Higgsino-like 
neutralinos~\cite{KitanoNomura}. In the second figure, the dashed curve
indicates the $m(\ell^+\ell^-)$ spectrum expected for gaugino-like neutralinos
with the same mass splitting.}
\label{fig:Hinchtwo}
\end{center}
\end{figure}

\begin{figure}
\begin{center}
\includegraphics[height=3.0in]{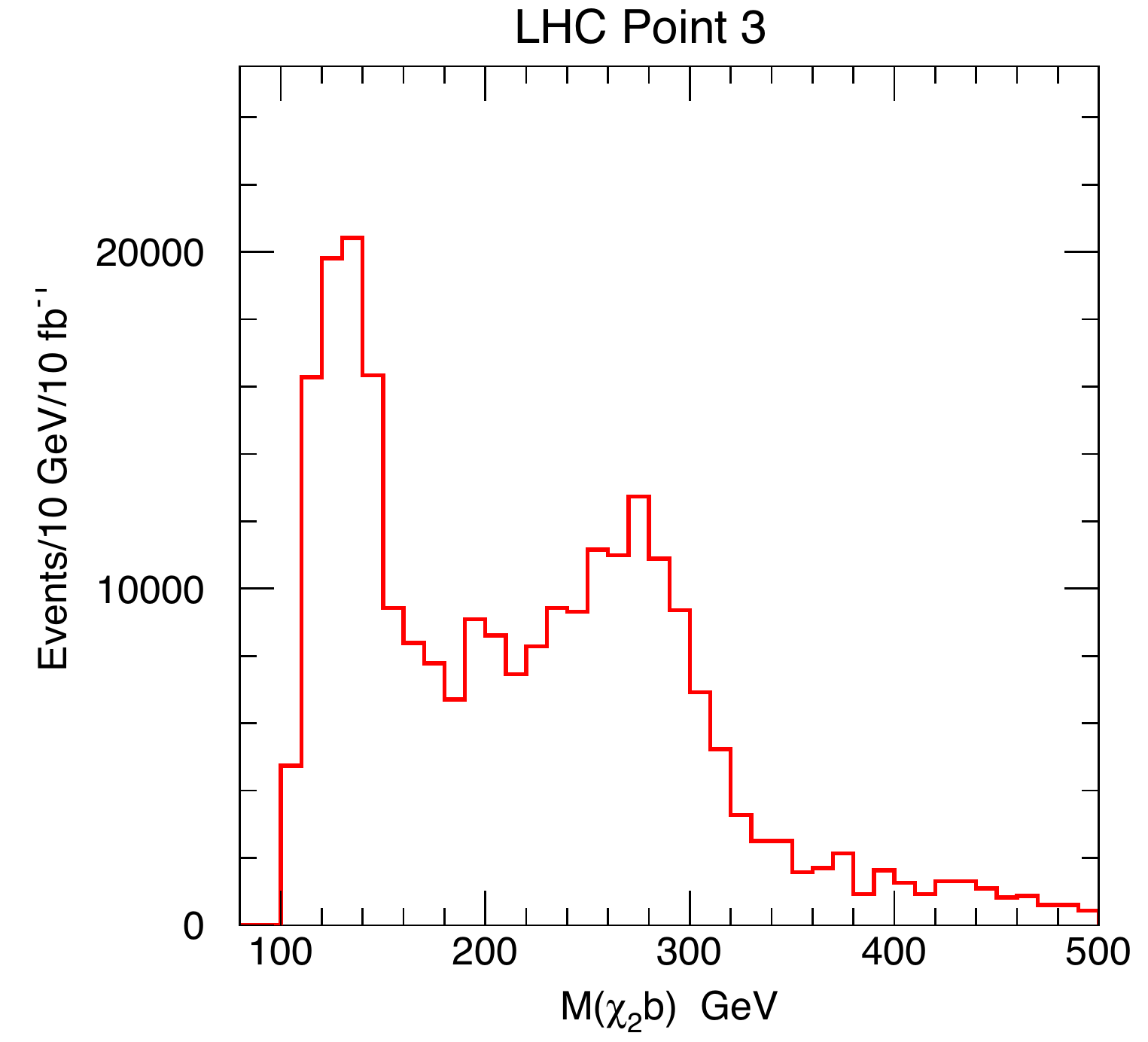}
\caption{Reconstruction of a squark  in the model of 
 Fig.~\ref{fig:Hinchtwo}(a) by combining a dilepton pair at the endpoint
of the $m(\ell^+\ell^-)$ distribution, the $\s N_1^0$ in the same frame with 
mass determined from kinematics, and a $b$-tagged quark jet.}
\label{fig:Hinchb}
\end{center}
\end{figure}

In the three-body case, the endpoint of the dilepton mass distribution is 
exactly
\beq
                 m(\s N_2) - m(\s N_1) \ ,
\eeq{endpointthree}
so the observable mass distribution gives a precise measurement of this 
SUSY mass difference.  The shape of the spectrum has more information.
For example, for heavy slepton masses, the shape is distinctly different for 
gaugino-like or Higgsino-like neutralinos.  Figure~\ref{fig:Hinchtwo}(a) shows
the dilepton mass distribution for an mSUGRA parameter set for which the 
lightest two neutralinos are gaugino-like~\cite{Hinch}.
  Figure~\ref{fig:Hinchtwo}(b) 
shows this distribution for a parameter set in which the two lightest
neutralinos are Higgsino-like~\cite{KitanoNomura}.

At the endpoint, the dilepton mass is maximal, and this requires that both 
the dilepton pair and the 
$N_1$ are at rest in the frame of the $N_2$.  By measuring the four-vectors 
of the leptons, we would then know the $N_1$ and $N_2$ four-vectors, up to 
knowledge of the $N_1$ mass.  It is possible to obtain this mass approximately
from 
other measurements, for example, from the kinematics of $\s{\bar q}$ 
decays directly to $N_1$.  With this information, we could determine the 
$N_2$ four-vector.  Now the problem of missing momentum is solved.  By 
adding observed jets to the $N_2$ four-vector, it is possible to find
squarks as resonances~\cite{Hinch}.   Figure~\ref{fig:Hinchb}
shows the result of such an analysis for the SUSY parameter set of 
Fig.~\ref{fig:Hinchtwo}.  The peak just below 300 GeV is a reconstructed
$\s{b}$ squark.

\begin{figure}
\begin{center}
\includegraphics[height=3.0in]{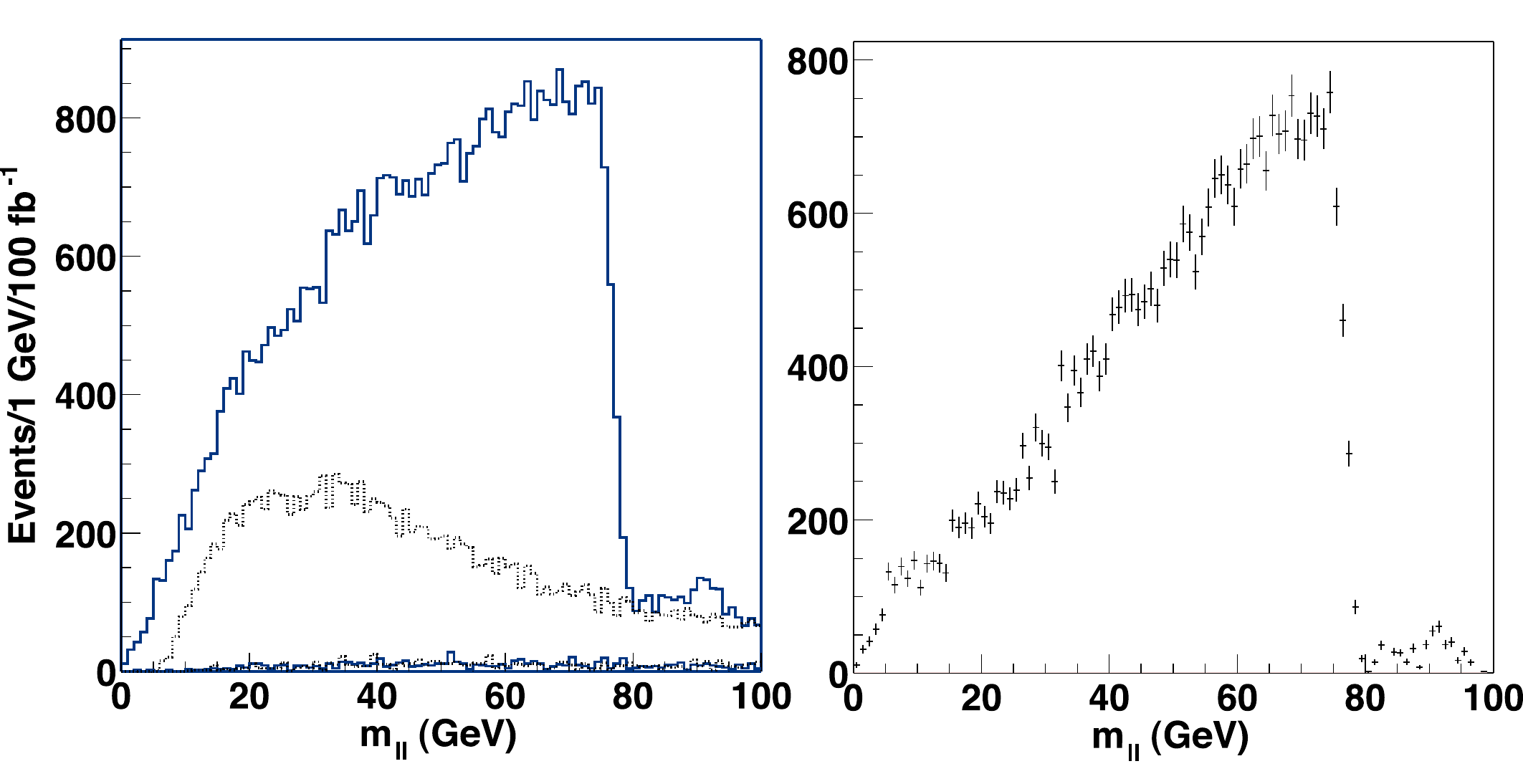}
\caption{Dilepton mass distribution in a model with two-body $\s N_2$
decays, from \cite{Weiglein}.  The left-hand plot shows the dilepton 
mass distributions for opposite-sign same-flavor dileptons (solid) and
for opposite-sign opposite-flavor dileptons (dashed).  The lower histograms
give the estimates of the Standard Model background.  The right-hand plot
shows the difference of the two distributions.}
\label{fig:twobodyll}
\end{center}
\end{figure}

\begin{figure}
\begin{center}
\includegraphics[height=3.0in]{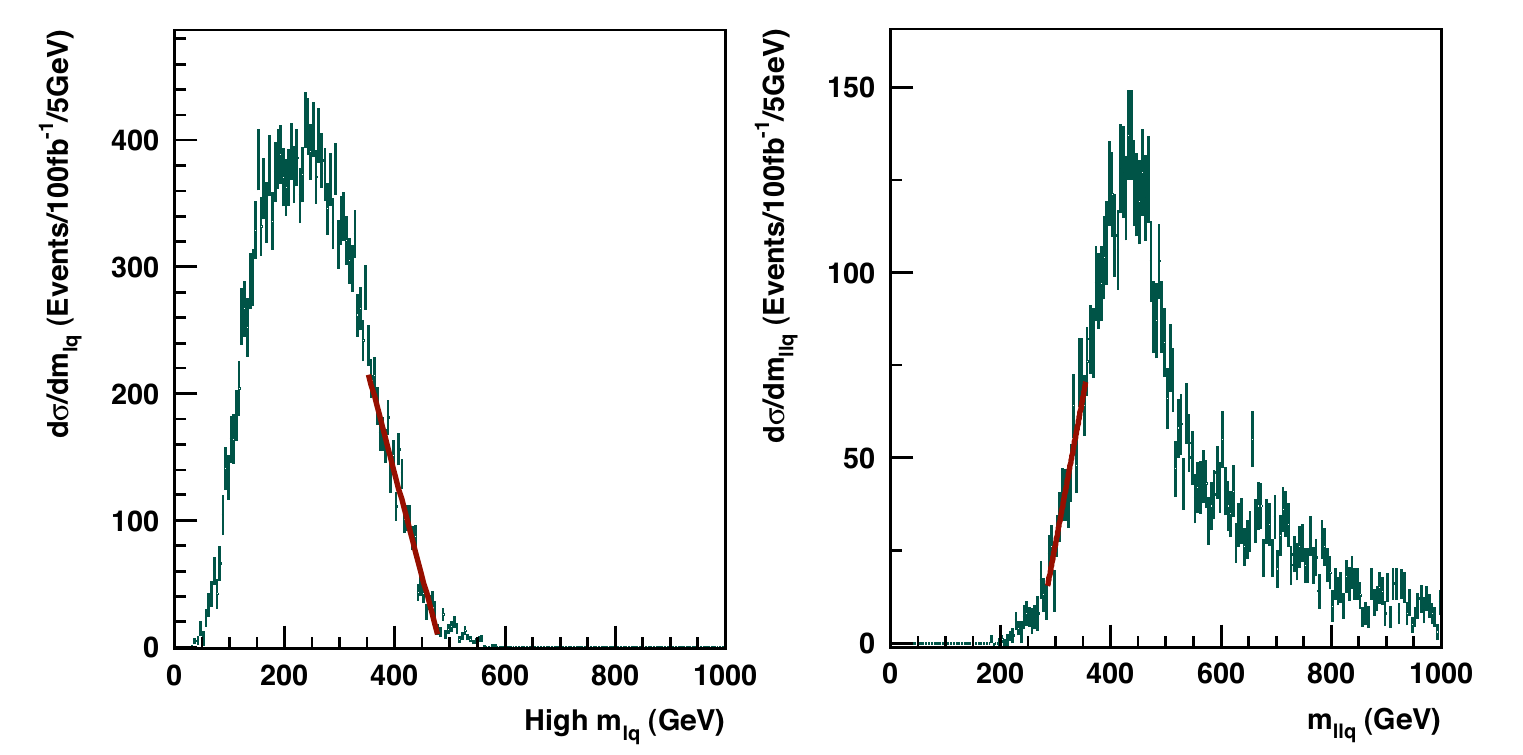}
\caption{Distributions of mass combinations of leptons and high-$p_T$ jets
       showing kinematic endpoints in the analysis of~\cite{Allanach}: (a.)
         the higher  $m(q\ell)$ combination; (b.) the $m(q\ell^+\ell^-)$
        distribution.}
\label{fig:qelldists}
\end{center}
\end{figure}

The two-body case of $\s N_2$ decay is even nicer.  In this case, we can see
from the right-hand figure in 
Fig.~\ref{fig:Dalitz}(b) that the endpoint of the dilepton mass 
distribution is not located at the mass difference \leqn{endpointthree}
but instead at the smaller value 
\beq
     m(\ell^+\ell^-) = m(\s N_2) \sqrt{1 - {m^2(\s \ell)\over m^2(\s N_2)}}
 \sqrt{1 - {m^2(\s N_1)\over m^2(\s \ell)}} \ .
\eeq{endpointtwo}
Figure~\ref{fig:twobodyll} shows an example of the dilepton spectrum from 
a SUSY parameter point in this region~\cite{Weiglein}  The decay 
$\s q \to q N_2$ is also a two-body decay, and there are similar kinematic
relations for the upper and lower endpoints of the $(q\ell)$ and $(q\ell\ell)$
invariant mass distributions.  These endpoints are likely to be visible
in the collider data. Figure~\ref{fig:qelldists} shows two jet-lepton
mass distributions from a similar analysis presented in~\cite{Allanach}.
  In that analysis, it was
possible to identify five well-measured kinematic endpoints, from which it 
was possible to solve (in an overdetermined way) for the 
four masses  $m(N_1)$, $m(\s \ell)$, $m(N_2)$, $m(\s q)$.

\begin{figure}
\begin{center}
\includegraphics[height=3.0in]{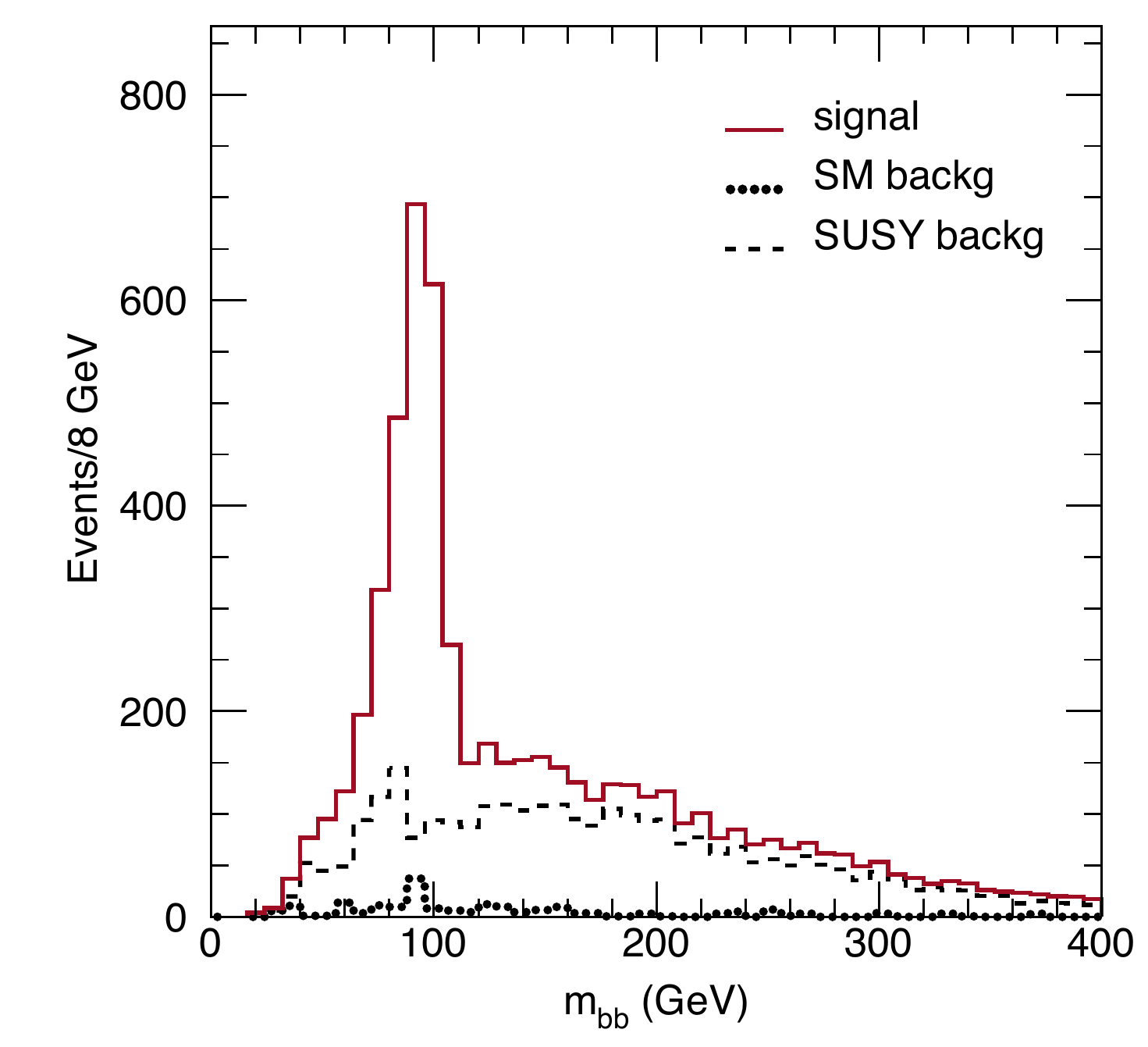}
\caption{The dijet mass distribution for 2 $b$-tagged jets at a point 
     in the SUSY parameter space where the decay $\s N_2^0 \to h^0 \s N_1^0$ is
       dominant, from~\cite{ATLAS}.}
\label{fig:HiggsfromSUSY}
\end{center}
\end{figure}

There is one more case of an $\s N_2\to \s N_1$ decay that should be mentioned.
If two-body decays of $\s N_2$ to sleptons are not kinematically allowed but
the decay to $\s N_1 h^0$ is permitted, this decay to a Higgs boson will be
the dominant $\s N_2$ decay.  In this case, supersymmetry can provide a copious
source of Higgs bosons.  Figure~\ref{fig:HiggsfromSUSY} shows an 
analysis of a SUSY model in this parameter region~\cite{ATLAS}. 
 Events with multijets
and missing transverse energy are selected.  In this sample, the mass
distribution of two $b$-quark-tagged jets is shown.  The signature of 
SUSY selects a sample of events in which the Higgs boson is visible in 
its dominant decay to $b\bar b$. 

There is much more to say about the measurement of SUSY parameters at the 
LHC.  Some more sophisticated sets of variables are introduced and applied
in \cite{Allanach,Lester}.  The question of measuring 
the spins of superparticles is discussed in 
\cite{Barr,Nojiri,Webber,Plehn}.
And, we have not touched on alternative possibilities for the realization 
of SUSY, with R-parity violation or charged superparticles that are observed
in the LHC experiments as stable particles.  A broader overview of 
SUSY phenomenology at the LHC can be found in the references cited at the 
beginning of this section.

\section{Electroweak Symmetry Breaking and Dark Matter in the MSSM}

\subsection{Electroweak Symmetry Breaking in the MSSM}

In Section 1.2, I motivated the introduction of SUSY with the claim that 
SUSY could give an explanation of electroweak symmetry breaking, and for 
the presence of weakly interacting dark matter in the universe.  Now that 
we have a detailed understanding of the structure of the 
MSSM, it is time to come back and discuss these issues.

To present the mechanism of electroweak symmetry breaking in the MSSM, I 
need to add a term to  one of the equations that I derived in Section~4.3.  
In \leqn{RGforMf}, I presented the RG equation for the soft SUSY breaking 
scalar mass parameters, including renormalization effects from gauge 
interactions.  I remarked that the contributions to this equation from 
Higgs Yukawa couplings are small for the scalars of the first and second
generations.  However, for the scalars of the third generation, these
corrections can plan an important role.

\begin{figure}
\begin{center}
\includegraphics[height=1.0in]{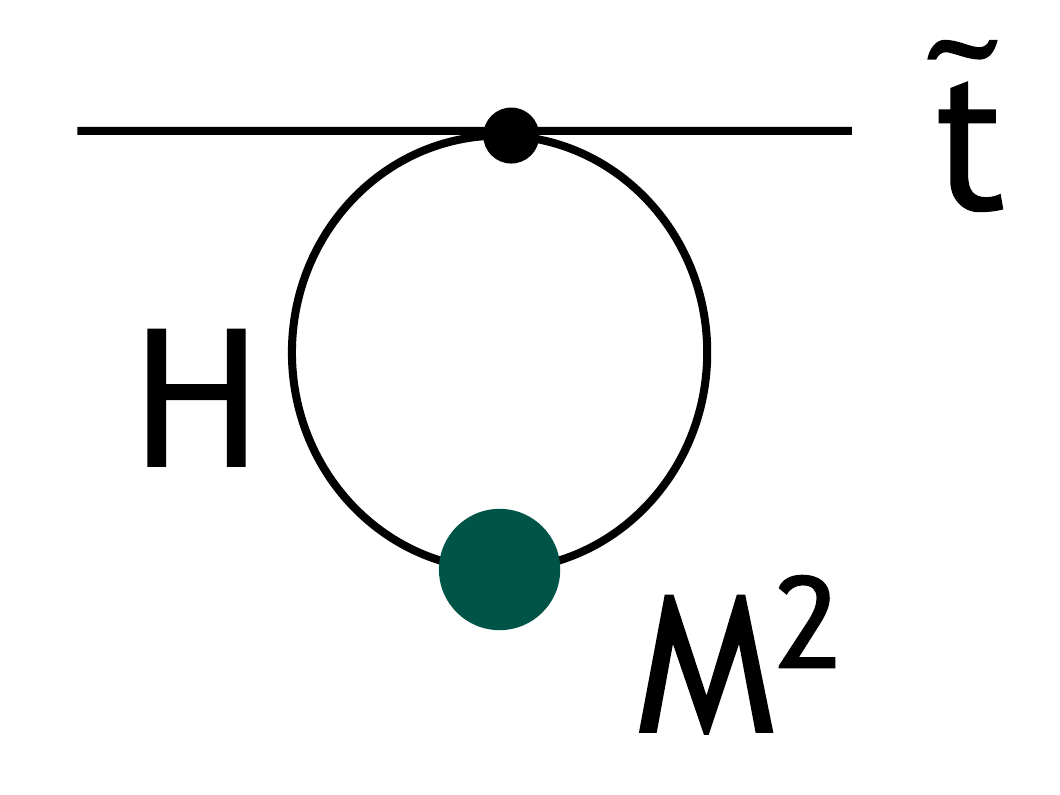}
\caption{Diagram contributing a term to the renormalization group
 equation for the soft mass parameter of $\s t$ proportional to the
 soft mass parameter for $H_u$.}
\label{fig:ytdiagram}
\end{center}
\end{figure}

The $F$-term interaction 
\beq
       \L = - \left| y_t H_u\cdot \s t\right|^2
\eeq{Lforyt}
leads to a contribution to the RG equations for $M_{t}$, the mass 
parameter of $\s t$, proportional to $M_{Hu}^2$, from the diagram shown
in Fig.~\ref{fig:ytdiagram}.  The value of the diagram is
\beq
       -i y_t^2 \int {d^4k\over (2\pi)^4} {i\over k^2} (- i M_{Hu}^2)
          {i\over k^2}  =   {i\over (4\pi)^2} y_t^2 M^2_{Hu} \log \Lambda^2\ .
\eeq{ytdiagramvalue}
A scalar self-energy diagram  is interpreted as $-i\delta m^2$, so this is
a {\it negative} contribution to $M_t^2$.  Each of the scalar fields
$(H_u, \s t, \s{\bar t})$ gives a similar contribution that renomalizes the
soft mass parameter of each of the others.  For each correction, there is a
counting factor from the number of color or $SU(2)$ degrees of freedom that
run around the loop.  There is also a correction to each of the scalar 
masses from the top quark $A$ term.  We must also remember that all of these
terms add to the positive mass correction from the gaugino loops in 
Fig.~\ref{fig:MfRG}, of which the gluino loop correction is the most important.

Taking all of these effects into account, we find for the RG equations of
the soft mass parameters of $H_u$, $ t$, and ${\bar t}$
\beqa
{d M_t^2\over d\log Q} & = & {2\over (4\pi)^2} \cdot 1 \cdot y_t^2
      [M_t^2 + M_{\bar t}^2 + M_{Hu}^2 + A_t^2] - {8\over 3\pi} \alpha_3
           m_3^2 + \cdots \CR
{d M_{\bar t}^2\over d\log Q} & = & {2\over (4\pi)^2} \cdot 2 \cdot y_t^2
      [M_t^2 + M_{\bar t}^2 + M_{Hu}^2 + A_t^2] - {8\over 3\pi} \alpha_3
           m_3^2 + \cdots \CR
{d M_{Hu}^2\over d\log Q} & = & {2\over (4\pi)^2} \cdot 3 \cdot y_t^2
      [M_t^2 + M_{\bar t}^2 + M_{Hu}^2 + A_t^2]
           m_3^2 + \cdots 
\eeqa{HttbarRGEs}
The structure is very interesting.  The three scalar fields  $H_u$, 
$\s t$, and $\s{\bar t}$ all receive negative corrections to their 
mass terms as these equations are integrated in the direction of 
decreasing $\log Q$.  If any of these mass terms were to become negative,  the 
corresponding field would have an instability to develop a vacuum expectation
value, and the 
symmetry of the MSSM would be spontaneously broken. The symmetry-breaking
we want is that associated with $\VEV{H_u} \neq 0$.  However, it seems equally 
possible that we could generate $\langle \s{\bar t}\rangle
\neq 0$, which would break 
color $SU(3)$, or $\langle\s t \rangle \neq 0$, which would break both $SU(2)$ 
and $SU(3)$.

If the three mass parameters have similar values at a high mass scale, 
they race toward negative values according to \leqn{HttbarRGEs}.  But
$H_u$ wins the race, and so the theory predicts the symmetry breaking
pattern that is the one observed.  In this way, the MSSM leads naturally to 
electroweak symmetry breaking and realizes the idea that electroweak 
symmetry breaking is connected to the large value of the top quark-Higgs
coupling.

\subsection{Higgs Boson Masses in the MSSM}

Once we expect that $M_u^2 < 0$ at the weak scale, we can work out the details
of the Higgs boson spectrum.  First, we should write the potential for the 
Higgs fields $H_u$, $H_d$.  As in the discussion of Sections~4.1 and 4.2, 
a number
of terms need to be collected from the various pieces of the Lagrangian.
The $F$ terms contriubute
\beq
    V_F = \mu^2 (H_u^{0 *} H_u^0 + H_d^{0*} H_d^0)
\eeq{VFtoHiggs}
The $D$ terms contribute 
\beq
    V_D = {g^2 + g^{\prime 2}\over 8} (H_u^{0 *} H_u^0 - H_d^{0 *} H_d^0)^2
\eeq{VDtoHiggs}
The soft SUSY breaking terms contribute
\beq
   V_{soft} = M^2_{Hu} H_u^{0 *} H_u^0 + M^2_{Hd} H_d^{0 *} H_d^0 
          - (B\mu H^0_u H^0_d + h.c.)
\eeq{VsofttoHiggs}
The sum of these terms gives the complete tree-level Higgs potential.
Differentiating this potential with respect to 
$H^0_u$ and $H^0_d$, we obtain the equations 
that determine the Higgs field vacuum 
expectation values. If we write these equations with the 
parametrization of the vacuum expectation values given in \leqn{Higgsvevs}, 
we find
\beqa
     \mu^2 + M^2_{Hu} &=& B\mu \cot \beta + \half \mz^2 \cos 2\beta\CR
     \mu^2 + M^2_{Hd} &=& B\mu \tan \beta - \half \mz^2 \cos 2\beta \  ,
\eeqa{mubetavequs}
where $\mz^2 =  (g^2 + g^{\prime 2}) v^2/4$.  This system of equations 
can be solved for $\mu$ to give
\beq
      \mu^2 = {M_{Hd}^2 - \tan^2\beta M_{Hu}^2\over \tan^2\beta - 1} - 
         \half \mz^2
\eeq{findmu}
This is, for example, the way that we would determine $\mu$ in the 
mSUGRA parameter space described in Section~5.2.

It is interesting to turn this equation around and write it as an equation
for $\mz$ in terms of the SUSY parameters,
\beq 
      \mz^2 = 2\,{M_{Hd}^2 - \tan^2\beta M_{Hu}^2\over \tan^2\beta - 1} - 
         2 \mu^2 \ .
\eeq{findmz}
From this equation, a small value of $\mz$ would require a cancellation 
between the Higgs soft mass parameters and $\mu$. The parameter $\mu$ 
sets the mass scale of the Higgsinos, and the Higgs soft mass parameters
might be related to other masses of the SUSY scalar particles.  Thus, if 
the masses of the charginos and neutralinos and, perhaps also, the sleptons
are not close to $\mz$, that disparity must be associated with an
apparently unnatural cancellation between different SUSY parameters.

If we prohibit a delicate cancellation in \leqn{findmz}, we put an 
upper bound on the SUSY partner masses.  To avoid cancellations in more 
than two decimal places, $\mu$ must be less than 700 GeV.  Similarly,
we find bounds on the Higgs soft masses, and on the parameters that 
contribute to these masses through the RG equation.  This consideration
turns out to give a constraint on the gluino mass, $m_3 < 800$ GeV.
Assuming gaugino universality, this becomes a condition $m_2 < 250$ GeV
that restricts the chargino and neutralino masses.  A variety of similar
naturalness arguments that constrain the SUSY scale can be found in 
\cite{ENV,BG,FMM}.  Though the logic is that of an estimate
rather than a rigorous bound, this analysis strongly supports the idea
that SUSY partners should be light enough to be discovered at the LHC 
and at the ILC.

Once we have the Higgs potential and the conditions for the Higgs vacuum
expectation values, we can work out the masses of the Higgs bosons by 
expanding the potential around its minimum.  A first step is to identify the 
combinations of Higgs fields that correspond to physical Higgs bosons.
Look first at the charged 
Higgs bosons.  There are two charged Higgs fields in the multiplets 
$H_u$, $H_d$.  One linear combination of these fields is the Goldstone
boson that is eaten by the $W$ boson as it obtains mass through the 
Higgs mechanism.  The orthogonal linear combination is a physical charged
scalar field.  If we decompose
\beqa
      H^+_u &=& \cos\beta H^+ + \sin\beta G^+ \CR 
      H^-_d &=& \sin\beta H^- + \sin\beta G^- 
\eeqa{Hplusdecomp}
where $H^- = (H^+)^*$, $G^- = (G^+)^*$, and $\beta$ is precisely the 
mixing angle in \leqn{Higgsvevs}, it can be seen that $G^\pm$ are the 
Goldstone bosons and $H^\pm$ are the physical scalar states.

A similar analysis applies to the neutral components of $H^0_u$ and 
$H^0_d$.  These are complex-valued fields.  It is appropriate to
 decomposed them as 
\beqa 
  H^0_u &=& {1\over \sqrt{2}} ( v \sin\beta + \sin\alpha H^0 + \cos \alpha
  h^0 + i \cos\beta A^0 +  i\sin\beta G^0) \CR
  H^0_d &=& {1\over \sqrt{2}} ( v \cos\beta + \cos\alpha H^0 - \sin\alpha
  h^0 + i \sin\beta A^0 -  i\cos\beta G^0) 
\eeqa{Hzerodecomp}
The components $H^0$, $h^0$ are even under CP; the fields $A^0$, $G^0$ are 
odd under CP.  The componet $G^0$ is the Goldstone boson eaten 
by the $Z^0$.  The other three fields create physical scalar particles.

Having identified these fields, we can compute their masses.  The formulae
for the Higgs masses take an especially simple form when they are 
expressed in terms of the mass of the $A^0$.  For the charged Higgs boson
\beq
          m^2_{H+} = m_A^2 + \mw^2 \ .
\eeq{mHplus}
For the CP-even scalars, one finds a mass matrix
\beq
 \pmatrix{   m_A^2 \sin^2\beta + \mz^2 \cos^2\beta & 
       - (m_A^2 + \mz^2) \sin\beta\cos\beta \cr
       - (m_A^2 + \mz^2) \sin\beta\cos\beta &
  m_A^2 \cos^2\beta + \mz^2 \sin^2\beta \cr }
\eeq{mHzeros}
The physical scalar masses $m_h^2$ and $m_H^2$ are the eigenvalues of this 
matrix, defined in such a way that $m_h^2 < m_H^2$. 
 The angle $\alpha$ in \leqn{Hzerodecomp} is the mixing angle 
that defines these eigenstates.  

Taking the trace of \leqn{mHzeros}, we
find the relation
\beq
        m_h^2 + m_H^2  = m_A^2 + \mz^2 \ .
\eeq{mhHrel}
We can also obtain an upper bound on the lighter Higgs mass $m_h^2$ by 
taking the matrix element of \leqn{mHzeros} in the state $(\cos\beta, 
\sin\beta)$.  The bound is a very strong one:
\beq
    m_h^2 \leq  \mz^2\cos^2\beta \ < \mz^2 \ .
\eeq{mhbound}
This seems inconsistent with lower bounds on the Higgs boson mass from LEP 2,
which exclude
$m_h < 114$ GeV for the Standard Model Higgs and for most 
scenarios of SUSY Higgs bosons~\cite{Tully}.\footnote{Some exceptional Higgs
decay schemes that escape these bounds are considered 
in~\cite{Dermisek,WeinerHiggs}.}  However, the 
one-loop corrections to the tree-level result \leqn{mHzeros} give a 
significant positive correction
\beq
   \delta m_h^2 = {3\over \pi} {\mt^4\over \mw^2} \sin^4\beta \, \log
     {m_{\s t} m_{\s{\bar t}}\over \mt^2} \ .
\eeq{correctmh}
This correction can move the mass of the $h^0$ up to about 130 GeV. 
The detailed summary of the radiative corrections to the $h^0$ mass
in the MSSM is presented in~\cite{DeGrassi}. A 
very clear and useful accounting of the major corrections  
can be found in~\cite{HaberHoang}.    

It is possible to raise the mass of the $h^0$  
by going outside the MSSM and adding additional
$SU(2)$ singlet superfields to the model.  However, this strategy is
limited by a  general constraint coming from grand unification.
The requirement that the Higgs couplings do not become strong up to the
grand unification scale limit the mass of the Higgs to about 
200~GeV~\cite{Petronzio}.  It is
possible to raise the mass of the Higgs further only by enlarging the Standard
Model gauge group or adding new thresholds that affect 
unification~\cite{Batra,fatHiggs}.

In the MSSM, we can easily have the situation in which $m_A \gg m_h$.  In
this limit, the couplings of the $h^0$ are very close to those of
the Standard Model Higgs boson, and the $H^0$, $A^0$, and $H^\pm$ are 
almost degenerate.  If $\tan\beta \gg 1$, the  heavy neutral Higgs bosons
decay dominantly to $b\bar b$ and $\tau^+\tau^-$. 

Much more about the phenomenology of Higgs bosons in supersymmetry can be
found in~\cite{HHunters,HaberCarena}.

\subsection{WIMP Model of Dark Matter}

Now we turn to the second problem highlighted in the Introduction, the problem
of dark matter in the universe.  It has been known from many astrophysical 
measurements that the universe contains enormous amounts of invisible, 
weakly interacting matter.  For an excellent review of the classic 
astrophysical evidence for this dark matter, see~\cite{Trimble}.  

In the 
past few years, measurements of the cosmic microwave background have 
given a new source of evidence for dark matter.  Since this 
data comes from an era in the early 
universe before the formation of any structure, it argues strongly that 
the invisible matter is not made of rocks or brown dwarfs but is actually 
a new, very weakly interacting form of matter.  These measurements also
determine quite accurately the overall amount of conventional and dark
matter in the universe.  Let  $\rho_b$, $\rho_N$, and $\rho_\Lambda$ be
the large-scale energy densities of the universe from baryons, dark matter,
and the energy of the vacuum.  The data from the microwave background 
tells us that $\rho_b + \rho_N  + \rho_\Lambda = \rho_c$, the `closure 
density' corresponding in general relativity to a flat universe, to about
1\% accuracy.  If $\Omega_i = \rho_i/\rho_c$, the most recent data from 
the WMAP experiment and other sources gives~\cite{WMAP,LLiddle}
\beq
   \Omega_b = 0.042\pm 0.003 \quad \Omega_N = 0.20 \pm 0.02 \quad
         \Omega_\Lambda = 0.74 \pm 0.02 \ . 
\eeq{ucomponents}
These results present a double mystery.  We do not know what particle 
the dark matter is made of, and we do not have any theory that explains
the observed magnitude of the vacuum energy or `dark energy'.

I believe that supersymmetry will eventually play an essential role in 
solving the problem of dark energy.  In ordinary quantum field theory,
the value of the vacuum energy is quartically divergent, so the problem 
of computing the vacuum energy is not even well-posed.  In supersymmetry,
there is at least a well-defined zero of the energy associated with 
exact supersymmetry, which implies $\bra{0} H \ket{0} = 0$.  Unfortunately,
in most of today's models of supersymmetry, the vacuum energy is set by the
SUSY breaking scale.  This gives 
$\Lambda \sim (10^{11}$~GeV$)^4$, about
80 orders of magnitude larger than the observed value of the vacuum energy.
From this starting point, $\Lambda$ must be fine-tuned to the scale of 
eV$^4$.   This is an important problem that needs new insights which, 
however, I will not provide here.

On the other hand, supersymmetry offers a very definite solution to the 
problem of the origin of dark matter.  We have already noted in Section~3.4
that it is straightforward to arrange that the lightest supersymmetric 
particle can be absolutely stable.  If this particle were produced in the 
early universe, some density of this type of matter should still be present.
In most, but not all, regions of parameter space, the lightest supersymmetric
particle is neutral.  Candidates 
include the lightest neutralino, the lightest sneutrino, and the
gravitino.  In the remainder of these lectures, I will concentrate on the 
case in which the lightest neutralino is the dark matter particle.
For a discussion of the other candidates, see~\cite{generalSUSYDM}.

To begin our discussion, I would like to estimate the cosmic density of 
dark matter in a more general context.  Let me make the following minimal
assumptions about the nature of dark matter, that the dark matter particle
is stable, neutral, and weakly interacting.  To these properties, I would
like to add one more, that dark matter particles can be created in 
pairs at sufficiently high temperature, and that, at some time in the 
early universe, dark matter particles were in thermal equilibrium.  I will
refer to a particle satisfying these assumptions as a `weakly interacting
massive particle' or WIMP.  The assumption of thermal equilibrium is a strong
one that is not satisfied even in many models of supersymmetric dark matter.
For some exceptions, see~\cite{RandallMoroi,KitanoNomuraDM}.  
However,
let us see what implications follow from these assumptions.

The assumption that WIMPs were once in thermal equilibrium provides a 
definite initial condition from which to compute the current density 
of dark matter.  In thermal equilibrium at temperture $T$, 
we have for the number 
density of dark matter particles
\beq
      n_{eq} = {g\over (2\pi)^{3/2}} (mT)^{3/2} e^{-m/T} \ .
\eeq{thermaleq}
where $g$ is the number of spin degrees of freedom of the massive particle.
As the universe expands, the temperature of the universe deccreases and
 the rate of WIMP pair production 
becomes very small.  But the rate of dark matter pair annihilation also 
becomes small as the WIMPs separate from one another. 

The expansion of 
the universe is governed by the Hubble constant $H = \dot a/a$,where 
$a$ is the scale factor.  Einstein's equations imply that 
\beq
      H^2 = {8\pi\over 3} {\rho\over m^2_\Pl} \ .  
\eeq{Hvalue}
In a radiation-dominated universe 
where $g_*$ is the number of relativistic degrees of freedom, 
$\rho = \pi^2 g_* T^4/30$.  Then $H$ is proportional to $T^2$.
In a radiation-dominated universe, the temperature red-shifts as the universe
expands, so that $T \sim a^{-1}$.  Combining this relation
with the equation $H = \dot a/a \sim T^2$, we find
$t\sim T^{-2} \sim a^2$, that is, $a \sim t^{1/2}$ or 
 $\dot a/a = 1/2t$. Setting this expression equal to the explict form of $H$
in \leqn{Hvalue}, we find a detailed formula for the 
time since the start of the radiation-dominated era for cooling to a 
temperature $T$,
\beq
       t =  \left({16 \pi^3 g_*\over 45}\right)^{-1/2} {m_\Pl\over T^2} \ .
\eeq{timeval}

 The evolution 
of the WIMP density is described by the Boltzmann equation
\beq
    {d n\over dt} = - 3 H n -  \VEV{\sigma v} (n^2 - n^2_{eq}) \ ,
\eeq{Boltzmann}
where $H$ is the Hubble constant, $\sigma$ is the $\s N\s N$ 
annihilation cross 
section---which appears thermally averaged with the relative velocity of
colliding WIMPs---and $n_{eq}$ is the equilibrium WIMP density 
\leqn{thermaleq}. 
Assume, just for the sake of argument, that the temperature $T$ is of
the order of 100~GeV.  At this temperature, the Hubble constant has
the magnitude $H \sim 10^{-17} T$, so the expansion of the universe is very
slow on the scale of typical elementary particle reactions. However, 
when $T$ becomes less than the WIMP mass $m$, the WIMP density is 
exponentially suppressed and so the collision term in the Boltzmann equation
is also very small.  These two terms are of the same size at the
{\it freezeout} temperature
$T_F$ satisfying
\beq
     e^{-m/T_F} \sim  {1\over m_\Pl m \VEV{\sigma v}} \ .
\eeq{fcondition}
At temperatures below $T_F$, we may neglect the production of WIMPs in 
particle collisions.  The WIMP density is then determined by the expansion
of the universe and the residual rate of WIMP pair annihilation.
Maybe it is more appropriate to think of $T_F$ as the temperature at which
a WIMP density is frozen {\it in}. To determine the freezeout temperature, 
we take the logarithm of the right-hand side of \leqn{fcondition}.
The result depends only on the order of magnitude of the annihilation 
cross section.  For any interaction of electroweak strength, 
\beq
           \xi_F = T_F/m \sim 1/25 \ . 
\eeq{xiFvalue}

This physical picture suggests a way to estimate the cosmic density of 
WIMP dark matter.  We can take as our initial condition the thermal
density of dark matter at freezeout.  We then integrate the Boltzmann
equation, ignoring the term proportional to $n_{eq}^2$ associated with 
the production of WIMP pairs~\cite{TurnerScherrer}.

In analyzing the Boltzmann equation, it is useful normalize the particle
density $n$ of dark matter to the density of entropy $s$.  Since the 
universe expands very slowly, this expansion is very close to adiabatic.
Then entropy is conserved,
\beq
            {d s\over d t} =  - 3 H s  \ .
\eeq{entropyeq}
In a radiation-dominated universe, $s = 2\pi^2 g_* T^3/45$.
Now define
\beq
         Y = {n\over s} \ ,\qquad   \xi = {T\over m} \ ,
\eeq{Yxidef}
the latter as in \leqn{xiFvalue}.  Using the expression \leqn{timeval}, we 
can convert the evolution in time to an evolution in temperature or in $\xi$.
 Applying these changes of variables
and dropping the $n_{eq}^2$ term, the Boltmann equation \leqn{Boltzmann}
rearranges to the form
\beq
     {d Y \over d\xi} =  C \VEV{\sigma v} Y^2 \ , 
\eeq{Yxieq}
where 
\beq
       C = \left({\pi g_*\over 45}\right)^{1/2} m m_\Pl \ .
\eeq{Cvalue}
Let $Y_F$ be the value of $Y$  at $\xi = \xi_F$.  If we assume that
$\VEV{\sigma v}$ is approximately constant, since we are at temperatures
close to threshold, it is straightforward
to integrate this equation to $\xi = 0$, corresponding to late times.
\beq
             Y^{-1} = Y_F^{-1}  + C \xi_F \VEV{\sigma v} \ .
\eeq{Yintegral}
The second term typically dominates the first.  Then we can put back the 
value of $C$ in \leqn{Cvalue} and write the final answer in terms of the 
ratio of the mass density of dark matter to the closure density
$\Omega_N = n m_N/\rho_c$.  In this way, we find
\beq
    \Omega_N = {s_0\over \rho_c} \left( {45\over \pi g_*}\right)^{1/2}
         {1\over \xi_F m_\Pl } {1\over \VEV{\sigma v} } \ ,
\eeq{TSvalue}
where $s_0$ is the current entropy density of the universe.
Turner and Scherrer observed that this formula gives a value of $\Omega_N$
that is usually within 10\% of the result from exact integration of 
 the Boltzmann equation~\cite{TurnerScherrer}.  If $\VEV{\ sigma v}$ has a 
significant dependence on temperature, the derivation is still correct
with the replacement
\beq
         \xi \VEV{\sigma v} \to  \int_0^{\xi_f} d\xi \VEV{\sigma v}(\xi)
\eeq{correctOmega}
in the denominator of the last term in \leqn{TSvalue}.

This is a remarkable relation.   Almost every factor in this relation is
known from astrophysical measurements.  The left-hand side is given 
by \leqn{ucomponents}. On the right-hand side, the entropy density of the
universe is dominated by the entropy of the microwave background photons and
can be computed from the microwave background temperature.  The closure density
is known from the measurement of the Hubble constant and the 
observation that the
universe is flat.  The parameters $g_*$ and $\xi_F$ are relatively insensitive
to the strength of the annihilation cross section, with values $g_*\sim 100$, 
$\xi_F \sim 1/25$.   The mass of the WIMP does not appear
explicitly in \leqn{TSvalue}.  We can then solve for $\VEV{\sigma v}$.
The result is
\beq
         \VEV{\sigma v } = 1 \ \mbox{pb} \ .
\eeq{sigmavvalue}
This is the value of a typical electroweak cross section at energies of a 
few hundred GeV.
If we convert this value to a mass $M$ of an exchanged particle using the 
formula
\beq
            \VEV{\sigma v} = {\pi \alpha^2 \over 8 M^2 } \ ,
\eeq{sigmatoM}
the value \leqn{sigmavvalue} corresponds to  $M = 100$~GeV.

I consider this a truly remarkable result.  From a purely astrophysical 
argument, relying on quite weak and general 
assumptions, we arrive at the conclusion 
that there must be new particles at the hundred GeV energy scale.  It is 
probably not a concidence that this argument leads us back to the 
mass scale of electroweak symmetry breaking.

In our study of supersymmetry, we have found an argument from the physics
of electroweak symmetry breaking that predicts the existence of dark matter.
As I discussed at the beginning of these lectures, models that explain
 electroweak symmetry breaking are complex.  They typically involve many 
new particles.  It is easily arranged that the lightest of the new particles
is neutral.  In supersymmetry, there is a reason why the new particles 
are likely to carry a conserved quantum number \leqn{Rparitydef}.  Other
models of electroweak symmetry breaking, such as the extra dimensional 
and little Higgs models discussed in Section 1.2, have their own reasons
to have a complex particle spectrum and discrete symmetries.  Then these 
models lead
in their own ways to WIMPs at the hundred GeV mass scale.

A slight extension of this argument adds more interest.  In supersymmetry,
the sector of new particles includes particles with QCD color.  Since the
top quark probably plays an essential role in the mechanism of electroweak
symmetry breaking, it is very likely that, in any model, some of the new
particles will carry color.  If these particles have masses below 1~TeV,
they have large (10 pb) pair-production cross sections at the LHC.  These 
particles will then decay to the dark matter particle, producting complex
events with several hard jets and missing transverse momentum.  These
mild assumptions thus lead to the conclusion, from any model that follows
this general line of argument, that {\it we should expect
exotic events with multiple jets and missing transverse momentum
to appear with pb cross sections at the LHC}.

\subsection{Dark Matter Annihilation in the MSSM}

This argument of the previous section 
gives a very optimistic conclusion for the discovery of new 
physics at the LHC.  However, we have already discussed that the 
first observation of supersymmetry or another model of new physics will 
only be
the first step in a lengthy experimental program.  Once we know that 
superparticles
or other new particles exist, we will need to study them in detail to learn
their detailed interactions and, eventually, to work out the underlying 
Lagrangian that governs their behavior.  As we have already discussed in 
Section 3.5 and 4.3, this Lagrangian can give us a clue to the nature of the 
ultimate theory at very short distances.

The study of dark matter intersects this program in an interesting way. 
In principle, once we have discovered supersymmetric particles, we can try to 
measure their properties and see if these coincide with the properties required
from astrophysical detections of dark matter. As we have seen in Section 5.3, 
the LHC experiments expect to measure the mass of the LSP to about 10\% 
accuracy.  These measurements can hopefully be compared to mass 
measurements at 
the 20\% level that can be expected from astrophysical dark matter detection 
experiments~\cite{mingamma,minCDMS}.   We would also wish to find 
out whether the 
annihilation cross section $\VEV{\sigma v}$
 that is predicted from the supersymmetry
parameters measured at colliders agrees with the value 
\leqn{sigmavvalue} required
to predict the observed WIMP relic density.  This comparison 
turns out to depend
in a complex way on the parameters of the underlying supersymmetry theory.

\begin{figure}
\begin{center}
\includegraphics[height=1.0in]{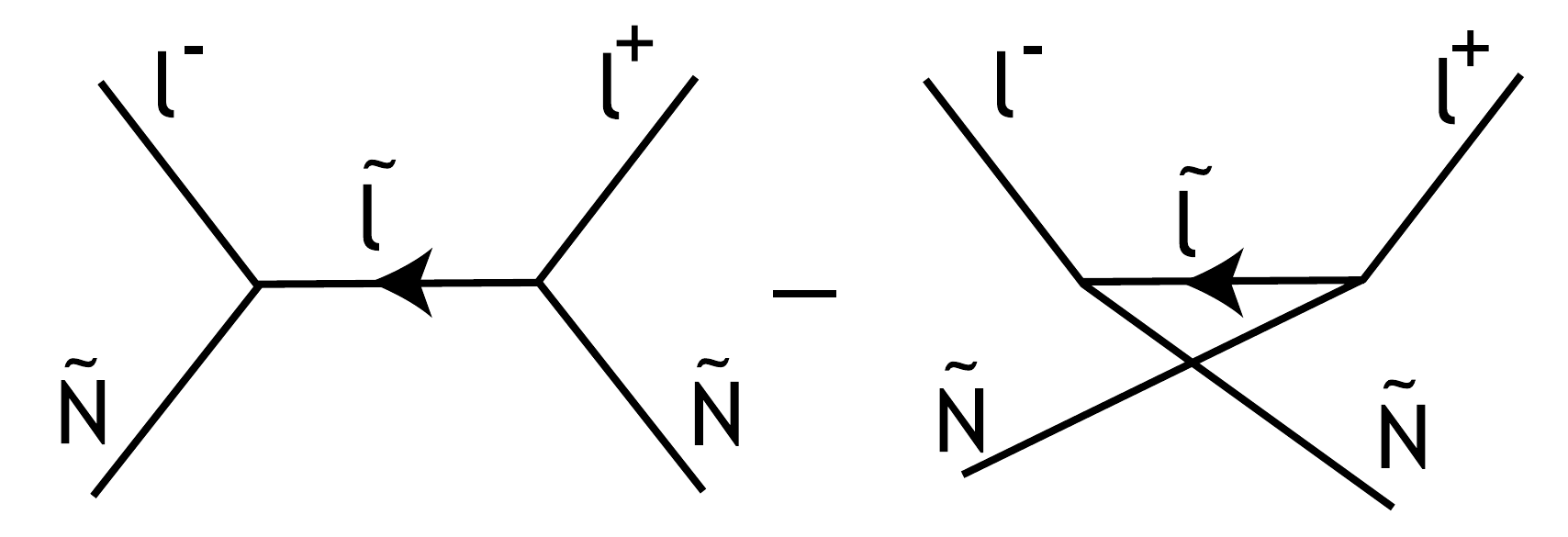}
\caption{Diagrams giving the simplest scheme of neutralino pair annihilation, 
leading to the annihilation cross section \leqn{sigmavvalsl}.}
\label{fig:sleptonannih}
\end{center}
\end{figure}

To begin our discussion of the annihilation cross section, we can make a
simple model of neutralino annihilation and see how well it works.
We have seen in Section~4.3 that the right-handed 
sleptons are often the lightest 
charged particles in the supersymmetry spectrum.  Consider, then, an
idealized parameter set in which the neutralino is a pure bino and
pair annihilation is dominated by the slepton exchange diagrams shown in 
Fig.~\ref{fig:sleptonannih}. (Away from the pure bino case, there are
also s-channel diagrams with $Z^0$, $h^0$, $H^0$, $A^0$.)
In this special limit,  the annihilation cross section is 
given by 
\beq
 v {d\sigma\over d \cos\theta} = {\pi \alpha^2} m_N^2   
   \left|{ 1\over c_w}\right|^2\ \left| {1\over m_{\s \ell}^2 -t } - 
  {1\over m_{\s\ell}^2 - u } \right|^2 \ ,
\eeq{sigmavvalsl}
where $m_N$ is the $\s N_1$ mass.
The relative velocity $v$ appears due to the flux factor in the 
cross section; this factor cancels in $\sigma v$.
I have ignored the lepton masses.  This expression is of the order of
\leqn{sigmatoM} with $M \sim m_N$, except for one unfortunate feature:  
At threshold,  $t = u$ and the cross section vanishes.  This leads to a 
severe suppression, by a factor of 
\beq
    v^2 \cdot \left|  {m_N^2 \over m_{\s\ell}^2 + m_N^2} \right|^4 \ ,
\eeq{NNllsuppression}
which is at least of order $\xi_f/16 $.  So the relic density estimated
in this simple way is too large by about a factor of 10.

\begin{figure}
\begin{center}
\includegraphics[height=0.7in]{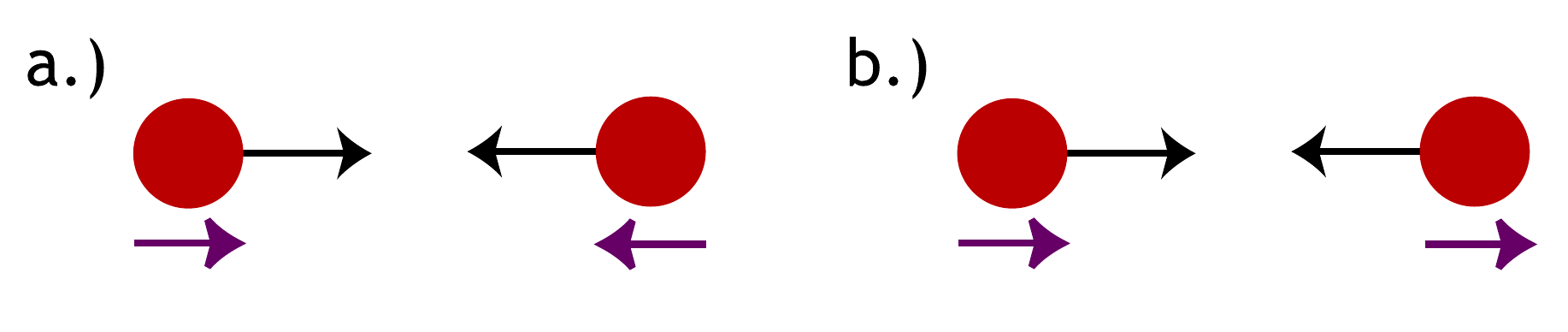}
\caption{Two possible spin configurations for neutralino annihilation: (a.)
       spin 0; (b.) spin 1.  Because of Fermi statistics, the latter 
      state does not exist in the S-wave.}
\label{fig:Nspinstates}
\end{center}
\end{figure}

There is an interesting physics explanation for the vanishing of this 
cross section at threshold~\cite{Goldberg}.   Neutralinos are spin-$\half$
fermions, and we might guess from this that, near threshold,
 they would annihilate in the 
S-wave either in a spin~0 or in a spin~1 state.  The two spin 
configurations are shown in Fig.~\ref{fig:Nspinstates}.   However,
because the neutralino is a Majorana fermion and therefore its own
antiparticle, an S-wave state of two neutralinos must be antisymmetric
in spin.  Hence, the spin~1 S-wave state does not exist
 However, as we know from pion decay, a spin~0 state can 
convert to a pair of light leptons only with a helicity flip.
Thus, there is an annihilation cross section from the spin~0 S-wave
only when lepton masses are included, and even then with 
the suppression factor 
$m^2_\ell/m^2_N$, which is $10^{-4}$ even for $\tau^+\tau^-$ final states.

\begin{figure}
\begin{center}
\includegraphics[height=1.5in]{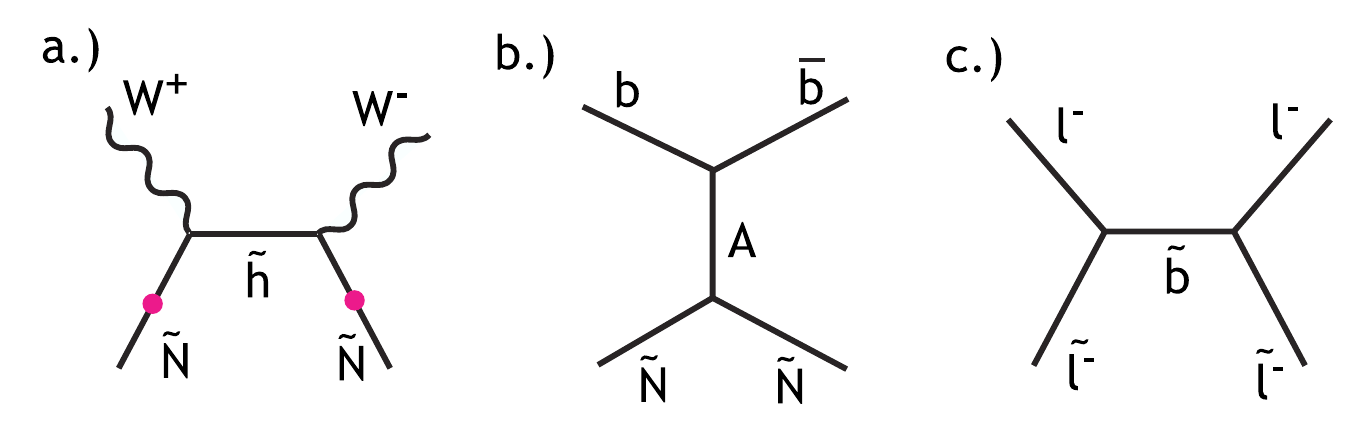}
\caption{Three mechanisms for obtaining a sufficiently large annihilation
 cross section to give the observed density of neutralino dark matter:
  (a.) gaugino-Higgsino mixing, opening the annihilation channels to 
       $W^+W^-$ and $Z^0Z^0$, (b.) resonance annihilation through the 
      Higgs boson $A^0$, (c.) co-annihilation with another supersymmetric
       particle, here taken to be a $\s \ell$.}
\label{fig:threeways}
\end{center}
\end{figure}

To obtain a realistic value for the neutralino relic density, we have to 
bring in more complicated mechanisms of neutralino annihilation.
These mechanisms are not difficult to find in various regions of the 
large supersymmetry paramet
er space~\cite{EllisOlive,Gondolo,Baer}.  We need 
to look for annihilation processes that can proceed in the S-wave with 
full strength.  Three possible mechanisms are shown in 
Fig.~\ref{fig:threeways}.

Pairs of neutralinos can annihilate in the S-wave into vector bosons.  The
bino does not couple to $W$ or $Z$ pairs, but if the lightest neutralino
has Higgsino or wino content, this reaction can be important. For 
charginos of mass about 200 GeV, this annihilation cross section can be
50 pb for a pure wino or Higgsino, so only a modest content of these 
states is needed to give a cross section of 1 pb.  

The s-channel
exchange of a Higgs boson can provide a mechanism for neutralino annihilation
in the spin~0 S-wave.  Because this state is CP-odd, it is the boson $A^0$
that is relevant here. If $m_A$ is close to the neutralino threshold $2m_N$,
the cross section has a resonant enhancement. 
Note that the $\s N_1$ annihilation vertex to $A$ arises as a 
Higgs-Higgsino-gaugino Yukawa term, so this vertex is nonzero only if  
 $\s N_1$ has both gaugino and Higgsino 
content.  If $m_A = 2 m_N$, the resonance enhancement is at full 
strength and the cross section can be as large as 50 pb.  Thus, it is 
$A$ boson masses about 20 GeV above or below the threshold that give the
desired
cross section \leqn{sigmavvalue}.

The final mechanism shown in the figure is {\it coannihilation}. As we have
discussed, the freezeout of the $\s N_1$ occurs at a temperature given 
by $T/m_N \sim 1/25$.  So if there is another particle in the supersymmetry 
spectrum that is within 4\% of the $\s N_1$ mass, this state will have 
a number density that remains in equilibrium with the number density of
the $\s N_1$.  If this particle has S-wave annihilation reactions, those
reactions can be the dominant mechanisms for the annihilation of supersymmetric
particles.  For a light slepton, the reactions
\beq
       \s \ell^- + \s N_1^0 \to \ell^- + \gamma\ , 
\qquad \s\ell^- + \s\ell^- \to   \ell^- + \ell^-
\eeq{extraannihs}
can give significant S-wave annihilation. In~\cite{EllisOlive,TAMU}, the 
lighter stau is invoked as the coannihilating particle.  In~\cite{FNALstop},
the lighter top squark is invoked as the coannihilating state.   If the 
lightest neutralinos and charginos are Higgsino-like, chargino coannihilation
can also be important.

It is, then, a complex matter to predict the neutralino relic density 
from microscopic physics.   We will first need to learn what particles in 
the supersymmetry spectrum play the dominant role as particle exchanged in 
annihilation reactions or as coannihilating species.  We will then need 
to measure the couplings and mixing angles of the important particles, since
the dominant annihilation diagrams depend sensitively on these.

\begin{figure}
\begin{center}
\includegraphics[height=3.0in]{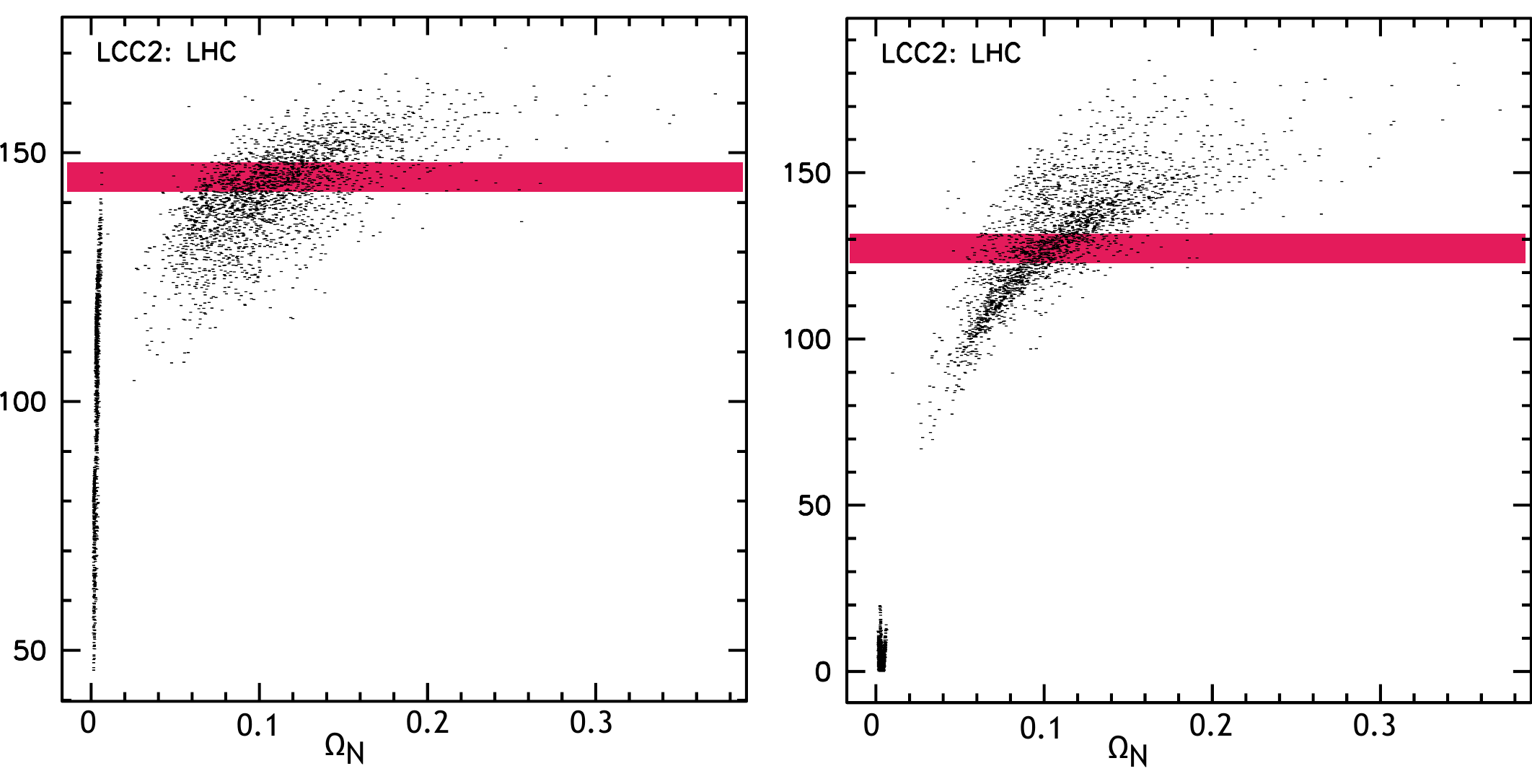}
\caption{Scatter plot of SUSY parameter points consistent with data from 
   the LHC in the analysis of the parameter set LCC2 from~\cite{Baltz}.
 The horizontal axis show the value of $\Omega_N$ at each parameter point.
   The vertical axes show polarized-beam cross sections measurable at the ILC,
in fb:
   (a.) $\sigma(e^-_Re^+_L\to \s C^+_1 \s C^-_1$), (b.) 
  $\sigma(e^-_Re^+_L\to \s N^0_2 \s N^0_3$).  The colored bands show the 
  $\pm 1 \sigma$ region allowed after the ILC cross section measurements.}
\label{fig:LCCtwocss}
\end{center}
\end{figure}

\begin{figure}
\begin{center}
\includegraphics[height=3.0in]{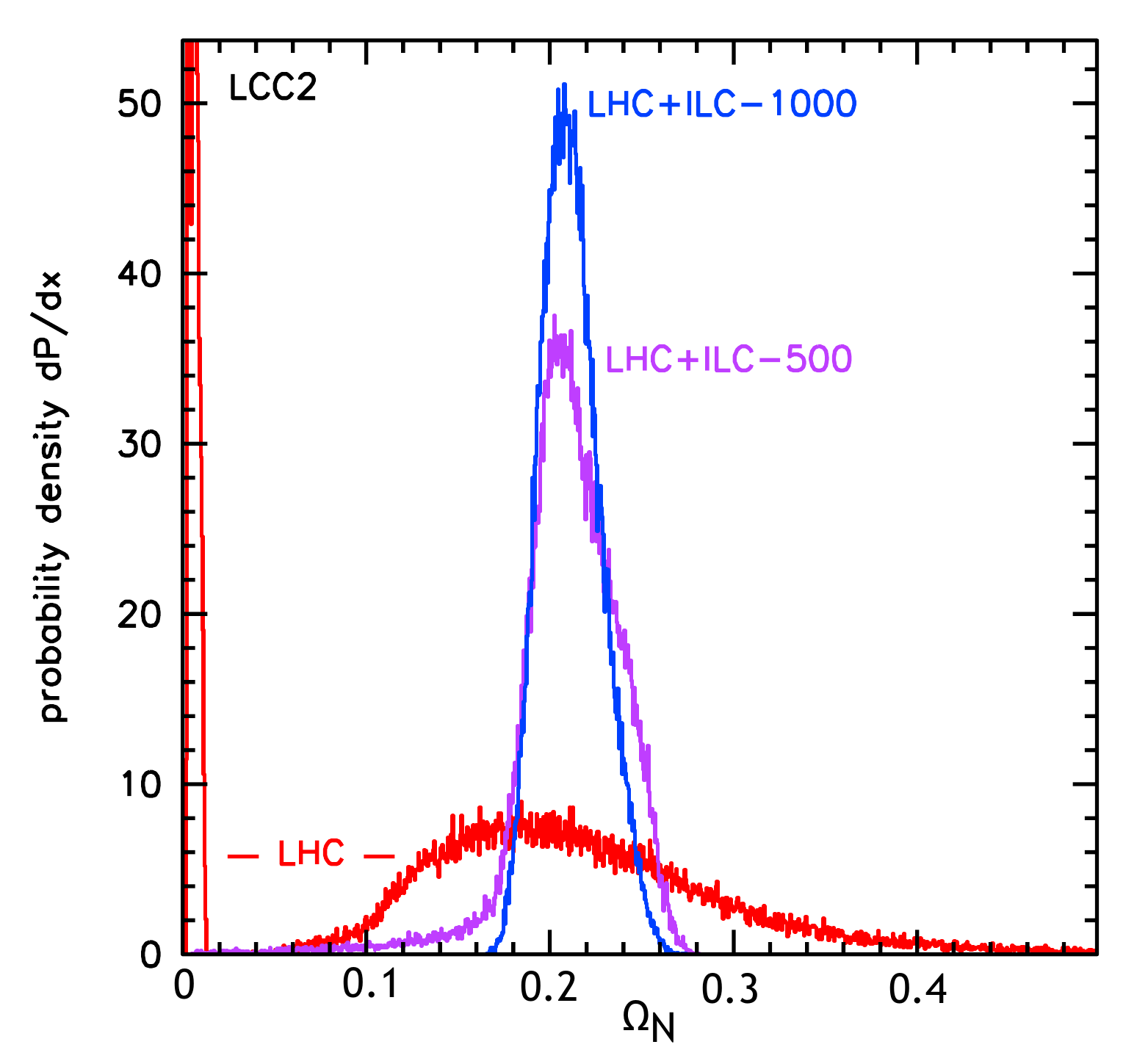}
\caption{Summary plot for the prediction of $\Omega_N$ from collider data
   for the SUSY parameter set LCC2 considered in~\cite{Baltz}.  The three
    curves show the likelihood distributions for the prediction of $\Omega_N$
     using data from the LHC, the ILC at 500~GeV, and the ILC at 1000~GeV.}
\label{fig:LCCtwosummary}
\end{center}
\end{figure}

Some examples of how measurements at the LHC and ILC can accumulate the 
relevant information are described in~\cite{Baltz}.  
Figure~\ref{fig:LCCtwocss}  shows a part of the analysis of this paper for 
a particular SUSY model in which the dominant annihilation reactions
are $\s N_1 \s N_1 \to W^+W^-, Z^0 Z^0$. As a first step, the authors 
constructed numerous supersymmetry parameter sets that were consistent 
with the mass spectrum of this model as it would be measured at the 
LHC.  These parameter sets included a variety of models in which the 
LSP was dominantly  bino and wino.   The figure shows  scatter
plots of the predictions of these models with ILC cross sections for 
neutralino and chargino pair production on the vertical axis and $\Omega_N$
on the horizontal axis.  The two cross sections clearly separate the 
bino- and wino-like solutions.  The second of these cross sections is the
polarized reaction of chargino pair production for which the cross section 
is displayed in Fig.~\ref{fig:CCmixing}.   The horizontal lines 
represent the accuracy of the measurements of these cross sections 
expected at the ILC.  These measurements select the bino solution and 
also play an important role in fixing the bino-Higgsino mixing angle
which is a crucial input to the annihilation cross sections. 
In Fig.~\ref{fig:LCCtwosummary}, I show the distribution of predictions
for $\Omega_N$ expected for this model, in the analysis of \cite{Baltz}, 
from the  data on SUSY particles that would be obtained from the LHC, 
from the ILC at a center-of-mass energy of 500 GeV, and from the ILC 
at a center-of-mass energy of 1000 GeV.


\begin{figure}
\begin{center}
\includegraphics[height=3.0in]{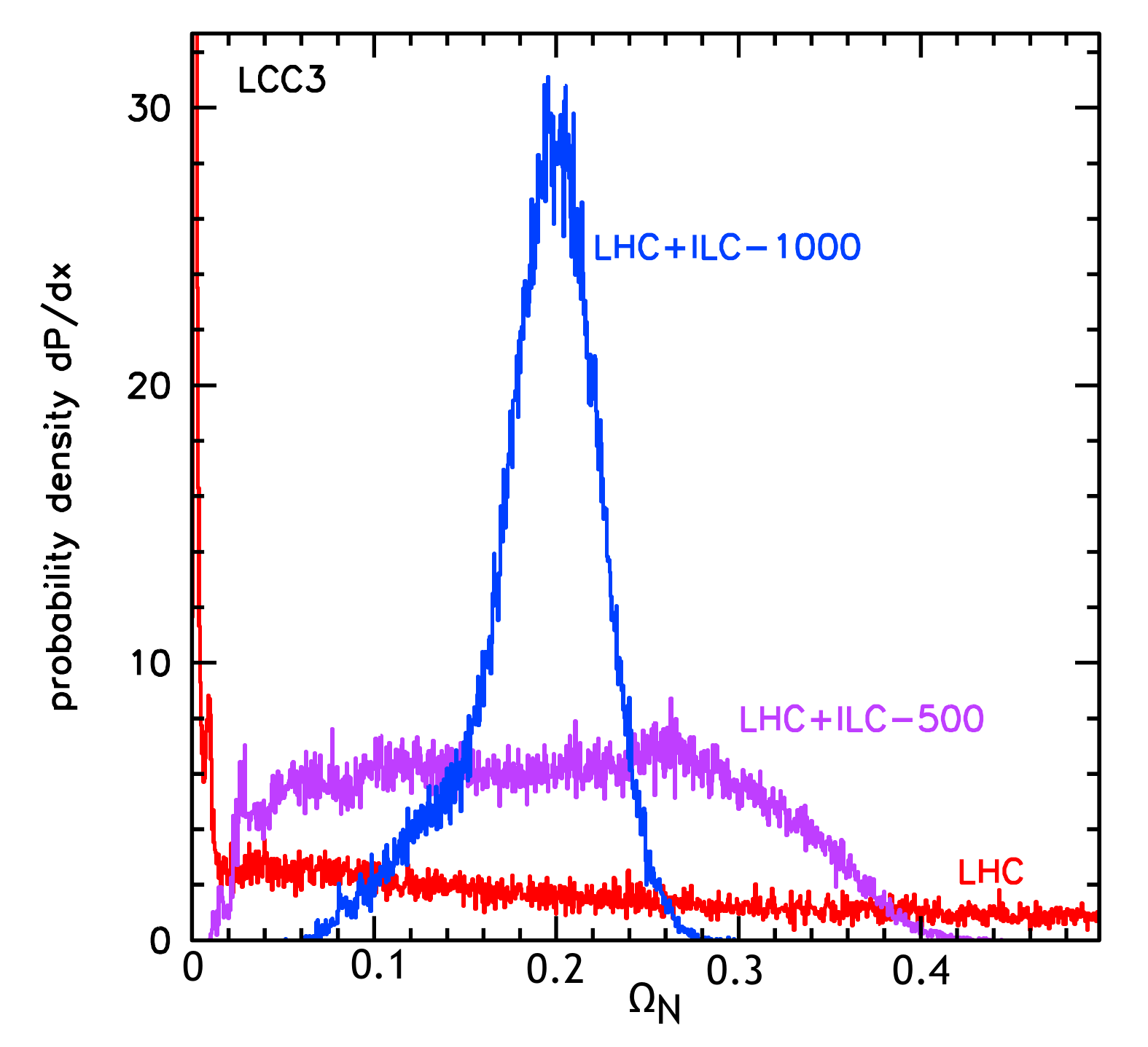}
\caption{Summary plot for the prediction of $\Omega_N$ from collider data
   for the SUSY parameter set LCC3 of~\cite{Baltz}. The notation is as 
  in Fig.~\ref{fig:LCCtwosummary}.}
\label{fig:LCCthree}
\end{center}
\end{figure}

The similar summary plot for another of the models considered in 
\cite{Baltz} is shown in Fig.~\ref{fig:LCCthree}.  The model considered
in this analysis is one in which the neutralino relic density is set by 
stau coannihilation.  In this model, the stau would be discovered at the 
LHC, and the stau-neutralino mass difference would be measured to 
about 10\% accuracy at the 500 GeV ILC.  However, the annihilation reactions
also depend on mixing angles and on the value of $\tan\beta$.  In this 
scenario, these are determined only by ILC measurements of some of the 
heavier states of the SUSY spectrum.

Collider measurements of the SUSY spectrum can also be used to constrain 
cross sections of the WIMP that are important for experiments that seek to
detect dark matter, for example, the neutralino-proton cross section and 
the cross section for neutralino pair annihilation to gamma rays.  If we 
can accurately predict these cross sections from collider data, the 
information about the SUSY spectrum that we learn from colliders will feed
back into the astrophysics of dark matter.   Some numerical examples that 
illustrate this are presented in \cite{Baltz}.

\section{Conclusions}

In these lectures, I have given an overview of supersymmetry and its
application to elementary particle physics.    In the early sections of 
this review, I presented the formalism of SUSY and explained the rules for 
constructing supersymmetric Lagrangians.  Our discussion then became more 
concrete, focusing on the mass spectrum of the MSSM and the properties
of the particle states of the MSSM spectrum.  This led us to a discussion
of the experimental probes of this spectrum and the possibility of 
measurement of the 
parameters of the supersymmetric Lagrangian.

This possibility is now coming very near.   As I have discussed in the 
last sections of this review, supersymmetry gives concrete answers
to the major questions about elementary particle physics that we expect
to be addressed at the hundred GeV scale---the questions of the origin
of electroweak symmetry breaking and the identity of cosmic dark matter.
In the next year, the LHC will begin to explore the physics of this 
mass scale.   Supersymmetry is one candidate for what will be found.  I 
hope that, after studying these lectures, you will agree that the 
picture provided by supersymmetry is highly plausible and even compelling.

Whatever explanations we will learn from the LHC data,  our investigation 
of it will follow the general paradigm that I have described here.
In successive stages, we will use data from the LHC and the ILC to learn 
the mass spectrum of new particles that are revealed at the LHC, to 
determine their quantum numbers and couplings, and to reconstruct their
underlying Lagrangian. On the basis of the detailed studies of this 
program that have been carried out for the MSSM, we have the expectation
that we will be able to learn the underlying theory of the new particles
and to test the specific explanations that this theory gives for 
the mysteries of the  fundamental interactions.

Is supersymmetry just an attractive theory, or is it a part of the 
true description of elementary particles?  We are about to find out.

\Acknowledgements

I am grateful to 
 Sally Dawson, Rabi Mohapatra and, especially, to K.~T.~Mahanthappa for 
organizing the 2006 TASI Summer School at which these lectures were presented.
I thank Howard Haber and Thomas Dumitrescu
 for instructive comments on the manuscript.
This work was supported by the US Department of Energy under
                     contract DE--AC02--76SF00515.

\end{document}